\documentstyle[11pt]{article}
\makeatletter \@addtoreset{equation}{section} \makeatother

\newcommand{\noi}{\vspace{12pt}\noindent}
\newcommand{\beq}{\begin{equation}}
\newcommand{\eeq}{\end{equation}}
\newcommand{\bea}{\begin{eqnarray}}
\newcommand{\eea}{\end{eqnarray}}

\newcommand{\e}[1]{{(\ref{#1})}}
\newcommand{\eq}[1]{{eq.\ (\ref{#1})}}
\newcommand{\es}[2]{{(\ref{#1}) and (\ref{#2})}}
\newcommand{\eqs}[2]{{eqs.\ (\ref{#1}) and (\ref{#2})}}
\newcommand{\Ref}[1]{{Ref.~\cite{#1}}}
\newcommand{\Refs}[2]{{Refs.~\cite{#1} and \cite{#2}}}
\newcommand{\mb}[1]{{\mbox{${#1}$}}}

\newcommand{\ie}{{${ i.e.\ }$}}
\newcommand{\eg}{{${ e.g.\ }$}}

\newcommand{\cf}{{cf.\ }}
\newcommand{\wrt}{{with respect to }}
\newcommand{\wtho}{{with the help of }}
\newcommand{\lhs}{{left-hand side }}
\newcommand{\rhs}{{right-hand side }}
\newcommand{\aka}{{also known as }}

\newcommand{\omegazero}{\Omega_{0}} 
\newcommand{\bomegazero}{\bar{\Omega}_{0}}
\newcommand{\omegadelta}{\Omega}
\newcommand{\bomegadelta}{\bar{\Omega}}
\newcommand{\omegaone}{\Omega_{1}}
\newcommand{\omegaonemin}{\Omega_{1,\min}}
\newcommand{\tomegaone}{\tilde{\Omega}_{1}}
\newcommand{\gbb}{{\cal G}} %%%%{{G_{\B\B^{*}}}}
\newcommand{\orig}{{\rm or}}
\newcommand{\brLambda}{\breve{\Lambda}}

\newcommand{\B}{B}
\newcommand{\cB}{{\cal C}}
\newcommand{\bcB}{\bar{\cal C}}
\newcommand{\tcB}{\tilde{\cal C}}

\newcommand{\C}{C}

\newcommand{\bC}{\bar{C}}
\newcommand{\bP}{\bar{P}}
\newcommand{\bcP}{\bar{\cal P}}
\newcommand{\cP}{{\cal P}}
\newcommand{\tbcP}{\tilde{\bar{\cal P}}{}}

\newcommand{\cE}{{\cal E}}
\newcommand{\tcE}{\tilde{\cal E}}
\newcommand{\tcF}{{\cal F}}
\newcommand{\brG}{\breve{G}{}}
\newcommand{\cH}{{\cal H}}

\newcommand{\cQ}{{\cal Q}}
\newcommand{\cR}{{\cal R}}
\newcommand{\tcR}{\tilde{\cal R}}
\newcommand{\tcS}{\tilde{\cal S}}

\newcommand{\cI}{{\cal I}}
\newcommand{\eI}{I}
\newcommand{\cM}{{\cal M}}
\newcommand{\eM}{M}
\newcommand{\cS}{{\cal S}}
\newcommand{\eS}{S}

\newcommand{\T}{T}
\newcommand{\bT}{\bar{T}}
\newcommand{\cT}{{\cal T}}

\newcommand{\brcT}{\breve{\cal T}}
\newcommand{\tcT}{\tilde{\cal T}}

\newcommand{\U}{U}
\newcommand{\bU}{\bar{U}}
\newcommand{\cU}{{\cal U}}
\newcommand{\brcU}{\breve{\cal U}}
\newcommand{\tcU}{\tilde{\cal U}}

\newcommand{\cV}{{\cal V}}

\newcommand{\X}{X}
\newcommand{\cX}{{\cal X}}

\newcommand{\tcX}{\tilde{\cal X}}

\newcommand{\Y}{Y}

\newcommand{\tcY}{\tilde{\cal Y}}

\newcommand{\Z}{Z}
\newcommand{\bZ}{\bar{Z}}
\newcommand{\cZ}{{\cal Z}}

\newcommand{\brcZ}{\breve{\cal Z}}
\newcommand{\tcZ}{\tilde{\cal Z}}

\newcommand{\eps}{\varepsilon^{}}
\newcommand{\plus}{{{+}^{}}}
\newcommand{\m}{{m}^{}}

\newcommand{\gh}{{\rm gh}}
\newcommand{\ngh}{{\rm ngh}}

\newcommand{\sgn}{{\rm sgn}}

\newcommand{\rank}{{\rm rank}}

\newcommand{\Hf}{{1 \over 2}}

\newcommand{\Ih}{{i \over \hbar}}

\newcommand{\twobyone}[2]{\left[\begin{array}{c}{#1} \cr
                                {#2} \end{array} \right]}
\newcommand{\onebytwo}[2]{\left[\begin{array}{ccc}{#1}&{#2}
                                        \end{array} \right]}
\newcommand{\twobytwo}[4]{\left[\begin{array}{ccc}{#1}&{#2} \cr
                                  {#3} & {#4} \end{array} \right]}

\newcommand{\deder}[1]{{ 
 {\stackrel{\raise.1ex\hbox{$\leftarrow$}}{\delta^r}   } 
\over {   \delta {#1}}  }}
\newcommand{\dedel}[1]{{ 
 {\stackrel{\lower.3ex \hbox{$\rightarrow$}}{\delta^{\ell}}   }
 \over {   \delta {#1}}  }}
\newcommand{\papar}[1]{{ 
 {\stackrel{\raise.1ex\hbox{$\leftarrow$}}{\partial^r}   } 
\over {   \partial {#1}}  }}
\newcommand{\papal}[1]{{ 
 {\stackrel{\lower.3ex \hbox{$\rightarrow$}}{\partial^{\ell}}   }
 \over {   \partial {#1}}  }}
\newcommand{\ddr}[1]{{ 
 {\stackrel{\raise.1ex\hbox{$\leftarrow$}}{\delta^r}   } 
\over {   \delta {#1}}  }}
\newcommand{\ddl}[1]{{ 
 {\stackrel{\lower.3ex \hbox{$\rightarrow$}}{\delta^{\ell}}   }
 \over {   \delta {#1}}  }}

\newcommand{\proofbox}{\begin{flushright}
${\,\lower0.9pt\vbox{\hrule \hbox{\vrule
height 0.2 cm \hskip 0.2 cm \vrule height 0.2 cm}\hrule}\,}$
\end{flushright}}

\begin{document}
\thispagestyle{empty}
\title{\Large{\bf Reducible Gauge Algebra of BRST-Invariant Constraints}}
\author{{\sc I.A.~Batalin}$^{\dag}$\\I.E.~Tamm Theory Division\\
P.N.~Lebedev Physics Institute\\Russian Academy of Sciences\\
53 Leninisky Prospect\\Moscow 119991\\Russia\\~\\and\\~\\
{\sc K.~Bering}$^{\ddag}$\\Institute for Theoretical Physics \& Astrophysics
\\Masaryk University\\Kotl\'a\v{r}sk\'a 2\\CZ-611 37 Brno\\Czech Republic}
\maketitle
\vfill
\begin{abstract}
We show that it is possible to formulate the most general first-class gauge
algebra of the operator formalism by only using BRST-invariant constraints. In
particular, we extend a previous construction for irreducible gauge algebras
to the reducible case. The gauge algebra induces two nilpotent, Grassmann-odd,
mutually anticommuting BRST operators that bear structural similarities with 
BRST/anti-BRST theories but with shifted ghost number assignments. In both
cases we show how the extended BRST algebra can be encoded into an operator
master equation. A unitarizing Hamiltonian that respects the two BRST
symmetries is constructed with the help of a gauge-fixing Boson. Abelian
reducible theories are shown explicitly in full detail, while non-Abelian
theories are worked out for the lowest reducibility stages and ghost momentum
ranks.
\end{abstract}

\vfill

\begin{quote}
PACS number(s): 04.60.Ds; 11.10.-z; 11.15.-q. 
\hrule width 5.cm \vskip 2.mm \noindent 
$^{\dag}${\small E-mail:~{\tt batalin@lpi.ru}} \hspace{10mm}
$^{\ddag}${\small E-mail:~{\tt bering@physics.muni.cz}} \\ 
\end{quote}

\newpage

\tableofcontents

\section{Introduction}
\label{secintro}

\noi
In the quantization of a classical system with first-class constraints
\mb{\T_{\alpha}},
\beq
 \{\T_{\alpha}, \T_{\beta}\}_{PB}^{}
~=~\U_{\alpha\beta}{}^{\gamma}\T_{\gamma}~,
\label{involutionrel}
\eeq
it is well-known that there may appear quantum anomalies in the expression for
the operator commutator \mb{[\T_{\alpha},\T_{\beta}]} at quadratic (or higher)
orders in \mb{\hbar}.  A useful approach to circumvent this obstacle, is, to
replace the initial constraints \mb{\T_{\alpha}} with BRST-invariant
constraints \mb{\cT_{\alpha}}. In practice, it turns out that the BRST
invariance of \mb{\cT_{\alpha}} protects the commutator
\mb{[\cT_{\alpha},\cT_{\beta}]} against anomalies. Our goal is to investigate
this fact systematically in a general setting. {}For irreducible first class
constraint such an investigation is undertaken in \Ref{BatTyu03}. Here we will
develop that construction and extend it to reducible gauge algebras.

\noi
The use of BRST-invariant constraints is best illustrated by a motivating
example. Consider the critical open Bosonic string theory in \mb{D=26}
dimensions with an intercept/normal-ordering constant \mb{a=1}, \cf
\Ref{Hwa83}, \Ref{GreSchWit87} and Appendix~\ref{appbosonstring}. The matter
Virasoro generators
\beq
\T_{m}~=~\Hf \eta_{\mu\nu}\sum_{n}:\alpha_{m-n}^{\mu}\alpha_{n}^{\nu}:
-\hbar a\delta_{m}^{0}
\label{tvirasoroconstraints}
\eeq
form a Virasoro algebra with non-zero central extension
\bea
[\T_{m},\T_{n}]&=&\hbar(m\!-\!n)\T_{m+n}+\hbar^{2}A_{m}\delta_{m+n}^{0}~, \\
A_{m}&\equiv&\frac{Dm(m^{2}-1)}{12}+2ma~.
\eea
The remedy is well-known: When one adds the appropriate ghost contributions 
\mb{\T_{m}^{(c)}} to the Virasoro constraints,
\beq
\cT_{m}~=~\T_{m}+\T_{m}^{(c)}~,~~~~~~~~~~~~
\T_{m}^{(c)}~=~\sum_{n}(m\!-\!n):b_{m+n}c_{-n}:
\label{ctvirasoroconstraints}
\eeq
the new constraints \mb{\cT_{m}} obey a Virasoro algebra with no central
extension, \aka a Witt algebra, precisely for \mb{D=26} and \mb{a=1},
\bea
 [\cT_{m},\cT_{n}]&=&\hbar(m\!-\!n)\cT_{m+n}
+\hbar^{2}{\cal A}_{m}\delta_{m+n}^{0}~, \\
{\cal A}_{m}&\equiv&\frac{(D-26)m^{3}+(24a+2-D)m}{12}~=~0~.
\label{twentysixdim}
\eea
How does the BRST formulation \cite{BecRouSto74,Tyu75} fit into this? 
Surprisingly, the new anomaly-free constraints \mb{\cT_{m}} happen to be
BRST-superpartners of the ghosts momenta \mb{b_{m}},
\beq
\cT_{m}~=~\frac{1}{\hbar}[b_{m},\omegazero]~,
\label{brstinvconstr}
\eeq
where \mb{\omegazero} denotes the BRST operator.
As a simple consequence of \eq{brstinvconstr}, the new constraints 
\mb{\cT_{m}} are BRST-invariant,
\beq
 [\cT_{m},\omegazero]~=~\frac{1}{\hbar}[[b_{m},\omegazero],\omegazero]
~=~\frac{1}{2\hbar}[b_{m},[\omegazero,\omegazero]]~=~0~.
\label{ctmbrstinv}
\eeq
Here we argue that \eq{brstinvconstr} should be viewed as the rule rather than
the exception. {}First of all, \eq{brstinvconstr} respects ghost number
conservation, since \mb{\omegazero}, \mb{\cT_{m}} and \mb{b_{m}} carry ghost
number \mb{1}, \mb{0} and \mb{-1}, respectively. In general, ghost number
conservation severely restricts what can happen.
Secondly, it follows from \eq{brstinvconstr}, using quite broad assumptions,
that the \mb{[\cT_{m},\cT_{n}]} commutator cannot develop a quantum anomaly,
which is a strong hint. {}For instances, guided by ghost number conservation, 
it is reasonable to expect that the commutator \mb{[\cT_{m},b_{n}]} between
the BRST-invariant constraints \mb{\cT_{m}} and the ghost momenta \mb{b_{n}}
again is proportional to the ghost momenta. In detail, let us assume that
\beq
[\cT_{m},b_{n}]~=~\hbar \sum_{p}\U_{mn}{}^{p}b_{p}
\label{jubiiansatz}
\eeq
for some BRST-invariant structure functions \mb{\U_{mn}{}^{p}}.
Then the proof goes as follows:
\beq
 [\cT_{m},\cT_{n}]~=~\frac{1}{\hbar}[\cT_{m},[b_{n},\omegazero]]
~=~\frac{1}{\hbar}[[\cT_{m},b_{n}],\omegazero]
~=~\sum_{p}\U_{mn}{}^{p}[b_{p},\omegazero]
~=~\hbar \sum_{p}\U_{mn}{}^{p}\cT_{p}~,
\eeq
\ie there are no terms of order \mb{{\cal O}(\hbar^{2})}.
Therefore the \eqs{brstinvconstr}{jubiiansatz} imply that the 
\mb{[\cT_{m},\cT_{n}]} commutator is anomaly-free. Such an assurance is a rare
commodity in a full-fledged quantum theory, and this is why we would like to
systematically seek for relationships of the form \e{brstinvconstr}.
(In string theory much more is known: The ghost momentum
\mb{b(z)=\sum_{m}b_{m}z^{-m-2}} is a primary field of conformal weight \mb{2},
which means that \mb{[\cT_{m},b_{n}]=\hbar(m\!-\!n)b_{m+n}}, thereby
confirming the Ansatz \e{jubiiansatz} with
\mb{\U_{mn}{}^{p}=(m\!-\!n)\delta_{n+m}^{p}}. However, we emphasize the
versatility of the argument.) Later in the construction a new 
generation of ghosts \mb{\cB^{A}} is introduced, and the BRST-doublets 
\beq
\cT_{A}~=~\{\cT_{m};b_{m}\}
\label{brstinvnewconstraintsintro}
\eeq
are turned into twice as many first-class constraints for a new BRST operator 
called \mb{\omegaone\!=\!\cB^{A}\cT_{A}+\ldots}, \cf
\eq{newconstraintssupersplit} below.

\noi
We stress that choosing BRST-invariant constraints
\mb{\cT_{\alpha}} is not a miraculous cure that turns anomalous theories into
anomaly-free ones. Rather it is a useful tool in overcoming a poor initial
choice of constraint basis \mb{\T_{\alpha}} for otherwise sound theories. 
{}For instance, in the case of critical open Bosonic strings,
the BRST formulation requires -- for starters -- that the BRST charge
\mb{\omegazero} is nilpotent,
\beq
0~=~[\omegazero,\omegazero]~=~
\sum_{m,n}([\cT_{m},\cT_{n}]
-\hbar(m\!-\!n)\cT_{m+n})c_{-m}c_{-n}
~=~\hbar^{2}\sum_{m}{\cal A}_{m}c_{-m}c_{m}~,
\label{bosonstringnilp}
\eeq
which by itself implies the familiar \mb{D=26} and \mb{a=1}, \cf 
\eqs{twentysixdim}{bosonstringomega}. In general, we shall simply assume as a
starting point that an anomaly-free nilpotent BRST operator \mb{\omegazero}
has been given. ({}For a typical physical model with infinitely many degrees of
freedom, it takes a fair amount of mathematical analysis to rigorously
establish this beyond the formal level \cite{FraVil75,BatVil77,FraFra78}.)

\noi
The paper is organized as follows. In Section~\ref{secoldbrst} the standard
Hamiltonian BRST construction
\cite{FraVil75,BatVil77,FraFra78,BatFra83a,BatFra83b} is reviewed.
In Section~\ref{secnewbrst} a new generation of ghosts and BRST symmetry is
introduced. An anti-BRST operator is used to infer cohomologically the
existence of an \mb{\cS} operator, which satisfies an operator master 
equation \cite{BatMar98a}, 
\beq
[\cS,[\cS,\omegadelta]]~=~(i\hbar)[\cS,\omegadelta]~, \label{omeintro}
\eeq
\cf \eq{ome} below. The \mb{\cS} operator contains the so-called
tilde constraints \mb{\tcT_{A}}, which are the analogues of the above string
ghost momenta \mb{b_{m}}, and which descend via the relation
\beq
\cT_{A_{0}}~=~\cV_{A_{0}}{}^{B_{0}}~\tcT_{B_{0}}
+\Ih [\tcT_{A_{0}},\omegazero]~,
\label{brstinvdoubleconstrintro}
\eeq
to the new \mb{\cT_{A}} constraints \e{brstinvnewconstraintsintro}, \cf
\eq{brstinvdoubleconstr} below. Eq.\ \e{brstinvdoubleconstrintro} generalizes
\eq{brstinvconstr} and is the heart of the construction. It guarantees that the
new \mb{\cT_{A_{0}}} constraints are BRST-invariant up to the first term on the
left-hand side, which depends on ghost momenta and vanishes in the unitary 
limit. In short, Section \ref{secnewbrst} presents the 
various pieces and definitions that go into the construction.  
In Section~\ref{secassemble} we assemble all the pieces and show how to
gauge-fix, \ie construct the unitarizing Hamiltonian. One of our main new 
points is that the construction shares strikingly many similarities with
BRST/anti-BRST symmetric models. In particular, it has two nilpotent,
Grassmann-odd, mutually anticommuting BRST operators, although their ghost
number assignments are different.
Here the first BRST operator \mb{\omegadelta} is a ghost-deformed version of
the ordinary BRST operator \mb{\omegazero}, and the second BRST operator is the
new BRST operator \mb{\omegaone}, associated with a next generation of ghosts
\mb{\cB^{A}}, \cf \eq{brstinvnewconstraintsintro}. 
Similar to BRST/anti-BRST theories, the gauge-fixing will depend on a
gauge-fixing Boson, and the unitarizing Hamiltonian will respect both BRST 
symmetries. 

\noi
In Section~\ref{secabel} the Abelian case is treated in full details. The
Abelian case, which besides being an important example in its own right,
establishes at the theoretical level the existence of the whole construction.
Later, in Section~\ref{secalgconstr} we discuss algebraic properties of the
constraints. Section~\ref{secconcl} briefly states our conclusions. 
The paper also includes five appendices: Appendix~\ref{appbosonstring}
gives an elementary derivation of the conformal anomaly for the critical
open Bosonic string. Appendix~\ref{appsuperfield} and Appendix~\ref{appmatrix}
discuss an optional superfield and matrix formulation, respectively.
The superfield formulation explores a (perfectly consistent) ghost number
deficit among the two superpartners of new constraints \mb{\cT_{A}}, as
already evident from \eq{brstinvnewconstraintsintro}. {}Finally, 
Appendix~\ref{appome} analyzes possible candidates for the operator master 
\eq{omeintro}, and Appendix~\ref{apptomegaone} reformulates certain additional 
involution relations as a nilpotency condition for a BRST operator.

\subsection{Operator Ordering}

\noi
String theory is most often formulated using elementary Fock space creation and
annihilation operators, \mb{\alpha_{m}^{\mu}}, \mb{b_{m}} and \mb{c_{m}}, also
referred to as Wick operators. The ordering prescription is Wick (=normal)
ordering, \cf \eqs{bosonstringnormal1}{bosonstringnormal2}; and the Hermitian
conjugation is implemented through opposite modes \mb{m\leftrightarrow -m},
\beq
 (\alpha_{m}^{\mu})^{\dagger}~=~\alpha_{-m}^{\mu},~~~~~
b_{m}^{\dagger}~=~b_{-m}~,~~~~~c_{m}^{\dagger}~=~c_{-m}~.
\label{bosonstringherm1} 
\eeq
However for a general theory, one does not want to make unnecessary assumptions
about the index structure of the fundamental operators, and one would therefore
have to write the Hermitian conjugation ``\mb{\dagger}'' explicitly, thereby
producing quite lengthy formulas.
To avoid this, we take from now on the fundamental operators to be Hermitian or
anti-Hermitian, and the ordering prescription to be \mb{qp}-ordering, \ie the 
coordinates \mb{q} ordered to the left of the momenta \mb{p}.
(In the sectors of the theory that is not written out explicitly, one is of
course free to assume any ordering.)
{}For more on the Wick formalism and the string example, see 
Subsection~6.B of \Ref{BatFra88} and Subsection~6.2 of \Ref{BatTyu03}.

\subsection{Notation}

\noi
Square brackets
\beq
[A,B]~\equiv~AB-(-1)^{\eps_{A}\eps_{B}}BA~,~~~~~~~~~
[A,B]_\plus~\equiv~AB+(-1)^{\eps_{A}\eps_{B}}BA~,
\label{supercom}
\eeq
denote the supercommutator and the anti-supercommutator of two operators \mb{A}
and \mb{B} of Grassmann parities \mb{\eps_{A}} and \mb{\eps_{B}}.
As we already saw in the Introduction, it becomes quite tedious to write out 
every \mb{\hbar}, so from now on we shall let a curly bracket  
\beq
\{A,B\}~\equiv~\frac{1}{i\hbar}[A,B]
\label{normsupercom}
\eeq
denote a normalized supercommutator. As usual, the normalized commutator
\e{normsupercom} becomes a Poisson bracket \mb{\{A,B\}_{PB}^{}} of commuting
functions \mb{A} and \mb{B} in the classical limit \mb{\hbar\to 0} by the
well-known correspondence principle of quantum mechanics. However, we stress
that the normalized commutator \e{normsupercom} is an {\em operator} for
\mb{\hbar\neq 0}. The notation \e{normsupercom} enables us to 1) almost
eliminate appearances of \mb{i\hbar}, 2) provide a full-fledged quantum
operator formalism and 3) at the same time give classical expressions. 
Note that \mb{\{A;B\}} separated by a semicolon ``;'' instead of a comma 
``,'' will denote a set of two elements \mb{A} and \mb{B}.

\noi
As a rule of thumb, calligraphic letters are associated with the new ghost
sector, \cf Subsection~\ref{secngh}. A bar ``\mb{-}'' over a quantity refers to
negative ghost number, while a tilde ``\mb{\sim}'' (resp.\ breve
``\mb{\smile}'') over a quantity means that it is associated with \mb{\cS_{1}}
(resp.\ \mb{\cS_{2}}), \cf Subsection~\ref{secessone} (resp.\ \ref{secesstwo}).

\section{Old BRST Algebra}
\label{secoldbrst}

\noi
In this Section~\ref{secoldbrst} we review the construction of a general BRST
operator algebra \cite{FraVil75,BatVil77,FraFra78,BatFra83a,BatFra83b,KugOji79,
Nish84,Hen85,HenTei92} with the operator ordering prescription taken to be
\mb{qp}-ordering. The review is partially to fix notation and partially to
motivate later constructions. The BRST algebra will only serve as a starting
point, and we shall refer to it as the ``old'' or the ``ordinary'' system to
distinguish it from newer sectors to be added in later Sections.

\subsection{Old Ghosts}
\label{secoldghost}

\noi
Besides the original phase variables
\mb{(q^{i},p_{j})}, which almost always appear only implicitly in formulas,
the ordinary phase space consists of \mb{L\!+\!1} stages of ordinary ghosts
\beq
\C^{\alpha}~=~\{\C^{\alpha_{s}}|s=0,\ldots, L\}~,
\eeq
and their momenta
\beq
\bP_{\alpha}~=~\{\bP_{\alpha_{s}}|s=0,\ldots, L\}~,
\eeq
with canonical commutation relations
\beq
\{\C^{\alpha},\C^{\beta}\}~=~0~,~~~~
\{\C^{\alpha},\bP_{\beta}\}~=~\delta^{\alpha}_{\beta}~=~
(-1)^{\eps_{\beta}}\{\bP_{\beta},\C^{\alpha}\}~,~~~~
\{\bP_{\alpha},\bP_{\beta}\}~=~0~.
\label{ccr}
\eeq
Here the index \mb{\alpha_{s}} runs over the set \mb{\{1,\ldots, \m_{s}\}},
the subscript \mb{s} is associated with stage \mb{s\in\{0,\ldots, L\}}, 
while the index \mb{\alpha} is a shorthand for all the stages 
\mb{\alpha_{0},\ldots,\alpha_{L}}, \ie the index \mb{\alpha} runs over 
\mb{\{1,\ldots, \sum_{s=0}^{L}\m_{s}\}}, and so forth. 
The Grassmann parity and ghost number of the ordinary ghosts are
\beq
\begin{array}{rccclcrcccl}
\eps(\C^{\alpha})&=& \eps_{\alpha}\!+\!1&=&\eps(\bP_{\alpha})~,&&
\eps(\C^{\alpha_{s}})&=& \eps_{\alpha_{s}}\!+\!s\!+\!1
&=&\eps(\bP_{\alpha_{s}})~,  \\
&&&&&&\gh(\C^{\alpha_{s}})&=&s\!+\!1&=&-\gh(\bP_{\alpha_{s}})~.
\end{array}
\eeq

\subsection{Old BRST Operator}
\label{secoldcharge}

\noi
The ordinary BRST operator \mb{\omegazero\!=\!\omegazero(q,p;\C,\bP)} is 
Grassmann-odd, nilpotent and with ghost number \mb{1},
\beq
\{ \omegazero,\omegazero\} ~=~ 0~,~~~~~~
\eps(\omegazero)~=~1~,~~~~~~
\gh(\omegazero)~=~1~.
\label{oldbrstcharge}
\eeq
It is a power series expansion in the ordinary ghosts \mb{\C} and \mb{\bP}, 
and ordered according to the \mb{\C\bP}-ordering prescription,
\bea
\omegazero&=&\sum_{s=0}^{L}\C^{\alpha_{s}}\T_{\alpha_{s}}
+\Hf\!\!\!\!\!\!\!\!\sum_{\footnotesize
\begin{array}{c}r,s\geq 0\cr r\!+\!s\leq L\end{array}}
\!\!\!\!\!\!\!\!\C^{\beta_{s}}\C^{\alpha_{r}}~
\U_{\alpha_{r}\beta_{s}}{}^{\gamma_{r+s}}\bP_{\gamma_{r+s}}
(-1)^{\eps_{\beta_{s}}+\eps_{\gamma_{r+s}}+r} \cr
&&+\Hf\!\!\!\!\!\!\!\!\sum_{\footnotesize
\begin{array}{c}r,s\geq 0\cr r\!+\!s\!+\!2\leq L\end{array}}
\!\!\!\!\!\!\!\!\C^{\gamma_{r+s+2}}~
\U_{\gamma_{r+s+2}}{}^{\beta_{s}\alpha_{r}}
\bP_{\alpha_{r}}\bP_{\beta_{s}}(-1)^{\eps_{\alpha_{r}}+r}\cr
&&+\frac{1}{4}\!\!\!\!\!\!\!\!\sum_{\footnotesize
\begin{array}{c}0\leq r,s,t,u\leq L\cr r\!+\!s\!=\!t\!+\!u\!+\!1\end{array}}
\!\!\!\!\!\!\!\!\C^{\beta_{s}}\C^{\alpha_{r}}~
\U_{\alpha_{r}\beta_{s}}{}^{\gamma_{t}\delta_{u}}
\bP_{\delta_{u}}\bP_{\gamma_{t}}
(-1)^{\eps_{\beta_{s}}+s+\eps_{\delta_{u}}+u}
+{\cal O}(\C^{3}\bP,\C\bP^{3})~,
\label{omegazeroexpan}
\eea
where \mb{\U_{\alpha\ldots}{}^{\beta\ldots}
=\U_{\alpha\ldots}{}^{\beta\ldots}(q,p)} denote
structure functions of the original phase space.
The \mb{\omegazero} operator starts with the original first-class constraints 
\mb{\T_{\alpha_{0}}\!=\!\T_{\alpha_{0}}(q,p)}, and in the reducible case one 
also introduces higher-stage constraints, which depend linearly on the ghost
momenta,  
\beq
\T_{\alpha_{s+1}}~\equiv~
\Z_{\alpha_{s+1}}{}^{\beta_{s}}~\bP_{\beta_{s}}
(-1)^{\eps_{\beta_{s}}+s}
~,~~~~s\in\{0,\ldots,L\!-\!1\}~.\label{tees}
\eeq
Note that the higher-stage constraints \mb{\T_{\alpha}} do not necessarily
commute with the \mb{\C^{\beta}} ghosts. The Grassmann parity and ghost
number are
\beq
\eps(\T_{\alpha})~=~\eps_{\alpha}~,~~~~~~
\eps(\T_{\alpha_{s}})~=~\eps_{\alpha_{s}}\!+\!s~,~~~~~~
\gh(\T_{\alpha_{s}})~=~-s~,
\eeq
which are opposite of the corresponding ordinary \mb{\C} and \mb{\bP} ghosts. 
The reducible structure functions
\mb{\Z_{\alpha_{s}}{}^{\beta_{s-1}}\!=\!\Z_{\alpha_{s}}{}^{\beta_{s-1}}(q,p)}
have Grassmann parity and ghost number given by
\beq
\eps(\Z_{\alpha_{s}}{}^{\beta_{s-1}})
~=~\eps_{\alpha_{s}}\!+\!\eps_{\beta_{{s-1}}}~,~~~~~~
\gh(\Z_{\alpha_{s}}{}^{\beta_{s-1}})~=~0~.
\eeq
To be systematic, let \mb{\m_{-1}=\gamma_{-1}\!+\!\gamma_{0}} denote half the
number of original phase variables \mb{(q,p)}, \ie the original phase space
consists effectively of \mb{2\gamma_{-1}} physical degrees of freedom,
\mb{\gamma_{0}} irreducible gauge-generating constraints, and \mb{\gamma_{0}}
irreducible gauge-fixing constraints. 
The logic behind reducible gauge algebras, is, that due to requirements of
locality and symmetries, such as \eg Lorentz symmetry, it is impossible to pick
an irreducible set of constraints. At each stage \mb{s} (except the last
stage) one keeps overshooting the number of remaining gauge symmetries.
In detail, the \mb{s}'th stage consists of 
\mb{\m_{s}=\gamma_{s}\!+\!\gamma_{s+1}} ghosts, which corresponds to
\mb{\gamma_{s}} independent gauge symmetries, and \mb{\gamma_{s+1}} redundant
symmetries, which in turn become gauge symmetries for the next stage
\mb{s\!+\!1}. If \mb{L<\infty} is finite, the process stops at the \mb{L}'th
stage with \mb{\gamma_{L+1}=0}.
\bea 
\gamma_{0}&=&\rank(\T_{\alpha_{0}})~>~0~,\label{rankcond1} \\
\gamma_{s}&=&\sum_{r=s}^{L}(-1)^{r-s}\m_{r}~=~
\left\{\begin{array}{ll} 0&{\rm for}~s\!\geq\!L\!+\!1~, \cr 
\rank(\Z_{\alpha_{s}}{}^{\beta_{s-1}})>0&{\rm otherwise}~.\end{array} \right.
\label{rankcond2}
\eea 
This leads to a rank condition on the \mb{\omegazero} operator. It should
contain as many non-trivial constraints as possible, \ie a quarter of the total
number of non-physical degrees of freedom,
\beq
\sum_{s=0}^{\infty}\gamma_{s}~=~\Hf\sum_{s=0}^{\infty}\m_{s}
~=~\sum_{r=0}^{\infty}\m_{2r}~=~\sum_{r=0}^{\infty}\m_{2r+1}~.
\label{rankcond3}
\eeq
Equivalently, the antibracket \mb{(\cdot,\cdot)_{\omegazero}} should have
half rank in the non-physical sector.
Here the (normalized, operator-valued) antibracket\footnote{The antibracket
in \Ref{BatTyu03} is based on the bare commutator \e{supercom}, while we here
use the normalized commutator \e{normsupercom}, so that
\mb{(\cdot,\cdot)_{\Omega}^{\rm there}\equiv(i\hbar)^{2}
(\cdot,\cdot)_{\omegazero}^{\rm here}}.} is defined as
\cite{BatMar98c,BatMar99c,BatMar99b,Ber06}
\beq
(A,B)_{\omegazero}~\equiv~
\Hf \{A,\{\omegazero,B \}\}+\Hf \{\{ A,\omegazero\},B \}~,
\label{operatorantibracket}
\eeq
where \mb{A} and \mb{B} are arbitrary operators. We remark that this
antibracket always satisfies the pertinent Jacobi identity modulo BRST-exact
terms, and that the Jacobi identity holds strongly within a {\em Dirac
subalgebra} \cite{Ber06}, which is by definition a maximal subalgebra that is
1) Abelian \wrt the commutator and 2) stabile under the antibracket
\e{operatorantibracket}.

\noi
The {\em rank} \mb{R} of a theory is defined as the highest power of ghost
momenta \mb{\bP_{\alpha}} in \mb{\omegazero}.
Some of the consequences of the \mb{\omegazero}-nilpotency \e{oldbrstcharge}
are
\bea
\Z_{\alpha_{1}}{}^{\beta_{0}}~\T_{\beta_{0}}&=&0~,\label{oldreductees} \\
\{\T_{\alpha_{0}},\T_{\beta_{0}}\}
&=&\U_{\alpha_{0}\beta_{0}}{}^{\gamma_{0}}~\T_{\gamma_{0}}~,
\label{oldinvolution} \\
\{\T_{\alpha_{0}}, \Z_{\beta_{s+1}}{}^{\gamma_{s}}\}
&=&\U_{\alpha_{0}\beta_{s+1}}{}^{\delta_{s+1}}\Z_{\delta_{s+1}}{}^{\gamma_{s}}
+(-1)^{\eps_{\alpha_{0}}(\eps_{\beta_{s+1}}+s)}
\Z_{\beta_{s+1}}{}^{\delta_{s}}~\U_{\delta_{s}\alpha_{0}}{}^{\gamma_{s}} \cr
&&-\U_{\alpha_{0}\beta_{s+1}}{}^{\gamma_{s}\delta_{0}}~\T_{\delta_{0}}
+\left(\rule[-1ex]{0ex}{3ex}
{\cal O}(\hbar)~{\rm terms,~if}~R\geq 2\right)~,\label{oldteez} \\
\Z_{\alpha_{s+2}}{}^{\gamma_{s+1}}\Z_{\gamma_{s+1}}{}^{\beta_{s}}
&=&\U_{\alpha_{s+2}}{}^{\beta_{s}\gamma_{0}}~\T_{\gamma_{0}}
-\frac{i\hbar}{2}\sum_{r=0}^{s}
\U_{\alpha_{s+2}}{}^{\delta_{s-r}\gamma_{r}}~
\U_{\gamma_{r}\delta_{s-r}}{}^{\beta_{s}} \cr
&&+\left({\cal O}(\hbar^{2})~{\rm terms,~if}~R\geq 3~{\rm and}~s\geq 1\right)~.
\label{oldzzrelation}
\eea
One recognizes the involution relation \e{oldinvolution} for the first class
constraints \mb{\T_{\alpha_{0}}}, and the reducibility relations
\es{oldreductees}{oldzzrelation}. Eq.\ \e{oldreductees} shows that the
zeroth-stage constraints \mb{\T_{\alpha_{0}}} are reducible if 
\mb{\rank(\Z_{\alpha_{1}}{}^{\beta_{0}})\neq 0}. One can re-express the
\eq{oldzzrelation} in the \mb{s\!=\!0} case as
\beq
 \Z_{\alpha_{2}}{}^{\beta_{1}}\Z_{\beta_{1}}{}^{\gamma_{0}}
~=~\Hf\U_{\alpha_{2}}{}^{\delta_{0}\beta_{0}}
\Z_{\beta_{0}\delta_{0}}{}^{\gamma_{0}}~, \label{oldzz210relation}
\eeq
where \mb{\Z_{\alpha_{0}\beta_{0}}{}^{\gamma_{0}}} is defined as
\beq
\Z_{\alpha_{0}\beta_{0}}{}^{\gamma_{0}}
~\equiv~\T_{\alpha_{0}}\delta_{\beta_{0}}^{\gamma_{0}}
-\frac{i\hbar}{2}\U_{\alpha_{0}\beta_{0}}{}^{\gamma_{0}}
-(-1)^{\eps_{\alpha_{0}}\eps_{\beta_{0}}}
(\alpha_{0}\leftrightarrow \beta_{0})~.
\label{zabc}
\eeq
If one takes the classical limit \mb{\hbar\to0} and go on-shell \wrt the
constraints \mb{\T_{\alpha_{0}}}, the quantities
\mb{\Z_{\alpha_{0}\beta_{0}}{}^{\gamma_{0}}} vanish, so that the structure
functions \mb{\Z_{\alpha_{2}}{}^{\beta_{1}}} become left zero-eigenvectors for
the matrix \mb{\Z_{\beta_{1}}{}^{\gamma_{0}}}, \cf \eq{oldzz210relation}. 
At the quantum level and off-shell, the quantities
\mb{\Z_{\alpha_{0}\beta_{0}}{}^{\gamma_{0}}} fulfill
\beq
\Z_{\alpha_{0}\beta_{0}}{}^{\gamma_{0}}~\T_{\gamma_{0}}
~=~[\T_{\alpha_{0}},\T_{\beta_{0}}]
-i\hbar\U_{\alpha_{0}\beta_{0}}{}^{\gamma_{0}}~\T_{\gamma_{0}}~=~0~,
\eeq
due to \eq{oldinvolution}.

\subsection{BRST-Improved Hamiltonian}
\label{secoldbrstimprovedham}

\noi
We mention for completeness that the original Hamiltonian 
\mb{H_{\orig}\!=\!H_{\orig}(q,p)}, which commutes weakly with the first-class
constraints
\beq
 \{\T_{\alpha_{0}}, H_{\orig} \}
~=~V_{\alpha_{0}}{}^{\beta_{0}}~\T_{\beta_{0}}~,
\eeq
is supposed to be BRST-improved,
\beq
H_{0}~=~H_{\orig}
+\C^{\alpha}V_{\alpha}{}^{\beta}\bP_{\beta}(-1)^{\eps_{\beta}}
+{\cal O}(\C^{2}\bP,\C\bP^{2})~,
\label{oldhamshort}
\eeq
by letting the improved Hamiltonian \mb{H_{0}\!=\!H_{0}(q,p;\C,\bP)} depend
on the ghosts \mb{\C^{\alpha}} and \mb{\bP_{\beta}} in such a way that it
becomes BRST-invariant,
\beq
\{\omegazero,H_{0}\}~=~0~,~~~~~~~~\eps(H_{0})~=~0~,~~~~~~~~~\gh(H_{0})~=~0~.
\label{oldhamquantumnumbers}
\eeq
Contrary to the standard approach
\cite{FraVil75,BatVil77,FraFra78,BatFra83a,BatFra83b} one does not introduce
non-minimal variables in the old sector. Instead, the idea is roughly speaking
to supply the gauge-generating constraints \mb{T_{\alpha}\!=\!0} with more 
gauge-generating constraints, which kill the ordinary ghost momenta
\mb{\bP_{\alpha}\!=\!0}, and supply the gauge-fixing constraints
\mb{\chi^{\alpha}\!=\!0} with more gauge-fixing constraints, which kill the
ordinary ghosts \mb{\C^{\alpha}\!=\!0}.
The complete gauge-fixing procedure will be mediated through non-minimal
variables in a new ghost sector, see Subsection~\ref{secnonmin}.

\subsection{Anti-BRST algebra}

\noi
At this point we mention an interesting possibility to include an anti-BRST
operator \mb{\bomegazero\!=\!\bomegazero(q,p;\C,\bP)}. Although anti-BRST
algebras are not an essential part of this paper (\ie one is free to simply
let \mb{\bomegazero=0}), they nevertheless constitute an important topic that
is worthwhile mentioning. In general, the anti-BRST operator \mb{\bomegazero}
is required to satisfy
\beq
\{\bomegazero,\bomegazero\}~=~ 0~,~~~~~~
\{\omegazero,\bomegazero\} ~=~ 0~,~~~~~~
\eps(\bomegazero)~=~1~,~~~~~~ 
\gh(\bomegazero)~=~-1~.
\label{boldbrstcharge}
\eeq 
We caution that the pertinent anti-BRST algebra \cite{BatGri03} will here be
{\em different} from the traditional notion of an anti-BRST algebra
\cite{CurFer76,Oji80,BatLavTyu90}. Recall that in the traditional setting,
the anti-BRST algebra explores an \mb{Sp(2)} duality among the Faddeev-Popov
ghost and antighost \mb{\C^{\alpha} \leftrightarrow \bC^{\alpha}}. Roughly
speaking, the traditional anti-BRST operator (which are closely related to
a co-BRST operator \cite{KalHol91}) amounts to substitute
\mb{\C^{\alpha}\rightarrow \bC^{\alpha}} in \mb{\omegazero}, while keeping
the constraints \mb{\T_{\alpha}} the same. However in our current setup, there
is no Faddeev-Popov antighost within the minimal framework; only a ghost
momenta \mb{\bP_{\alpha}}. (Instead the Faddeev-Popov antighost typically
belongs to a non-minimal sector.) The crucial difference is that the
traditional \mb{Sp(2)} doublets \mb{(\C^{\alpha};\bC^{\beta})} of Faddeev-Popov
ghost pairs commute \mb{[\C^{\alpha},\bC^{\beta}]=0}, while the canonical pairs
\mb{(\C^{\alpha};\bP_{\beta})} do not commute. Also the anti-constraints 
\mb{\bT^{\alpha_{0}}} in \mb{\bomegazero} will in general be different from the
first-class constraints \mb{\T_{\alpha_{0}}}, \cf \eq{completesolutees1} below.
In detail, the \mb{\bomegazero} operator is a power series expansion in
ordinary ghosts \mb{\C} and \mb{\bP}, and ordered according to the
\mb{\C\bP}-ordering prescription,
\bea
\bomegazero&=&\sum_{s=0}^{L}\bT^{\alpha_{s}}\bP_{\alpha_{s}}
(-1)^{\eps_{\alpha_{s}}+s}
+\Hf\!\!\!\!\!\!\!\!\sum_{\footnotesize
\begin{array}{c}r,s\geq 0\cr r\!+\!s\leq L\end{array}}
\!\!\!\!\!\!\!\!\C^{\gamma_{r+s}}~
\bU_{\gamma_{r+s}}{}^{\beta_{s}\alpha_{r}}
\bP_{\alpha_{r}}\bP_{\beta_{s}}(-1)^{\eps_{\alpha_{r}}+r}\cr
&&+\Hf\!\!\!\!\!\!\!\!\sum_{\footnotesize
\begin{array}{c}r,s\geq 0\cr r\!+\!s\!+\!2\leq L\end{array}}
\!\!\!\!\!\!\!\!\C^{\beta_{s}}\C^{\alpha_{r}}~
\bU_{\alpha_{r}\beta_{s}}{}^{\gamma_{r+s+2}}\bP_{\gamma_{r+s+2}}
(-1)^{\eps_{\beta_{s}}+\eps_{\gamma_{r+s+2}}+r} \cr
&&+\frac{1}{4}\!\!\!\!\!\!\!\!\sum_{\footnotesize
\begin{array}{c}0\leq r,s,t,u\leq L\cr r\!+\!s\!+\!1\!=\!t\!+\!u\end{array}}
\!\!\!\!\!\!\!\!\C^{\beta_{s}}\C^{\alpha_{r}}~
\bU_{\alpha_{r}\beta_{s}}{}^{\gamma_{t}\delta_{u}}
\bP_{\delta_{u}}\bP_{\gamma_{t}}
(-1)^{\eps_{\beta_{s}}+s+\eps_{\delta_{u}}+u}
+{\cal O}(\C^{3}\bP,\C\bP^{3})~.
\label{bomegazeroexpan}
\eea
The \mb{\bomegazero} operator starts with the anti-constraints 
\mb{\bT^{\alpha_{0}}\!=\!\bT^{\alpha_{0}}(q,p)}, and in the reducible case one 
also introduces higher-stage anti-constraints, which depend linearly on the
ghosts,  
\beq
\bT^{\alpha_{s}}~\equiv~
\C^{\beta_{s-1}}\bZ_{\beta_{s-1}}{}^{\alpha_{s}}
~,~~~~~~~~~~s\in\{1,\ldots,L\}~.\label{btees}
\eeq
The anti-constraints \mb{\bT^{\alpha}} carry the same Grassmann parity as the
corresponding constraints \mb{\T_{\alpha}},
\beq
\eps(\bT^{\alpha})~=~\eps_{\alpha}~,~~~~~~~~~~~~
\eps(\bT^{\alpha_{s}})~=~\eps_{\alpha_{s}}\!+\!s~,~~~~~~~~~~~~
\gh(\bT^{\alpha_{s}})~=~s~.
\eeq 
The reducible structure functions
\mb{\bZ_{\alpha_{s}}{}^{\beta_{s+1}}\!=\!\bZ_{\alpha_{s}}{}^{\beta_{s+1}}(q,p)}
have Grassmann parity and ghost number given by
\beq
\eps(\bZ_{\alpha_{s}}{}^{\beta_{s+1}})
~=~\eps_{\alpha_{s}}\!+\!\eps_{\beta_{{s+1}}}~,~~~~~~~~~~~~
\gh(\bZ_{\alpha_{s}}{}^{\beta_{s+1}})~=~0~.
\eeq
Some of the consequences of the \mb{\bomegazero}-nilpotency \e{boldbrstcharge}
are
\bea
\bT^{\alpha_{0}}~\bZ_{\alpha_{0}}{}^{\beta_{1}}&=&0~,\label{boldreductees} \\
\{\bT^{\alpha_{0}},\bT^{\beta_{0}}\}
&=&\bT^{\gamma_{0}}~\bU_{\gamma_{0}}{}^{\alpha_{0}\beta_{0}}~,
\label{boldinvolution} \\
\{\bZ_{\gamma_{s}}{}^{\beta_{s+1}},\bT^{\alpha_{0}}\}
&=&\bZ_{\gamma_{s}}{}^{\delta_{s+1}}~
\bU_{\delta_{s+1}}{}^{\beta_{s+1}\alpha_{0}}
+(-1)^{\eps_{\alpha_{0}}(\eps_{\beta_{s+1}}+s)}
\bU_{\gamma_{s}}{}^{\alpha_{0}\delta_{s}}\bZ_{\delta_{s}}{}^{\beta_{s+1}} \cr
&&-\bT^{\delta_{0}}~\bU_{\delta_{0}\gamma_{s}}{}^{\beta_{s+1}\alpha_{0}}
+\left(\rule[-1ex]{0ex}{3ex}
{\cal O}(\hbar)~{\rm terms,~if}~R\geq 2\right)~,\label{boldteez} \\
\bZ_{\alpha_{s}}{}^{\gamma_{s+1}}\bZ_{\gamma_{s+1}}{}^{\beta_{s+2}}
&=&\bT^{\gamma_{0}}~\bU_{\gamma_{0}\alpha_{s}}{}^{\beta_{s+2}}
-\frac{i\hbar}{2}\sum_{r=0}^{s}
\bU_{\alpha_{s}}{}^{\delta_{s-r}\gamma_{r}}~
\bU_{\gamma_{r}\delta_{s-r}}{}^{\beta_{s+2}} \cr
&&+\left({\cal O}(\hbar^{2})~{\rm terms,~if}~R\geq 3~{\rm and}~s\geq 1\right)~.
\label{boldzzrelation}
\eea
Some of the consequences of the compatibility \e{boldbrstcharge} of the
BRST and the anti-BRST operator, \ie the fact that they commute, are
\bea
\bT^{\alpha_{0}}~\T_{\alpha_{0}}&=&0~, \label{tees1} \\
\{\T_{\alpha_{0}},\bT^{\beta_{0}}\}
+\bZ_{\alpha_{0}}{}^{\gamma_{1}}\Z_{\gamma_{1}}{}^{\beta_{0}}
&=&\bT^{\gamma_{0}}~\U_{\gamma_{0}\alpha_{0}}{}^{\beta_{0}}
+\bU_{\alpha_{0}}{}^{\beta_{0}\gamma_{0}}~\T_{\gamma_{0}}
-\frac{i\hbar}{2}\bU_{\alpha_{0}}{}^{\delta_{0}\gamma_{0}}~
\U_{\gamma_{0}\delta_{0}}{}^{\beta_{0}}+{\cal O}(\hbar^{2}) \cr
&=&\bT^{\gamma_{0}}~\U_{\gamma_{0}\alpha_{0}}{}^{\beta_{0}}
+\Hf\bU_{\alpha_{0}}{}^{\gamma_{0}\delta_{0}}~
\Z_{\delta_{0}\gamma_{0}}{}^{\beta_{0}}
+{\cal O}(\hbar^{2})~,\label{tees2} \\
\Z_{\alpha_{s}}{}^{\gamma_{s-1}}~\bZ_{\gamma_{s-1}}{}^{\beta_{s}}
+\bZ_{\alpha_{s}}{}^{\gamma_{s+1}}~\Z_{\gamma_{s+1}}{}^{\beta_{s}}
&=&\bT^{\gamma_{0}}~\U_{\gamma_{0}\alpha_{s}}{}^{\beta_{s}}
+\bU_{\alpha_{s}}{}^{\beta_{s}\gamma_{0}}~\T_{\gamma_{0}}
-\frac{i\hbar}{2}\sum_{r=0}^{s}
\bU_{\alpha_{s}}{}^{\delta_{s-r}\gamma_{r}}~
\U_{\gamma_{r}\delta_{s-r}}{}^{\beta_{s}} \cr
&&-\frac{i\hbar}{2}\sum_{r=0}^{s-2}
\U_{\alpha_{s}}{}^{\delta_{s-2-r}\gamma_{r}}~
\bU_{\gamma_{r}\delta_{s-2-r}}{}^{\beta_{s}}
+{\cal O}(\hbar^{2})~.\label{zzbarrelation}
\eea

\subsection{The \mb{\eS_{-2}} Operator}
\label{secessminustwo}

\noi
We have up until now mostly reviewed the notions of BRST and anti-BRST symmetry
in the Hamiltonian setting. In the next
Subsections~\ref{secessminustwo}-\ref{secomezero} we shall make a series of new
observations concerning the cohomology of \mb{\omegazero}, which lead to an 
operator master equation, \cf \eq{omezero} below. There is a parallel operator
master \eq{ome} for the main construction, to be presented in 
Section~\ref{secnewbrst}.

\noi
Normally it is assumed that the cohomology of \mb{\{\omegazero,\cdot\}} is
trivial in sectors of non-vanishing ghost number. Then the
\mb{\omegazero}-closeness \e{boldbrstcharge} of \mb{\bomegazero} implies that
\mb{\bomegazero} is \mb{\omegazero}-exact, \ie there exists a Bosonic operator
\mb{\eS_{-2}} such that
\beq
\bomegazero~=~\{\omegazero,\eS_{-2}\}
~,~~~~~~~~~~~~\eps(\eS_{-2})~=~0~,~~~~~~~~~~~~\gh(\eS_{-2})~=~-2~.
\eeq
In detail, the \mb{\eS_{-2}} operator is a power series expansion in the
ordinary ghosts \mb{\C} and \mb{\bP},
\bea
\eS_{-2}&=&\Hf  A^{\beta_{0}\alpha_{0}} \bP_{\alpha_{0}}\bP_{\beta_{0}}
(-1)^{\eps_{\alpha_{0}}}
+A^{\alpha_{1}} \bP_{\alpha_{1}}(-1)^{\eps_{\alpha_{1}}}
+ {\cal O}(\C\bP)~,
\eea
where \mb{A^{\alpha_{0}\beta_{0}}=
-(-1)^{\eps_{\alpha_{0}}\eps_{\beta_{0}}}A^{\beta_{0}\alpha_{0}}}
and \mb{A^{\alpha_{1}}} are arbitrary operators. {}For instance, the complete
\mb{\bT^{\alpha_{0}}} solution to \eq{tees1} alone, is
\beq
\bT^{\gamma_{0}}~=~\Hf A^{\beta_{0}\alpha_{0}}
\Z_{\alpha_{0}\beta_{0}}{}^{\gamma_{0}}
+A^{\alpha_{1}}\Z_{\alpha_{1}}{}^{\gamma_{0}}~,
\label{completesolutees1}
\eeq
where \mb{\Z_{\alpha_{0}\beta_{0}}{}^{\gamma_{0}}} is defined in \eq{zabc}.

\subsection{The \mb{\eS_{-4}} Operator}
\label{secessminusfour}

\noi
Similarly, one may define an \mb{\eS_{-4}} operator as follows.
{}First note that the antibracket \mb{(\eS_{-2},\eS_{-2})_{\omegazero}^{}} of 
\mb{\eS_{-2}} with itself can be written as the commutator of \mb{\eS_{-2}}
and \mb{\bomegazero}, 
\beq
(\eS_{-2},\eS_{-2})_{\omegazero}^{}~=~\{\eS_{-2},\{\omegazero,\eS_{-2}\}\}
~=~\{\eS_{-2},\bomegazero\}~.
\label{s1s1antibracketzero}
\eeq 
The antibracket \mb{(\eS_{-2},\eS_{-2})_{\omegazero}^{}} is
\mb{\omegazero}-closed, as the following calculation shows,
\beq
\{\omegazero,(\eS_{-2},\eS_{-2})_{\omegazero}\}
~=~\{\omegazero,\{\eS_{-2},\bomegazero\}\}
~=~\{\{\omegazero,\eS_{-2}\},\bomegazero\}
~=~\{\bomegazero,\bomegazero\}~=~0~,
\eeq
since \mb{\bomegazero} is \mb{\omegazero}-closed and nilpotent, \cf 
\eq{boldbrstcharge}. Hence the antibracket
\mb{(\eS_{-2},\eS_{-2})_{\omegazero}^{}} is \mb{\omegazero}-exact, \ie 
there exists a quantity \mb{\eS_{-4}}, such that
\beq
\Hf(\eS_{-2},\eS_{-2})_{\omegazero}^{}~=~\{\omegazero,\eS_{-4}\}
~,~~~~~~~~~~~~\eps(\eS_{-4})~=~0~,~~~~~~~~~~~~\gh(\eS_{-4})~=~-4~.
\label{sminus4exists}
\eeq

\subsection{Operator Master Equation}
\label{secomezero}

\noi
The process of finding higher and higher cohomology relations may be automated
by introducing an operator master equation
\beq
-(\eS,\eS)_{\omegazero}^{}~\equiv~\{\eS,\{\eS,\omegazero\}\}~=~\omegazero~.
\label{omezero}
\eeq
To define the operator \mb{\eS}, one first let a quantity \mb{\eS_{0}} be the 
ghost operator, 
\beq
 \eS_{0}~=~G~,
\eeq
see \eq{gee} below. {}Furthermore, one may show by mathematical induction that 
there exist quantities \mb{\eS_{-2k}} for each positive integer \mb{k\ge 1}, 
and with quantum numbers
\beq
\eps(\eS_{-2k})~=~0~,~~~~~~~~~\gh(\eS_{-2k})~=~-2k~,
\eeq
such that the sum
\beq
 \eS~\equiv~\sum_{k=0}^{\infty}\eS_{-2k}~,~~~~~~~~~\eps(\eS)~=~0~,
\eeq
of indefinite ghost number, satisfies the above operator master \eq{omezero},
\cf Appendix~\ref{appome}. The master \eq{omezero} is really an infinite tower
of equations. The first non-trivial equation is \eq{sminus4exists}.

\noi
Although we shall not pursue this here, let us mention that a stronger
formulation of the anti-BRST algebra would impose rank conditions on
\mb{\bomegazero} (\ie the analogues of conditions \e{rankcond1}-\e{rankcond3}
for \mb{\omegazero}), and the anti-BRST algebra would contain non-trivial
information, which allows for a complexification the whole BRST algebra.
{}For starters, the dimension \mb{m_{s}} must then be even at each stage
\mb{s\in\{0,\ldots,L\}}. In contrast, there are no such requirements for a
traditional \mb{Sp(2)} algebra.

\section{New BRST algebra}
\label{secnewbrst}

\noi
In this Section~\ref{secnewbrst} we present all the ingredients of the main
construction. 

\subsection{New Ghosts}
\label{secngh}

\noi
The new BRST algebra depends on \mb{L\!+\!1} stages of new ghosts
\beq
\cB^{A}~=~\{\cB^{A_{s}}|s=0,\ldots, L\}~,~~~~~
\eeq
and their momenta
\beq
\bcP_{A}~=~\{\bcP_{A_{s}}|s=0,\ldots, L\}~,
\eeq
with canonical commutation relations
\beq
\{\cB^{A},\cB^{B}\}~=~0~,~~~~~~~~~~
\{\cB^{A},\bcP_{B}\}~=~\delta^{A}_{B}~=~
(-1)^{\eps_{B}}\{\bcP_{B},\cB^{A}\}~,~~~~~~~~~~
\{\bcP_{A},\bcP_{B}\}~=~0~.
\eeq
In addition there are non-minimal variables, which we will often not write 
explicitly in formulas for the sake of simplicity, \cf 
Subsection~\ref{secnonmin}. Technically, the various expansions will be kept
under control by a bi-graded mesh of integer-graded conservation laws, which
(besides the usual ghost number conservation) includes a new ghost number
grading, denoted ``\mb{\ngh}''. The Grassmann parity and new ghost number of
the new ghosts are
\beq
\begin{array}{rccclcrcccl}
\eps(\cB^{A})&=& \eps_{A}\!+\!1&=&\eps(\bcP_{A})~,&&
\eps(\cB^{A_{s}})&=& \eps_{A_{s}}\!+\!s\!+\!1&=&\eps(\bcP_{A_{s}})~, \\
&&&&&&\ngh(\cB^{A_{s}})&=&s\!+\!1&=&-\ngh(\bcP_{A_{s}})~.
\end{array}
\eeq
In detail, each of the new ghosts consist of two superpartners
\beq\begin{array}{rccclcrcl}
\cB^{A}&\equiv&\{\cB_{0}^{\alpha};~\cB_{1}^{\alpha}\}
&\equiv&\{\B^{\alpha};~
(-1)^{\eps_{\alpha}+1}\Pi_{*}^{\alpha}\}~, &&
\cB^{A_{s}}&\equiv&\{\B^{\alpha_{s}};~
(-1)^{\eps_{\alpha_{s}}+s+1}\Pi_{*}^{\alpha_{s}}\}~,  \\
\bcP_{A}&\equiv&\{\bcP^{0}_{\alpha};~\bcP^{1}_{\alpha}\}
&\equiv&\{\Pi_{\alpha};~\B^{*}_{\alpha}\}~, &&
\bcP_{A_{s}}&\equiv&\{\Pi_{\alpha_{s}};~\B^{*}_{\alpha_{s}}\}~,
\end{array} 
\label{ghostsplit}
\eeq
Note that there are two equivalent notations 
\mb{\cB_{0}^{\alpha}\!\equiv\!\B^{\alpha}},
\mb{\cB_{1}^{\alpha}\!\equiv\!(-1)^{\eps_{\alpha}+1}\Pi_{*}^{\alpha}},
\mb{\bcP^{0}_{\alpha}\equiv\Pi_{\alpha}} and
\mb{\bcP^{1}_{\alpha}\equiv\B^{*}_{\alpha}} for each of the superpartners. 
The latter notation stresses a relationship with the antifield formalism,
\cf \eq{abelomegadeltarankcond} below. The canonical commutation relations
read
\bea
\{\cB_{0}^{\alpha},\bcP^{0}_{\beta}\}~=~
\{\B^{\alpha},\Pi_{\beta}\}&=&\delta^{\alpha}_{\beta}~, \\
\{\cB_{1}^{\alpha},\bcP^{1}_{\beta}\}~=~
\{\B^{*}_{\beta},\Pi_{*}^{\alpha}\}&=&\delta^{\alpha}_{\beta}~,
\eea
with the remaining canonical commutation relations being zero. 
There are twice as many new ghosts \mb{(\cB^{A},\bcP_{A})} as old ghosts
\mb{(\C^{\alpha},\bP_{\alpha})}. In other words, the index \mb{A_{s}}
corresponds to two copies of index \mb{\alpha_{s}}, \mb{s\in\{0,\ldots, L\}}.
Moreover, the index \mb{A} is a shorthand for all stages
\mb{A_{0},\ldots,A_{L}}. The Grassmann parity and the old ghost
number are shifted among the ghost superpartners as follows
\beq
\begin{array}{rccclcrcccl}
\eps(\cB_{0}^{\alpha})&=&\eps_{\alpha}\!+\!1
&=&\eps(\bcP^{0}_{\alpha})~, &&
\eps(\cB_{0}^{\alpha_{s}})&=&\eps_{\alpha_{s}}\!+\!s\!+\!1
&=&\eps(\bcP^{0}_{\alpha_{s}})~,  \\
\eps(\cB_{1}^{\alpha})&=&\eps_{\alpha}
&=&\eps(\bcP^{1}_{\alpha})~,&&
\eps(\cB_{1}^{\alpha_{s}})&=&\eps_{\alpha_{s}}\!+\!s
&=&\eps(\bcP^{1}_{\alpha_{s}})~,\\
&&&&&&\gh(\cB_{0}^{\alpha_{s}})&=&s\!+\!1&=&-\gh(\bcP^{0}_{\alpha_{s}})~, \\
&&&&&&\gh(\cB_{1}^{\alpha_{s}})&=&s\!+\!2&=&-\gh(\bcP^{1}_{\alpha_{s}})~, \\
\eps(\B^{\alpha})&=&\eps_{\alpha}\!+\!1
&=&\eps(\Pi_{\alpha})~, &&
\eps(\B^{\alpha_{s}})&=&\eps_{\alpha_{s}}\!+\!s\!+\!1
&=&\eps(\Pi_{\alpha_{s}})~,  \\
\eps(\B^{*}_{\alpha})&=&\eps_{\alpha}
&=&\eps(\Pi_{*}^{\alpha})~,&&
\eps(\B^{*}_{\alpha_{s}})&=&\eps_{\alpha_{s}}\!+\!s
&=&\eps(\Pi_{*}^{\alpha_{s}})~,\\
&&&&&&\gh(\B^{\alpha_{s}})&=&s\!+\!1&=&-\gh(\Pi_{\alpha_{s}})~, \\
&&&&&&\gh(\Pi_{*}^{\alpha_{s}})&=&s\!+\!2&=&-\gh(\B^{*}_{\alpha_{s}})~.
\end{array} 
\label{newstatistics}
\eeq
The shifts \e{newstatistics} in Grassmann and ghost
statistics will play a crucial role in the following. It is compelling that the
\mb{2\times2} ghost superpartners \e{ghostsplit} can be incorporated into two
\mb{N\!=\!1} superfields of definite Grassmann and ghost statistics, where the
offsets in statistics have been compensated by the appearance of a 
\mb{\theta}-parameter; see Appendix~\ref{appsuperfield} for details.

\subsection{Ghost Operators}
\label{secghostoper}

\noi
The old (resp.\ new) ghost number ``\mb{\gh(A)}'' (resp.\ ``\mb{\ngh(A)}'') of
an arbitrary operator \mb{A} may be implemented through a corresponding ghost
operator \mb{G} (resp.\ \mb{\gbb}) as follows\footnote{
The new ghost number should not be confused with the notion of new ghost number
in \mb{Sp(2)} theories \cite{BatLavTyu90}, although the underlying motivation
is basically the same: namely, to tame the ghost expansions. 
\Ref{BatTyu03} calls the new ghost number for degree: 
\mb{\deg^{\rm there}\equiv\ngh^{\rm here}}.}
\beq
\{G,A\}~=~\gh(A)A~,~~~~~~~~~~~\{\gbb,A\}~=~\ngh(A)A~,
\eeq
Here the ghost operators read
\bea
G&\equiv&G_{\C}+\gbb
-\Hf \sum_{s=0}^{L}[\bcP^{1}_{\alpha_{s}},\cB_{1}^{\alpha_{s}}]_\plus
~=~G_{\C}+\gbb
+\Hf \sum_{s=0}^{L}[\Pi_{*}^{\alpha_{s}},\B^{*}_{\alpha_{s}}]_\plus~,
\label{gee}\\
G_{\C}&\equiv& -\Hf\sum_{s=0}^{L} (s\!+\!1) 
[\bP_{\alpha_{s}},\C^{\alpha_{s}}]_\plus~,\label{geec} \\
\gbb&\equiv&
-\Hf\sum_{s=0}^{L} (s\!+\!1) [\bcP_{A_{s}},\cB^{A_{s}}]_\plus
~=~ -\sum_{s=0}^{L} (s\!+\!1) \bcP_{A_{s}}\cB^{A_{s}} \cr
&=& \Hf\sum_{s=0}^{L}(s\!+\!1) 
([\Pi_{*}^{\alpha_{s}},\B^{*}_{\alpha_{s}}]_\plus
-[\Pi_{\alpha_{s}},\B^{\alpha_{s}}]_\plus)~.\label{geebb}
\eea
The anti-supercommutators \e{supercom} in the above formulas \e{gee}-\e{geebb}
ensure the Hermiticity of the ghost operators. 
Whereas the \eqs{geec}{geebb} contain no surprises, note that the last term in
\eq{gee} implements a crucial additional shift in the ghost number assignments
for the antifields \mb{\B^{*}_{\alpha}} and \mb{\Pi_{*}^{\alpha}}, \cf the
last line of \eq{newstatistics}.

\subsection{Improved BRST Operator \mb{\omegadelta}}

\noi
The next step is to improve the old BRST charge 
\mb{\omegazero\!=\!\omegazero(q,p;\C,\bP)} so that it also probes the new 
ghost sector \mb{(\cB,\bcP)}. This is needed for covariance of the theory
under \mb{(\cB,\bcP)}-dependent unitary transformations.
In detail, one introduces an improved charge
\mb{\omegadelta\!=\!\omegadelta(q,p;\C,\bP;\cB,\bcP)} as a \mb{\cB\bcP}-ordered
power series expansion in the new ghosts \mb{\cB} and \mb{\bcP}.
{}First of all, \mb{\omegadelta} should meet the boundary condition
\beq
\omegadelta~=~\omegazero+{\cal O}(\cB\bcP)~.
\label{omegazeroshort}
\eeq
Moreover, the quantum numbers of \mb{\omegadelta} should be the same as for 
\mb{\omegazero}. Therefore \mb{\omegadelta} should satisfy
\beq
\{\omegadelta,\omegadelta\}~=~0~,~~~~~~\eps(\omegadelta)~=~1~,~~~~~~
\gh(\omegadelta)~=~1~,~~~~~~\ngh(\omegadelta)~=~0~.
\label{omegadeltaquantumnumbers}
\eeq
In more detail, the \mb{\omegadelta} operator has the form\footnote{
\Ref{BatTyu03} uses a different notation for the various BRST operators. In 
general, \mb{\Omega^{\rm there}\equiv\omegazero^{\rm here}},
\mb{\Delta^{\rm there}\equiv\omegadelta^{\rm here}},
\mb{\bar{\Delta}^{\rm there}\equiv\bomegadelta^{\rm here}} and
\mb{\Sigma_{1}^{\rm there}\equiv\omegaone^{\rm here}}.}
\bea
\omegadelta&=&\omegazero+\sum_{s=0}^{L}\cB^{A_{s}}~\cV_{A_{s}}{}^{B_{s}}~
\bcP_{B_{s}}(-1)^{\eps_{B_{s}}+s}
+\Hf\!\!\!\!\!\!\!\!\sum_{\footnotesize
\begin{array}{c}r,s\geq 0\cr r\!+\!s\!+\!1\leq L\end{array}}
\!\!\!\!\!\!\!\!\cB^{B_{s}}\cB^{A_{r}}~
\cV_{A_{r}B_{s}}{}^{C_{r+s+1}}~\bcP_{C_{r+s+1}}
(-1)^{\eps_{B_{s}}+\eps_{C_{r+s+1}}+r+1} \cr
&&+\Hf\!\!\!\!\!\!\!\!\sum_{\footnotesize
\begin{array}{c}r,s\geq 0\cr r\!+\!s\!+\!1\leq L\end{array}}
\!\!\!\!\!\!\!\!\cB^{C_{r+s+1}}~
\cV_{C_{r+s+1}}{}^{B_{s}A_{r}}~\bcP_{A_{r}}\bcP_{B_{s}}(-1)^{\eps_{A_{r}}+r}\cr
&&+\frac{1}{4}\!\!\!\!\!\!\!\!\sum_{\footnotesize
\begin{array}{c}0\leq r,s,t,u\leq L\cr r\!+\!s\!=\!t\!+\!u\end{array}}
\!\!\!\!\!\!\!\!\cB^{B_{s}}\cB^{A_{r}}~
\cV_{A_{r}B_{s}}{}^{C_{t}D_{u}}~\bcP_{D_{u}}\bcP_{C_{t}}
(-1)^{\eps_{B_{s}}+s+\eps_{D_{u}}+u}+{\cal O}(\cB^{3}\bcP,\cB\bcP^{3}) \cr
&=&\omegazero+\onebytwo{\B^{\alpha}}{(-1)^{\eps_{\alpha}+1}\Pi_{*}^{\alpha}}
\twobytwo{\cV_{\alpha 0}^{0\beta}}{\cV_{\alpha 1}^{0 \beta}}
{\cV_{\alpha 0}^{1\beta}}{\cV_{\alpha 1}^{1\beta}}
\twobyone{(-1)^{\eps_{\beta}}\Pi_{\beta}}{(-1)^{\eps_{\beta}+1}\B^{*}_{\beta}}
+{\cal O}(\cB^{2}\bcP,\cB\bcP^{2})~,
\label{omegadeltaexpan}
\eea
where \mb{\cV_{A\ldots}{}^{B\ldots}=\cV_{A\ldots}{}^{B\ldots}(q,p;\C,\bP)} 
denote structure functions of the old phase space. {}Finally, \mb{\omegadelta}
should contain as many non-trivial constraints as possible, \ie a quarter of
the total number of non-physical phase variables. One implements this by
demanding that the southwestern quadrant \mb{\cV_{\alpha 0}^{1\beta}} (\ie the
block below the diagonal) of the matrices \mb{\cV_{A}{}^{B}} is invertible, 
\beq
\rank~\cV_{\alpha 0}^{1\beta}~=~m~\equiv~\sum_{s=0}^{L}\m_{s}~.
\label{omegadeltarankcond1}
\eeq
Equivalently, the antibracket \mb{(\cdot,\cdot)_{\omegadelta}} should have 
maximal rank in the \mb{\B\B^{*}}-sector, because
\beq
  \cV_{\alpha 0}^{1\beta}
~=~(-1)^{\eps_{\alpha}+1}(\B^{*}_{\alpha},\B^{\beta})^{}_{\omegadelta}~,
\label{omegadeltarankcond2}
\eeq
\cf \eq{operatorantibracket}. (Again, we have for simplicity notationally
suppressed that \mb{\omegadelta} is supposed to have quadratic dependence on 
half the non-minimal variables to allow for gauge-fixing, \cf 
\eq{nonminomegadelta} below.) Some of the consequences of the
\mb{\omegadelta}-nilpotency \e{omegadeltaquantumnumbers} are
\bea
\{\omegazero,\omegazero\}&=&0~,\\
\{\cV_{A_{s}}{}^{B_{s}},\omegazero \}(-1)^{\eps_{B_{s}}+s}
&=&\cV_{A_{s}}{}^{C_{s}}~\cV_{C_{s}}{}^{B_{s}}
+\frac{i\hbar}{2}\sum_{r=0}^{s-1}
\cV_{A_{s}}{}^{D_{s-r-1}C_{r}}~\cV_{C_{r}D_{s-r-1}}{}^{B_{s}} \cr
&&+\left({\cal O}(\hbar^{2})~{\rm terms,~if}~s\geq 2\right)~.
\eea

\subsection{Improved Anti-BRST Operator \mb{\bomegadelta}}

\noi
There is also an improved version
\beq
\bomegadelta~=~\bomegazero+{\cal O}(\cB\bcP)~,
\label{bomegazeroshort}
\eeq
of the anti-BRST operator \mb{\bomegazero} such that
\beq
\{\bomegadelta,\bomegadelta\}~=~0~,~~~~~~\eps(\bomegadelta)~=~1~,~~~~~~
\gh(\bomegadelta)~=~-1~,~~~~~~\ngh(\bomegadelta)~=~0~.
\label{bomegadeltaquantumnumbers}
\eeq
Perhaps surprisingly, one does {\em not} demand that the improved BRST and 
anti-BRST operators commute. Instead the normalized commutator of the two
improved charges is assumed to be equal to the new ghost operator \mb{\gbb},
\beq
 \{\omegadelta,\bomegadelta\}~=~\gbb~.\label{ooghomotopy}
\eeq
In this way, the \mb{\bomegadelta} charge becomes a homotopy operator for the
BRST complex of \mb{\omegadelta}. As a result the cohomology of
\mb{\{\omegadelta,\cdot\}} is trivial in sectors of non-zero new ghost number.
{\em Proof}: If \mb{A} is a \mb{\omegadelta}-closed operator, \ie
\mb{\{\omegadelta,A\}=0}, and if \mb{A} has non-zero new ghost number, it is
possible to rewrite \mb{A} as a \mb{\omegadelta}-exact operator,
\beq
A~=~ g^{-1}\{\gbb,A\} 
~=~g^{-1}\{ \{\omegadelta,\bomegadelta\},A\} 
~=~g^{-1} \{\omegadelta,\{\bomegadelta,A\}\}
~=~ \{\omegadelta,\{\bomegadelta,g^{-1}A\}\}~,
\eeq 
because the operator \mb{g\equiv\{\gbb, \cdot \}} commutes with both
\mb{\{\omegadelta, \cdot \}} and \mb{\{\bomegadelta, \cdot \}}.

\noi
On the other hand, since ultimately one requires that the physical model in
question is \mb{\omegadelta}-invariant, the homotopy \eq{ooghomotopy} implies
that one can no longer maintain the anti-BRST symmetry of the model.
In this sense, the anti-BRST algebra plays here only a secondary r\^ole.
Similarly, to be consistent, we will not require that the BRST-improved
Hamiltonian obeys the anti-BRST symmetry, \cf 
Subsection~\ref{secnewbrstimprovedham} below.

\subsection{New BRST Operator \mb{\omegaone}}

\noi
The \mb{\omegadelta} operator has no net new ghost charge,
\mb{\ngh(\omegadelta)=0}, and is merely a deformation of the old BRST operator
\mb{\omegazero}. In general, it is not adequate to control expansions in the
new ghosts. To this end, one introduces a new BRST generator \mb{\omegaone}
that is charged \wrt the new ghost number. In detail, the new BRST charge
\mb{\omegaone\!=\!\omegaone(q,p;\C,\bP;\cB,\bcP)} has properties 
\bea
\{\omegaone,\omegaone \}&=&0~,\label{omegaonenilp} \label{newbrstcharge} \\
\eps(\omegaone)&=&1~,~~~~~~\gh(\omegaone)~=~1~,~~~~~~\ngh(\omegaone)~=~1~.
\eea
It should also respect the symmetries of the improved charge \mb{\omegadelta},
\ie \mb{\omegaone} should be \mb{\omegadelta}-closed,
\beq
\{\omegadelta,\omegaone\}~=~0~.\label{omegaonedeltaclosed}
\eeq
The new BRST operator \mb{\omegaone} is a  \mb{\cB\bcP}-ordered power series 
expansion in the new ghosts \mb{\cB} and \mb{\bcP}. 
\bea
\omegaone&=&\sum_{s=0}^{L}\cB^{A_{s}}\cT_{A_{s}}
+\Hf\!\!\!\!\!\!\!\!\sum_{\footnotesize
\begin{array}{c}r,s\geq 0\cr r\!+\!s\leq L\end{array}}
\!\!\!\!\!\!\!\!\cB^{B_{s}}\cB^{A_{r}}~
\cU_{A_{r}B_{s}}{}^{C_{r+s}}~\bcP_{C_{r+s}}
(-1)^{\eps_{B_{s}}+\eps_{C_{r+s}}+r} \cr
&&+\Hf\!\!\!\!\!\!\!\!\sum_{\footnotesize
\begin{array}{c}r,s\geq 0\cr r\!+\!s\!+\!2\leq L\end{array}}
\!\!\!\!\!\!\!\!\cB^{C_{r+s+2}}~
\cU_{C_{r+s+2}}{}^{B_{s}A_{r}}~\bcP_{A_{r}}\bcP_{B_{s}}(-1)^{\eps_{A_{r}}+r}\cr
&&+\frac{1}{4}\!\!\!\!\!\!\!\!\sum_{\footnotesize
\begin{array}{c}0\leq r,s,t,u\leq L\cr r\!+\!s\!=\!t\!+\!u\!+\!1\end{array}}
\!\!\!\!\!\!\!\!\cB^{B_{s}}\cB^{A_{r}}~
\cU_{A_{r}B_{s}}{}^{C_{t}D_{u}}~\bcP_{D_{u}}\bcP_{C_{t}}
(-1)^{\eps_{B_{s}}+s+\eps_{D_{u}}+u}+{\cal O}(\cB^{3}\bcP,\cB\bcP^{3}) \cr
&=&\sum_{s=0}^{L}(\B^{\alpha_{s}}\cT_{\alpha_{s}}
-\Pi_{*}^{\alpha_{s}}\cX_{\alpha_{s}})
+{\cal O}(\cB^{2}\bcP,\cB\bcP^{2})~.
\label{omegaoneexpan}
\eea
It starts with the new constraints
\mb{\cT_{A_{0}}\!=\!\cT_{A_{0}}(q,p;\C,\bP)}, 
and in the reducible case one also introduces higher-stage constraints
\beq
\cT_{A_{s+1}}~\equiv~
\cZ_{A_{s+1}}{}^{B_{s}}~\bcP_{B_{s}}(-1)^{\eps_{B_{s}}+s}
~,~~~~~~~~~~s\in\{0,\ldots,L\!-\!1\}~. \label{ctees}
\eeq
The Grassmann parity and new ghost number are
\beq
\eps(\cT_{A})~=~\eps_{A}~,~~~~~~~~~~~~
\eps(\cT_{A_{s}})~=~\eps_{A_{s}}\!+\!s~,~~~~~~~~~~~~
\ngh(\cT_{A_{s}})~=~-s~.
\eeq
The reducible structure functions
\mb{\cZ_{A_{s}}{}^{B_{s-1}}\!=\!\cZ_{A_{s}}{}^{B_{s-1}}(q,p;\C,\bP)}
have Grassmann parity and new ghost number given by
\beq
\eps(\cZ_{A_{s}}{}^{B_{s-1}})~=~\eps_{A_{s}}\!+\!\eps_{B_{{s-1}}}~,~~~~~~~~~~~~
\ngh(\cZ_{A_{s}}{}^{B_{s-1}})~=~0~.
\eeq
Altogether, the construction is governed by the two mutually commuting 
Grassmann-odd nilpotent BRST operators \mb{\omegadelta} and \mb{\omegaone},
which form a \mb{Sp(2)}-like algebra,
\beq
\{\omegadelta,\omegadelta\}~=~0~,~~~~~~~~~~~
\{\omegadelta,\omegaone\}~=~0~,~~~~~~~~~~~\{\omegaone,\omegaone\}~=~0~,
\label{sp2doublet}
\eeq
\cf eqs.\ \e{omegadeltaquantumnumbers}, 
\es{newbrstcharge}{omegaonedeltaclosed}.
Some of the consequences of the \mb{\omegadelta}-closeness condition 
\e{omegaonedeltaclosed} for \mb{\omegaone} read
\bea
\{\cT_{A_{0}},\omegazero\}&=&-\cV_{A_{0}}{}^{B_{0}}~\cT_{B_{0}}~,
\label{brstinvnewconstr} \\
\{\cZ_{A_{s+1}}{}^{B_{s}},\omegazero \}(-1)^{\eps_{B_{s}}+s}
&=&\cZ_{A_{s+1}}{}^{C_{s}}~\cV_{C_{s}}{}^{B_{s}}+
\cV_{A_{s+1}}{}^{C_{s+1}}~\cZ_{C_{s+1}}{}^{B_{s}}
-\cV_{A_{s+1}}{}^{B_{s}C_{0}}~\cT_{C_{0}}\cr 
&&+\frac{i\hbar}{2}\sum_{r=0}^{s-1}
\cU_{A_{s+1}}{}^{D_{s-r-1}C_{r}}~\cV_{C_{r}D_{s-r-1}}{}^{B_{s}}
+\frac{i\hbar}{2}\sum_{r=0}^{s}
\cV_{A_{s+1}}{}^{D_{s-r}C_{r}}~\cU_{C_{r}D_{s-r}}{}^{B_{s}} \cr
&&+\left({\cal O}(\hbar^{2})
~{\rm terms,~if}~R\geq 3~{\rm and}~s\geq 1\right)~,\\
\{\cU_{A_{0}B_{0}}{}^{C_{0}},\omegazero \}(-1)^{\eps_{C_{0}}}
&=& -\{\cT_{A_{0}},\cV_{B_{0}}{}^{C_{0}}\}
+\Hf \cU_{A_{0}B_{0}}{}^{D_{0}}~\cV_{D_{0}}{}^{C_{0}}
-(-1)^{\eps_{B_{0}}}\cV_{A_{0}}{}^{D_{0}}~\cU_{D_{0}B_{0}}{}^{C_{0}}\cr  
&&-\frac{1}{4}\cV_{A_{0}B_{0}}{}^{D_{0}E_{0}}~\cZ_{E_{0}D_{0}}{}^{C_{0}}
-(-1)^{\eps_{A_{0}}\eps_{B_{0}}}(A_{0}\leftrightarrow B_{0})~, \\
\{\cU_{A_{0}B_{s}}{}^{C_{s}},\omegazero \}(-1)^{\eps_{C_{s}}+s}
&=& -\{\cT_{A_{0}},\cV_{B_{s}}{}^{C_{s}}\}
+\cU_{A_{0}B_{s}}{}^{D_{s}}~\cV_{D_{s}}{}^{C_{s}}
+(-1)^{\eps_{A_{0}}(\eps_{B_{s}}+s+1)}
\cV_{B_{s}}{}^{D_{s}}~\cU_{D_{s}A_{0}}{}^{C_{s}}\cr  
&&-\cV_{A_{0}B_{s}}{}^{C_{s}D_{0}}~\cT_{D_{0}}
+\left(\rule[-1ex]{0ex}{3ex} {\cal O}(\hbar)
~{\rm terms,~if}~R\geq 2\right)~,~~~~~~~~~s\neq 0~,
\eea
where
\mb{\cZ_{A_{0}B_{0}}{}^{C_{0}}} is defined as
\beq
\cZ_{A_{0}B_{0}}{}^{C_{0}}
~\equiv~\cT_{A_{0}}\delta_{B_{0}}^{C_{0}}
-\frac{i\hbar}{2}\cU_{A_{0}B_{0}}{}^{C_{0}}
-(-1)^{\eps_{A_{0}}\eps_{B_{0}}}(A_{0}\leftrightarrow B_{0})~.
\label{czabc}
\eeq
Here \mb{R} continues to denote the rank of the theory. In particular, 
\eq{brstinvnewconstr} shows that the new zeroth-stage constraints
\mb{\cT_{A_{0}}} are on-shell BRST-invariant. Some of the consequences of the
\mb{\omegaone}-nilpotency \e{newbrstcharge} are
\bea
\cZ_{A_{1}}{}^{B_{0}}\cT_{B_{0}}&=&0~,\label{newreductees} \\
\{\cT_{A_{0}},\cT_{B_{0}} \}&=&\cU_{A_{0}B_{0}}{}^{C_{0}}~\cT_{C_{0}}~,
\label{newinvolution} \\
\{\cT_{A_{0}}, \cZ_{B_{s+1}}{}^{C_{s}}  \}
&=&\cU_{A_{0}B_{s+1}}{}^{D_{s+1}}~\cZ_{D_{s+1}}{}^{C_{s}}
+(-1)^{\eps_{A_{0}}(\eps_{B_{s}}+s+1)}
\cZ_{B_{s+1}}{}^{D_{s}}~\cU_{D_{s}A_{0}}{}^{C_{s}} \cr
&&-\cU_{A_{0}B_{s+1}}{}^{C_{s}D_{0}}~\cT_{D_{0}}
+\left(\rule[-1ex]{0ex}{3ex} {\cal O}(\hbar)
~{\rm terms,~if}~R\geq 2\right)~, \label{newteez} \\
\cZ_{A_{s+2}}{}^{C_{s+1}}~\cZ_{C_{s+1}}{}^{B_{s}}
&=&\cU_{A_{s+2}}{}^{B_{s}C_{0}}~\cT_{C_{0}}
-\frac{i\hbar}{2}\sum_{r=0}^{s}
\cU_{A_{s+2}}{}^{D_{s-r}C_{r}}~\cU_{C_{r}D_{s-r}}{}^{B_{s}} \cr
&&+\left({\cal O}(\hbar^{2})~{\rm terms,~if}~R\geq 3~{\rm and}~s\geq 1\right)~.
\label{newzzrelation}
\eea
The analogy with the old sector \eq{oldreductees}-\e{oldzzrelation} is evident.
If the original theory is of rank \mb{R}, one may choose \mb{\omegadelta} and
\mb{\omegaone} to have at most \mb{R} powers of new ghost momenta
\mb{\bcP_{A}}. In the case of a rank \mb{R\!=\!1} theory, some of the
consequences of \eq{sp2doublet} can neatly be recast as a nilpotency condition
for a matrix \mb{\hat{\Omega}_{A}{}^{B}} with operator-valued entries, \cf
Appendix~\ref{appmatrix}.

\noi
The new constraints \mb{\cT_{A_{s}}} consist of two
superpartners\footnote{\Ref{BatTyu03} uses a different notation: 
\mb{\X_{\alpha_{0}}^{\rm there}
\equiv(-1)^{\eps_{\alpha_{0}}+1}\cX_{\alpha_{0}}^{\rm here}}.}
\beq
 \cT_{A}~\equiv~\{\cT^{0}_{\alpha};~\cT^{1}_{\alpha}\}
~\equiv~\{\cT_{\alpha};~
(-1)^{\eps_{\alpha}}\cX_{\alpha}\}~,~~~~~~
 \cT_{A_{s}}~\equiv~\{\cT_{\alpha_{s}};~
(-1)^{\eps_{\alpha_{s}}+s}\cX_{\alpha_{s}}\}~,
\label{newconstraintssupersplit}
\eeq
with boundary conditions
\beq
\cT_{\alpha_{0}}~=~\Lambda_{\alpha_{0}}{}^{\beta_{0}}~\T_{\beta_{0}}
+{\cal O}(\cB\bcP)~,
\label{constraintbc}
\eeq
where \mb{\Lambda_{\alpha_{0}}{}^{\beta_{0}}
=\Lambda_{\alpha_{0}}{}^{\beta_{0}}(q,p)} is an invertible matrix.
The zero-stage constraints \mb{\cT^{0}_{\alpha_{0}}\!\equiv\!\cT_{\alpha_{0}}}
are the new BRST-invariant constraints that one is seeking for. In fact, the
zero-stage components \mb{\cT^{0}_{\alpha_{0}}} and \mb{\cT^{1}_{\alpha_{0}}}
are the analogues of the BRST-invariant Virasoro constraints \mb{\cT_{m}} and
the string ghost momenta \mb{b_{m}} mentioned in the Introduction.
The Grassmann parity and the old ghost number are shifted among the two
superpartner constraints
\beq
\begin{array}{rccclcrcccl}
\eps(\cT^{0}_{\alpha})&=&\eps_{\alpha}
&=& \eps(\cT^{1}_{\alpha})\!+\!1~, &&
\eps(\cT^{0}_{\alpha_{s}})&=&\eps_{\alpha_{s}}\!+\!s
&=& \eps(\cT^{1}_{\alpha_{s}})\!+\!1~, \\
&&&&&&\gh(\cT^{0}_{\alpha_{s}})&=&-s&=&\gh(\cT^{1}_{\alpha_{s}})\!+\!1~, \\
\eps(\cT_{\alpha})&=&\eps_{\alpha}
&=& \eps(\cX_{\alpha})\!+\!1~, &&
\eps(\cT_{\alpha_{s}})&=&\eps_{\alpha_{s}}\!+\!s
&=& \eps(\cX_{\alpha_{s}})\!+\!1~, \\
&&&&&&\gh(\cT_{\alpha_{s}})&=&-s&=&\gh(\cX_{\alpha_{s}})\!+\!1~.
\end{array}
\eeq
Note that the new constraints \mb{\cT_{A}} have non-positive ghost numbers
\mb{\gh(\cT_{A})\leq 0} and \mb{\ngh(\cT_{A})\leq 0}, and they vanish if
the old contraints  \mb{\T_{\alpha}} and the 
ghost momenta \mb{\bP_{\alpha}} and \mb{\bcP_{A}} are put to zero,
\beq
\tcT_{A}~=~{\cal O}(\T,\bP,\bcP)~.
\eeq
In other words, if one takes the unitary limit, where the new ghost momenta 
\mb{\bcP_{A}\to 0} vanish, then the new zero-stage constraints 
\mb{\cT_{A_{0}}\to 0} imply that both the old zero-stage constraints
\mb{\T_{\alpha}\to 0} and the old zero-stage ghost momenta
\mb{\bP_{\alpha_{0}}\to 0} vanish, at least within the naive path integral
formulation, \cf \eq{unitarylimit} below.

\subsection{BRST-Improved Hamiltonian}
\label{secnewbrstimprovedham}

\noi
The old BRST-improved Hamiltonian \mb{H_{0}\!=\!H_{0}(q,p;\C,\bP)} is once
again BRST-improved (this time \wrt the new BRST structures), 
\beq
\cH~=~H_{0}+{\cal O}(\cB\bcP)~,
\label{newhamshort}
\eeq
by letting the improved Hamiltonian \mb{\cH\!=\!\cH(q,p;\C,\bP;\cB,\bcP)}
depend on the new ghosts \mb{\cB^{A}} and \mb{\bcP_{B})} in such a way that it
becomes BRST-invariant \wrt both \mb{\omegadelta} and \mb{\omegaone},
\beq
\{\omegadelta,\cH\}~=~0~,~~~~~~~~\{\omegaone,\cH\}~=~0~,~~~~~~~~
\eps(\cH)~=~0~,~~~~~~~~~\ngh(\cH)~=~0~,~~~~~~~~~\gh(\cH)~=~0~.
\label{newhamquantumnumbers}
\eeq
The reason why \mb{\cH} can be chosen to commute simultaneously with both 
BRST operators \mb{\omegadelta} and \mb{\omegaone} is that \mb{\omegadelta}
and \mb{\omegaone}, after all, convey the same BRST symmetry, originally
encoded in the \mb{\omegazero} operator.

\subsection{The \mb{\cS_{1}} Operator}
\label{secessone}

\noi
It is useful to think of the new BRST operators \mb{\omegadelta} and
\mb{\omegaone} as analogues of the old BRST/anti-BRST operators \mb{\omegazero}
and \mb{\bomegazero}, in particular because of the \mb{Sp(2)}-like algebra
\e{sp2doublet} although the ghost number assignments are different,
\cf Table~\ref{analogytable} below. We shall now widen this analogy 
by introducing a counterpart \mb{\cS_{1}} of the old \mb{\eS_{-2}} of
Subsection~\ref{secessminustwo}. To this end, note that \mb{\omegaone} is
\mb{\omegadelta}-closed \e{omegaonedeltaclosed} and has non-zero new ghost
number. Hence, it must be \mb{\omegadelta}-exact, \cf
\eq{ooghomotopy}, \ie there exists a quantity \mb{\cS_{1}} with quantum numbers
\beq
\eps(\cS_{1})~=~0~,~~~~~~~~~\ngh(\cS_{1})~=~1~,~~~~~~~~~\gh(\cS_{1})~=~0~,
\eeq
such that
\beq
\omegaone~=~\{\omegadelta,\cS_{1}\}~.
\label{s1ohs1}
\eeq
In detail, one may expand \mb{\cS_{1}} as follows:
\bea
\cS_{1}&=&\sum_{s=0}^{L}\cB^{A_{s}}\tcT_{A_{s}}
+\Hf\!\!\!\!\!\!\!\!\sum_{\footnotesize
\begin{array}{c}r,s\geq 0\cr r\!+\!s\leq L\end{array}}
\!\!\!\!\!\!\!\!\cB^{B_{s}}\cB^{A_{r}}~
\tcU_{A_{r}B_{s}}{}^{C_{r+s}}~\bcP_{C_{r+s}}
(-1)^{\eps_{B_{s}}+\eps_{C_{r+s}}+r} \cr
&&+\Hf\!\!\!\!\!\!\!\!\sum_{\footnotesize
\begin{array}{c}r,s\geq 0\cr r\!+\!s\!+\!2\leq L\end{array}}
\!\!\!\!\!\!\!\!\cB^{C_{r+s+2}}~
\tcU_{C_{r+s+2}}{}^{B_{s}A_{r}}~\bcP_{A_{r}}\bcP_{B_{s}}(-1)^{\eps_{A_{r}}+r}
+{\cal O}(\cB^{3}\bcP,\cB^{2}\bcP^{2},\cB\bcP^{3}) \cr
&=&\sum_{s=0}^{L}(\B^{\alpha_{s}}\tcX_{\alpha_{s}}
-\Pi_{*}^{\alpha_{s}}\tcY_{\alpha_{s}})+{\cal O}(\cB^{2}\bcP,\cB\bcP^{2})~.
\label{soneexpan}
\eea
It starts with the so-called tilde constraints
\mb{\tcT_{A_{0}}\!=\!\tcT_{A_{0}}(q,p;\C,\bP)}, 
and in the reducible case one also introduces higher-stage tilde constraints
\beq
\tcT_{A_{s+1}}~\equiv~
\tcZ_{A_{s+1}}{}^{B_{s}}~\bcP_{B_{s}}(-1)^{\eps_{B_{s}}+s}
~,~~~~~~~~~~s=0,\ldots,L\!-\!1~.\label{tctees}
\eeq
The \mb{\tcT_{A}} constraints carry the same Grassmann parity as the new
ghosts \mb{\cB} or \mb{\bcP}, or equivalently, the opposite Grassmann 
parity of \mb{\cT_{A}},
\beq
\eps(\tcT_{A})~=~\eps_{A}\!+\!1~,~~~~~~~~~~~
\eps(\tcT_{A_{s}})~=~\eps_{A_{s}}\!+\!s\!+\!1~,~~~~~~~~~~~
\ngh(\tcT_{A_{s}})~=~-s~.
\eeq 
The reducible structure functions
\mb{\tcZ_{A_{s}}{}^{B_{s-1}}\!=\!\tcZ_{A_{s}}{}^{B_{s-1}}(q,p;\C,\bP)}
have Grassmann parity and new ghost number given by
\beq
\eps(\tcZ_{A_{s}}{}^{B_{s-1}})
~=~\eps_{A_{s}}\!+\!\eps_{B_{{s-1}}}\!+\!1~,~~~~~~~~~~~~
\ngh(\tcZ_{A_{s}}{}^{B_{s-1}})~=~0~.
\eeq
The tilde constraints consist of two superpartners
\beq
 \tcT_{A}~\equiv~\{\tcT^{0}_{\alpha};~\tcT^{1}_{\alpha}\}
~\equiv~\{\tcX_{\alpha};~
(-1)^{\eps_{\alpha}}\tcY_{\alpha}\}~,~~~~~~
 \tcT_{A_{s}}~\equiv~\{\tcX_{\alpha_{s}};~
(-1)^{\eps_{\alpha_{s}}+s}\tcY_{\alpha_{s}}\}~.
\eeq
To lowest order, the zeroth-stage tilde constraints 
\mb{\tcT_{A_{0}}\equiv\{\tcT^{0}_{\alpha_{0}};\tcT^{1}_{\alpha_{0}}\}
\equiv\{\tcX_{\alpha_{0}};(-1)^{\eps_{\alpha_{0}}}\tcY_{\alpha_{0}} \}}
are just linear combinations of the old ghost momenta 
\mb{\{\bP_{\alpha_{0}};\bP_{\alpha_{1}} \}},
\beq
\tcX_{\alpha_{0}}~=~\X_{\alpha_{0}}{}^{\beta_{0}}~
\bP_{\beta_{0}}(-1)^{\eps_{\beta_{0}}+1}+{\cal O}(\cB\bcP)~,~~~~~~~~~~~
\tcY_{\alpha_{0}}~=~\Y_{\alpha_{0}}{}^{\beta_{1}}~\bP_{\beta_{1}}
(-1)^{\eps_{\beta_{1}}}+{\cal O}(\cB\bcP)~.
\label{tconstraintbc}
\eeq 
The Grassmann parity and the old ghost number are shifted among the two
superpartner constraints
\beq
\begin{array}{rccclcrcccl}
\eps(\tcT^{0}_{\alpha})\!+\!1&=&\eps_{\alpha}
&=& \eps(\tcT^{1}_{\alpha})~, &&
\eps(\tcT^{0}_{\alpha_{s}})\!+\!1&=&\eps_{\alpha_{s}}\!+\!s
&=& \eps(\tcT^{1}_{\alpha_{s}})~, \\
&&&&&&\gh(\tcT^{0}_{\alpha_{s}})&=&-s\!-\!1
&=&\gh(\tcT^{1}_{\alpha_{s}})\!+\!1~, \\
\eps(\tcX_{\alpha})\!+\!1&=&\eps_{\alpha}
&=& \eps(\tcY_{\alpha})~, &&
\eps(\tcX_{\alpha_{s}})\!+\!1&=&\eps_{\alpha_{s}}\!+\!s
&=& \eps(\tcY_{\alpha_{s}})~, \\
&&&&&&\gh(\tcX_{\alpha_{s}})&=&-s\!-\!1&=&\gh(\tcY_{\alpha_{s}})\!+\!1~.
\end{array}
\eeq
The tilde constraints \mb{\tcT_{A}} have non-positive ghost numbers
\mb{\gh(\tcT_{A})<0} and \mb{\ngh(\tcT_{A})\leq 0}, and they vanish if
the ghost momenta \mb{\bP_{\alpha}} and \mb{\bcP_{A}} are put to zero,
\beq
\tcT_{A}~=~{\cal O}(\bP,\bcP)~. \label{tctgoesdown}
\eeq
Eq.\ \e{tctgoesdown} shows that tilde constraints \mb{\tcT_{A}} indeed can be
viewed as bona-fide constraints in the unitary limit, where the new ghost
momenta \mb{\bcP_{A}\to 0} vanish, \cf \eq{unitarylimit} below. The vanishing
of the old zero-stage ghost momenta \mb{\bP_{\alpha_{0}}\to 0} is enforced via
the superpartners \mb{\cT^{1}_{\alpha_{0}}} of the new zero-stage constraints
\mb{\cT_{A_{0}}\equiv\{\cT^{0}_{\alpha_{0}};\cT^{1}_{\alpha_{0}}\}}, \cf 
\eq{newconstraintssupersplit}.

\subsection{The \mb{\cS_{2}} Operator}
\label{secesstwo}

\noi
Similarly, one may define a \mb{\cS_{2}} quantity as follows.
{}First note that the antibracket \mb{(\cS_{1},\cS_{1})_{\omegadelta}^{}} of 
\mb{\cS_{1}} with itself can be written as a commutator of \mb{\cS_{1}}
and \mb{\omegaone}, 
\beq
(\cS_{1},\cS_{1})_{\omegadelta}^{}~=~\{\cS_{1},\{\omegadelta,\cS_{1}\}\}
~=~\{\cS_{1},\omegaone\}~.
\label{s1s1antibracket}
\eeq 
The antibracket \mb{(\cS_{1},\cS_{1})_{\omegadelta}^{}} is
\mb{\omegadelta}-closed, as the following calculation shows,
\beq
\{\omegadelta,(\cS_{1},\cS_{1})_{\omegadelta}\}
~=~\{\omegadelta,\{\cS_{1},\omegaone\}\}
~=~\{\{\omegadelta,\cS_{1}\},\omegaone\}
~=~\{\omegaone,\omegaone\}~=~0~,
\eeq
since \mb{\omegaone} is \mb{\omegadelta}-closed and nilpotent, \cf 
\eqs{omegaonenilp}{omegaonedeltaclosed}.
Hence there exists a quantity \mb{\cS_{2}} with quantum numbers
\beq
\eps(\cS_{2})~=~0~,~~~~~~~~~\ngh(\cS_{2})~=~2~,~~~~~~~~~\gh(\cS_{2})~=~0~,
\eeq
such that 
\beq
(\cS_{1},\cS_{1})_{\omegadelta}~=~\{\cS_{2},\omegadelta\}~.
\label{s2exists}
\eeq
One may expand \mb{\cS_{2}} as follows:
\bea
\cS_{2}&=&\sum_{s=1}^{L}\cB^{A_{s}}~\brcT_{A_{s}}
+\Hf\cB^{B_{0}}\cB^{A_{0}}~\brcU_{A_{0}B_{0}}(-1)^{\eps_{B_{0}}} \cr
&&+\Hf\!\!\!\!\!\!\!\!\sum_{\footnotesize
\begin{array}{c}r,s\geq 0\cr 0\leq r\!+\!s\!-\!1\leq L\end{array}}
\!\!\!\!\!\!\!\!\cB^{B_{s}}\cB^{A_{r}}~
\brcU_{A_{r}B_{s}}{}^{C_{r+s-1}}~\bcP_{C_{r+s-1}}
(-1)^{\eps_{B_{s}}+\eps_{C_{r+s-1}}+r-1} \cr
&&+\Hf\!\!\!\!\!\!\!\!\sum_{\footnotesize
\begin{array}{c}r,s\geq 0\cr r\!+\!s\!+\!3\leq L\end{array}}
\!\!\!\!\!\!\!\!\cB^{C_{r+s+3}}~
\brcU_{C_{r+s+3}}{}^{B_{s}A_{r}}~\bcP_{A_{r}}\bcP_{B_{s}}(-1)^{\eps_{A_{r}}+r}
+{\cal O}(\cB^{3}\bcP,\cB^{2}\bcP^{2},\cB\bcP^{3})~.
\label{stwoexpan}
\eea
It starts with the breve constraints
\mb{\brcT_{A_{1}}\!=\!\brcT_{A_{1}}(q,p;\C,\bP)}, and in the \mb{L\geq 2} case
one also introduces higher-stage breve constraints
\beq
\brcT_{A_{s+2}}~\equiv~
\brcZ_{A_{s+2}}{}^{B_{s}}~\bcP_{B_{s}}(-1)^{\eps_{B_{s}}+s}
~,~~~~~~~~~~s=0,\ldots,L\!-\!2~.\label{brctees}
\eeq

\subsection{Operator Master Equation}
\label{secome}

\begin{table}[t] 
\caption{A dictionary between notions in the old and the new sector.}
\label{analogytable}
\begin{center}
\begin{tabular}{|c||c|c|} \hline
$\downarrow$~Notion~$\backslash$~Sector~$\rightarrow$&
\rule[0ex]{9ex}{0ex} Old \rule[0ex]{9ex}{0ex} &New \\ \hline\hline
Ghosts&$\C^{\alpha}$&$\cB^{A}\equiv\{\B^{\alpha};
(-1)^{\eps_{\alpha}+1}\Pi_{*}^{\alpha}\}$ \\
Ghost momenta&$\bP_{\alpha}$&
$\bcP_{A}\equiv\{\Pi_{\alpha};\B^{*}_{\alpha}\}$ \\ 
Ghost operator&$G=\eS_{0}$&$\gbb=\cS_{0}$ \\ 
Ghost number&$\gh=\{G,\cdot\}$&$\ngh=\{\gbb,\cdot\}$ \\ 
$1$st BRST operator &$\omegazero$&$\omegadelta$ \\ 
$2$nd BRST operator &$\bomegazero=\{\omegazero,\eS_{-2}\}$&
$\omegaone=\{\omegadelta,\cS_{1}\}$ \\ 
Homotopy operator &(Not introduced)&$\bomegadelta$ \\
&$\eS_{-2k}$&$\cS_{k}$ \\
&$\eS=\sum_{k=0}^{\infty}\eS_{-2k}$&$\cS=\sum_{k=0}^{\infty}\cS_{k}$ \\
Operator master eq. &$\{\eS,\{\eS,\omegazero\}\}=\omegazero$&
$\{\cS,\{\cS,\omegadelta\}\}=\{\cS,\omegadelta\}$ \\
\hline
\end{tabular}
\end{center}
\end{table}

\noi
The process of finding higher and higher cohomology relations may be automated
by introducing an operator master equation \cite{BatMar98a}
\beq
(\cS,\cS)^{}_{\omegadelta}~\equiv~\{\{\cS,\omegadelta\},\cS\}
~=~\{\omegadelta,\cS\}~.
\label{ome}
\eeq
The operator master equation \e{ome} expresses that \mb{\{\cS,\cdot\}} acts as
an idempotent on \mb{\omegadelta}, \cf \Ref{BatLyaMar02}. To define the
operator \mb{\cS}, one first let a quantity \mb{\cS_{0}} be the new ghost
operator, 
\beq
 \cS_{0}~=~ \gbb~.
\label{omebc}
\eeq
{}Furthermore, one may show by mathematical induction that there exist
quantities \mb{\cS_{k}} for each positive integer \mb{k\ge 1}, and with quantum
numbers
\beq
\eps(\cS_{k})~=~0~,~~~~~~~~~\ngh(\cS_{k})~=~k~,~~~~~~~~~\gh(\cS_{k})~=~0~,
\eeq
such that the sum
\beq
 \cS~\equiv~\sum_{k=0}^{\infty}\cS_{k}~,~~~~~~~~~\eps(\cS)~=~0
~,~~~~~~~~~\gh(\cS)~=~0~,
\eeq
of indefinite new ghost number, satisfies the above operator master \eq{ome},
\cf Appendix~\ref{appome}. The master \eq{ome} is really an infinite tower of
equations,
\beq
\sum_{j=1}^{k-1}(\cS_j,\cS_{k-j})^{}_{\omegadelta}
~=~(k\!-\!1) \{ \cS_k,\omegadelta \}~,~~~~~~~~~~ k\in\{0,1,2,\ldots\}~.
\label{ometower}
\eeq
The first non-trivial equation (with \mb{k\!=\!2}) is just 
\eq{s2exists}. The maximal arbitrariness of solutions \mb{\cS} to the operator
master \eq{ome} with the boundary conditions \es{s1ohs1}{omebc} is given by
canonical transformations of the form \cite{BatMar99b}
\beq
\cS^{\prime}~=~e^{-\Ih \{\omegadelta,\Psi\}} \cS e^{\Ih \{\omegadelta,\Psi\}}
~=~e^{\{\{\omegadelta,\Psi\},~\cdot~\}}\cS~,
\label{maximalarbitrariness}
\eeq
where the generating Fermionic operator
\beq
 \Psi~\equiv~\sum_{k=1}^{\infty}\Psi_{k}~,~~~~~~~~~
\eps(\Psi)~=~1~,~~~~~~~~~\gh(\Psi)~=~-1~,
\eeq
has indefinite new ghost number,
\beq
\eps(\Psi_{k})~=~1~,~~~~~~~~~\ngh(\Psi_{k})~=~k~,~~~~~~~~~\gh(\Psi_{k})~=~-1~.
\eeq
The first few components of the transformed solution \e{maximalarbitrariness} 
read
\bea
\cS^{\prime}_{0}&=&S_{0}~=~\gbb~,\label{maximalarbitrariness0} \\
\cS^{\prime}_{1}&=&S_{1}-\{\omegadelta,\Psi_{1}\}~,
\label{maximalarbitrariness1} \\
\cS^{\prime}_{2}&=&S_{2}+\{\{\omegadelta,\Psi_{1}\},\cS_{1}\}
-2\{\omegadelta,\Psi_{2}\}~, \label{maximalarbitrariness2} \\
\vdots && \nonumber
\eea
It is instructive to mention that the new ghost operator \mb{\gbb} by itself
is a trivial solution for \mb{\cS} to the operator master \eq{ome} and to the
first boundary condition \e{omebc}, but it fails to meet the second boundary
condition \e{s1ohs1}.

\noi
{\em Proof of properties related to \eq{maximalarbitrariness}:} It follows from
the Baker-Campbell-Hausdorff Theorem that the transformations
\e{maximalarbitrariness} form a group. It is also clear that the
transformations \e{maximalarbitrariness} preserve the operator master \eq{ome}
with the boundary conditions \es{s1ohs1}{omebc}, so it only remains to prove
that all solutions are connected this way. Let there be given two arbitrary
solutions \mb{\cS^{(1)}} and \mb{\cS^{(2)}}. {}For all \mb{k\geq 1}, call the 
difference for \mb{\cX_{k}\equiv\cS^{(2)}_{k}-\cS^{(1)}_{k}}.
Consider successively, for each \mb{k\geq 1}, starting with the smallest \mb{k}
and going up, whether the difference \mb{\cX_{k}} is \mb{\omegadelta}-exact. 
If this is the case, perform a suitable transformation \e{maximalarbitrariness}
of the second solution \mb{\cS^{(2)}}, generated by some \mb{\Psi_{ k}}, such
that the new difference \mb{\cX_{k}\!=\!0} becomes identically zero. One now
proceeds by indirect reasoning: Assume that this process stops at some step
\mb{k}, \ie
\mb{\cX_{1}\!=\!\cX_{2}\!=\!\ldots\!=\!\cX_{k-1}\!=\!0}, but \mb{\cX_{k}} is
not \mb{\omegadelta}-exact. 
On the other hand, the two \eqs{s1ohs1}{ometower} now guarantee that the
difference \mb{\cX_{k}} is \mb{\omegadelta}-closed in the two cases \mb{k=1} 
and \mb{k\geq 2}, respectively. Since \mb{\cX_{k}} has non-vanishing new ghost
number, \mb{\ngh(\cX_{k})=k \neq 0}, one then concludes from cohomology
considerations that \mb{\cX_{k}} is \mb{\omegadelta}-exact, \cf
\eq{ooghomotopy}. This is in contradiction with the above, and hence the
process did not stop after all.

\subsection{An Inverse Relation to \mb{\omegaone=\{\omegadelta,\cS_{1}\}}}
\label{secinvrel}

\noi
{}For any given solution \mb{\cS}, it is possible to create an interesting 
primed solution \mb{\cS^{\prime}} by a canonically transformation
\e{maximalarbitrariness} that has an \mb{\bomegadelta}-exact Fermionic
generator of the form
\beq
\Psi_{1}~=~\{\bomegadelta, \cS_{1}\}~,
\label{invpsione}
\eeq
and where the higher generators \mb{\Psi_{k\geq 2}} remain unspecified. Then an
application of the Jacobi identity 
\mb{\{\omegadelta,\{\bomegadelta, \cS_{1}\}\}=
\{\{\omegadelta,\bomegadelta\}, \cS_{1}\}
-\{\bomegadelta,\{\omegadelta, \cS_{1}\}\}}
yields
\beq
\cS^{\prime}_{1}
~=~\cS_{1}-\{\omegadelta,\Psi_{1}\}
~=~\cS_{1}-\{\gbb, \cS_{1}\}
+\{\bomegadelta,\{\omegadelta,\cS_{1}\}\}
~=~\{\bomegadelta,\omegaone\}~,
\label{invs1ohs1}
\eeq 
\cf eqs.\ \e{ooghomotopy}, \es{s1ohs1}{maximalarbitrariness1}.
Eq.\ \e{invs1ohs1} can be thought of as an inverse relation to \eq{s1ohs1}.
Whereas \eq{s1ohs1} gives the new \mb{\cT_{A_{0}}} constraints in terms of the
tilde constraints \mb{\tcT_{A_{0}}}, \cf \eq{brstinvdoubleconstr} below,
the inverse relation \e{invs1ohs1} gives the primed tilde constraints
\mb{\tcT^{\prime}_{A_{0}}} in terms of the \mb{\cT_{A_{0}}} constraints. 

\noi
{}Finally note that if one successively repeats the above transformation 
\e{invpsione} and transforms the primed solution \mb{\cS^{\prime}} with a
Fermionic generator
\beq
\Psi^{\prime}_{1}~=~\{\bomegadelta, \cS^{\prime}_{1}\}
~=~\{\bomegadelta,\{\bomegadelta,\omegaone\} \}~=~0~,
\label{invprimepsione}
\eeq
nothing happens, because of the \mb{\bomegadelta} nilpotency 
\e{bomegadeltaquantumnumbers}. So the primed solution \mb{\cS^{\prime}} is
stabile in this sense.

\section{Assembling the Pieces}
\label{secassemble}

\noi
Let us briefly summarize the construction so far. We are considering a physical
gauge system, which is ordinarily described by a BRST-improved Hamiltonian
\mb{H_{0}}, an ordinary BRST charge \mb{\omegazero} and an ordinary ghost
number ``\mb{\gh}''. We introduced twice as many new ghosts \mb{\cB^{A}} and 
an improved BRST operator \mb{\omegadelta}, which is the old BRST charge 
\mb{\omegazero} adapted to the new ghost sector. We then collected the old
constraints \mb{\T_{\alpha_{0}}} and the old ghost momenta
\mb{\bP_{\alpha_{0}}} into new constraints  \mb{\cT_{A_{0}}} and introduced a
new  BRST charge \mb{\omegaone\!=\!\cB^{A}\cT_{A}+\ldots}, and a
new ghost number called ``\mb{\ngh}''. We then lifted
cohomologically the new BRST operator \mb{\omegaone} to a hierarchy of
quantities \mb{\cS\!\equiv\!\sum_{k=0}^{\infty}\cS_{k}}, which satisfies an
operator master \eq{ome}. In this Section we will stress the main features
and show how to gauge-fix the Hamiltonian.

\subsection{The Main Relation}

\noi
We now display some of the consequences of the relation
\mb{\omegaone=\{\omegadelta,\cS_{1}\}} in more detail, \cf \e{s1ohs1}.
\bea
\cT_{A_{0}}&=&\cV_{A_{0}}{}^{B_{0}}\tcT_{B_{0}}-\{\tcT_{A_{0}},\omegazero\}~,
\label{brstinvdoubleconstr} \\
\cZ_{A_{s+1}}{}^{B_{s}}&=&\cV_{A_{s+1}}{}^{C_{s+1}}\tcZ_{C_{s+1}}{}^{B_{s}}
-\tcZ_{A_{s+1}}{}^{C_{s}}~\cV_{C_{s}}{}^{B_{s}}
+\{\tcZ_{A_{s+1}}{}^{B_{s}},\omegazero\}(-1)^{\eps_{B_{s}}+s} \cr
&&+\cV_{A_{s+1}}{}^{B_{s}C_{0}}\tcT_{C_{0}}(-1)^{\eps_{B_{s}}+s}
+\frac{i\hbar}{2}\sum_{r=0}^{s}\cV_{A_{s+1}}{}^{C_{r}D_{s-r}}~
\tcU_{D_{s-r}C_{r}}{}^{B_{s}} \cr
&&-\frac{i\hbar}{2}\sum_{r=0}^{s-1}\tcU_{A_{s+1}}{}^{C_{r}D_{s-1-r}}~
\cV_{D_{s-1-r}C_{r}}{}^{B_{s}}
+\left({\cal O}(\hbar^{2})~{\rm terms,~if}~R\geq 3~{\rm and}~s\geq 1\right)~,\\
\cU_{A_{0}B_{0}}{}^{C_{0}}
&=&-\cV_{A_{0}}{}^{D_{0}}~\tcU_{D_{0}B_{0}}{}^{C_{0}}(-1)^{\eps_{B_{0}}}
-\Hf\tcU_{A_{0}B_{0}}{}^{D_{0}}~\cV_{D_{0}}{}^{C_{0}}
+\Hf\cV_{A_{0}B_{0}}{}^{D_{1}}\tcZ_{D_{1}}{}^{C_{0}} \cr
&&+\Hf\{\tcU_{A_{0}B_{0}}{}^{C_{0}},\omegazero\}(-1)^{\eps_{C_{0}}}
-\{\tcT_{A_{0}},\cV_{B_{0}}{}^{C_{0}}\}(-1)^{\eps_{B_{0}}} \cr
&&+\Hf\cV_{A_{0}B_{0}}{}^{C_{0}D_{0}}~\tcT_{D_{0}}(-1)^{\eps_{C_{0}}}
+\frac{i\hbar}{4}\cV_{A_{0}B_{0}}{}^{D_{0}E_{0}}~\tcU_{E_{0}D_{0}}{}^{C_{0}}
-(-1)^{\eps_{A_{0}}\eps_{B_{0}}}(A_{0}\leftrightarrow B_{0})~. \label{cuabc} 
\eea
{}From \eq{brstinvdoubleconstr} follows that the new zeroth-stage constraints
\mb{\cT_{A_{0}}} are BRST-invariant up to the first term on the left-hand
side, which depends on ghost momenta and vanishes in the unitary limit, \cf
\eq{unitarylimit} below. Eq.\ \e{brstinvdoubleconstr} is the heart of the
construction. It may be regarded as a generalization of \eq{brstinvconstr} in
the Introduction, where the tilde constraints \mb{\tcT_{A_{0}}} generalize the
string ghost momenta \mb{b_{m}}.

\subsection{Non-Minimal Sector}
\label{secnonmin}

\noi
In this Subsection we show how to gauge-fix the extended BRST symmetries. We
shall for simplicity concentrate on the irreducible case \mb{L\!=\!0}. The 
reducible case \mb{L\!>\!0} will be discussed elsewhere. One first introduces 
new non-minimal variables 
\mb{(\cP^{A_{0}},\bcB_{B_{0}};\lambda^{A_{0}},\pi^{}_{B_{0}})} with canonical
commutation relations
\beq
   \{\cP^{A_{0}}, \bcB_{B_{0}} \}~=~\delta^{A_{0}}_{B_{0}}~,~~~~~~~~~~~~~
 \{\lambda^{A_{0}}, \pi^{}_{B_{0}} \}~=~\delta^{A_{0}}_{B_{0}}~,
\eeq
and where the remaining canonical commutation relations are zero. (Pay 
attention to the perhaps confusing but commonly used convention that
\mb{\cP^{A_{0}}} is a coordinate, while the Faddeev-Popov antighost
\mb{\bcB_{A_{0}}} is a momentum.) The Grassmann parity and new ghost number
are
\beq
\begin{array}{rccclcrcccl}
\eps(\cP^{A_{0}})&=& \eps_{A_{0}}\!+\!1&=&\eps(\bcB_{A_{0}})~, &~~~~~~&
\ngh(\cP^{A_{0}})&=& 1&=&-\ngh(\bcB_{A_{0}})~, \\
\eps(\lambda^{A_{0}})&=& \eps_{A_{0}}&=&\eps(\pi^{}_{A_{0}})~, &&
\ngh(\lambda^{A_{0}})&=&0&=&\ngh(\pi^{}_{A_{0}})~.
\end{array}
\eeq
The \mb{4} new non-minimal variables 
\mb{(\cP^{A_{0}},\bcB_{B_{0}};\lambda^{A_{0}},\pi^{}_{B_{0}})} are naturally
divided into \mb{2\times 4=8} superpartner fields. The Grassmann parity and
the old ghost number are shifted among the superpartners as follows
\beq
\begin{array}{rccclcrcccl}
\eps(\cP_{0}^{\alpha_{0}})&=&\eps_{\alpha_{0}}\!+\!1
&=&\eps(\bcB^{0}_{\alpha_{0}})~, &~~~~~~&
\gh(\cP_{0}^{\alpha_{0}})&=&1&=&-\gh(\bcB^{0}_{\alpha_{0}})~,  \\
\eps(\cP_{1}^{\alpha_{0}})&=&\eps_{\alpha_{0}}
&=&\eps(\bcB^{1}_{\alpha_{0}})~,&&
\gh(\cP_{1}^{\alpha_{0}})&=&2&=&-\gh(\bcB^{1}_{\alpha_{0}})~,  \\
\eps(\lambda_{0}^{\alpha_{0}})&=&\eps_{\alpha_{0}}
&=&\eps(\pi^{0}_{\alpha_{0}})~, &~~~~~~&
\gh(\lambda_{0}^{\alpha_{0}})&=&0&=&\gh(\pi^{0}_{\alpha_{0}})~,  \\
\eps(\lambda_{1}^{\alpha_{0}})&=&\eps_{\alpha_{0}}\!+\!1
&=&\eps(\pi^{1}_{\alpha_{0}})~,&&
\gh(\lambda_{1}^{\alpha_{0}})&=&1&=&-\gh(\pi^{1}_{\alpha_{0}})~.
\end{array} 
\label{nonminnewstatistics}
\eeq
The non-minimal extensions of the two BRST operators \mb{\omegadelta} and 
\mb{\omegaone} are
\bea
\omegadelta&=&\omegadelta_{\min}+\cP_{1}^{\alpha_{0}}\bcB^{0}_{\alpha_{0}}
-\lambda_{1}^{\alpha_{0}}\pi^{0}_{\alpha_{0}}~, 
\label{nonminomegadelta} \\
\omegaone&=&\omegaonemin+\cP^{A_{0}}\pi^{}_{A_{0}}~.\label{nonminomegaone}
\eea
This extension is consistent with the \mb{Sp(2)}-like algebra \e{sp2doublet}.
Similarly, the non-minimal extensions to the ghost operators \mb{G} and 
\mb{\gbb} and the anti-BRST operator \mb{\bomegadelta} read
\bea
G&=&G_{\min}-\Hf[\bcB^{0}_{\alpha_{0}},\cP_{0}^{\alpha_{0}}]_\plus
-[\bcB^{1}_{\alpha_{0}},\cP_{1}^{\alpha_{0}}]_\plus
-\Hf[\pi^{1}_{\alpha_{0}},\lambda_{1}^{\alpha_{0}}]_\plus~, 
\label{nonmingee} \\
\gbb&=&\gbb_{\min}-\Hf[\bcB_{A_{0}},\cP^{A_{0}}]_\plus
~=~\gbb_{\min}- \bcB_{A_{0}}\cP^{A_{0}}~, 
\label{nonmingeebb} \\
\bomegadelta&=&\bomegadelta_{\min}-\cP_{0}^{\alpha_{0}}\bcB^{1}_{\alpha_{0}}~.
\eea
The gauge-fixed (or unitarizing) Hamiltonian 
\beq
 \cH_{\Phi}~=~\cH+\{\omegaone,\{\omegadelta, \Phi\}\}~=~
\cH+\{\omegaone, \Psi_{\Phi}\}
\label{unitarizingham1}
\eeq
depends on a gauge Boson \mb{\Phi} through a \mb{\omegadelta}-exact gauge
{}Fermion \mb{\Psi_{\Phi}} of the form
\beq
\Psi_{\Phi}~\equiv~\{\omegadelta,\Phi\}~.
\eeq
Here the BRST-improved Hamiltonian \mb{\cH} should be invariant \wrt both the 
BRST operators \mb{\omegadelta} and \mb{\omegaone}, \cf
\eq{newhamquantumnumbers}. The quantum numbers for \mb{\Phi} and
\mb{\Psi_{\Phi}} are
\beq
\begin{array}{rclcrclcrcl}
\eps(\Phi)&=&0~,&~~~~~~~~~&\ngh(\Phi)&=&-1~,&~~~~~~~~~&\gh(\Phi)&=&-2~, \\
\eps(\Psi_{\Phi})&=&1~,&&\ngh(\Psi_{\Phi})&=&-1~,&&\gh(\Psi_{\Phi})&=&-1~.
\end{array}
\eeq
A simple gauge choice is 
\beq
 \Phi~=~\bcB^{1}_{\alpha_{0}}\chi_{0}^{\alpha_{0}}
+\bcP^{1}_{\alpha_{0}}\lambda_{0}^{\alpha_{0}}~,
\label{simplegaugechoice}
\eeq
where the Faddeev-Popov matrix 
\mb{\{\chi_{0}^{\alpha_{0}},\cT^{0}_{\beta_{0}}\}} carries maximal rank
\beq
\rank\{\chi_{0}^{\alpha_{0}},\cT^{0}_{\beta_{0}}\}~=~m_{0}~.
\eeq
Then the corresponding gauge fermion \mb{\Psi_{\Phi}} becomes
\beq
\Psi_{\Phi}~=~\{\omegaone, \Psi_{\Phi}\}
~=~\bcB^{0}_{\alpha_{0}}\chi_{0}^{\alpha_{0}}
-\bcB^{1}_{\alpha_{0}}\{\chi_{0}^{\alpha_{0}},\omegazero\}
+\bcP^{1}_{\alpha_{0}} \lambda_{1}^{\alpha_{0}}
+\{\omegadelta,\bcP^{1}_{\alpha_{0}}\} \lambda_{0}^{\alpha_{0}}~.
\eeq
This is consistent with a gauge fermion 
\mb{\Psi_{\Phi}} of the form
\beq
\Psi_{\Phi}~=~\bcB_{A_{0}}\chi^{A_{0}}+\cX_{A_{0}} \lambda^{A_{0}}
+{\cal O}(\lambda\cB\bcP^{2})~.
\eeq
where \mb{\chi^{A_{0}}\equiv\{\chi_{0}^{\alpha_{0}};\chi_{1}^{\alpha_{0}}\}}
are gauge-fixing conditions with
\beq
\begin{array}{rclcrcl}
\eps(\chi^{A_{0}})&=&\eps_{A_{0}}~,&~~~~~~~&\ngh(\chi^{A_{0}})&=&0~, \\
\eps(\chi_{0}^{\alpha_{0}})&=&\eps_{\alpha_{0}}~,&&
\gh(\chi_{0}^{\alpha_{0}})&=&0~,  \\
\eps(\chi_{1}^{\alpha_{0}})&=&\eps_{\alpha_{0}}\!+\!1~,&&
\gh(\chi_{1}^{\alpha_{0}})&=&1~.
\end{array}
\label{gaugefixingcond}
\eeq
The superpartners \mb{\chi_{1}^{\alpha_{0}}} are of the form
\beq
\chi_{1}^{\alpha_{0}}~=~-\{\chi_{0}^{\alpha_{0}},\omegazero\}
~=~-\{\chi_{0}^{\alpha_{0}},\T_{\beta_{0}}\}\C^{\beta_{0}}+{\cal O}(\C^{2}\bP)
~=~{\cal O}(\C)~.
\eeq
They have positive ghost number, and they vanish if and only if the old ghosts
\mb{\C^{\alpha_{0}}} are put to zero.

\noi
The \mb{\cX_{A_{0}}\equiv\{\cX^{0}_{\alpha_{0}};\cX^{1}_{\alpha_{0}}\}} are
superpartners to the new constraints \mb{\cT_{A_{0}}}, \ie one may
introduce a next generation of superpartners
\mb{\cT_{{\cal A}_{0}}\equiv\{\cT_{A_{0}};\cX_{A_{0}}\}\equiv
\{\cT^{0}_{\alpha_{0}},\cT^{1}_{\alpha_{0}};\cX^{0}_{\alpha_{0}},
\cX^{1}_{\alpha_{0}}\}}.
(This leads naturally to a third generation of BRST operators and ghost
numbers. Here we count the old BRST operator \mb{\omegazero} and the old
ghost number ``\mb{\gh}'' as belonging to the first generation, and the new
BRST operator \mb{\omegaone} and the new ghost number ``\mb{\ngh}'' as the
second generation. Therefore in this terminology the next generation is the
third generation. In principle, one may introduce infinitely many generations.)
One has
\beq
\eps(\cX_{A_{0}})~=~\eps_{A_{0}}\!+\!1~,~~~~~~\ngh(\cX_{A_{0}})~=~-1~.
\eeq
In fact,
\bea
\cX^{0}_{\alpha_{0}}&=&\cV_{\alpha_{0}}^{1}{}_{0}^{\beta_{0}}~
\bcP^{0}_{\beta_{0}}(-1)^{\eps_{\beta_{0}}}~, \label{bcpindisguise1} \\
\cX^{1}_{\alpha_{0}}&=&\bcP^{1}_{\alpha_{0}}~, \label{bcpindisguise2}
\eea 
where \mb{\cV_{\alpha_{0}}^{1}{}_{0}^{\beta_{0}}} is the invertible matrix
from \eq{omegadeltarankcond1}. The constraints \mb{\cX_{A_{0}}} have negative
ghost numbers \mb{\gh(\cX_{A_{0}})<0} and \mb{\ngh(\cX_{A_{0}})< 0}, and they
vanish if and only if the new ghost momenta \mb{\bcP_{A_{0}}} are put to zero,
\beq
\cX_{A_{0}}~=~{\cal O}(\bcP)~.\label{bcpindisguise3}
\eeq
Let us mention for completeness that the unitary limit is obtained through the
substitution 
\beq
\bcB_{A_{0}} ~\to~ \epsilon~ \bcB_{A_{0}}~,~~~~~~~
\pi^{}_{A_{0}} ~\to~ \epsilon~ \pi^{}_{A_{0}}~,~~~~~~~
\chi^{A_{0}} ~\to~ \frac{1}{\epsilon} ~\chi^{A_{0}}~,
\label{unitarylimit}
\eeq
and letting the parameter \mb{\epsilon\to 0}. In the naive path integral,
the kinetic terms for the non-original variables 
\beq
\bcP_{A_{0}}\dot{\cB}^{A_{0}}
+\bcB_{A_{0}}\dot{\cP}^{A_{0}}
+\pi^{}_{A_{0}}\dot{\lambda}^{A_{0}}~\rightarrow~0
\label{kineticterms}
\eeq
vanish in this limit. The unitarizing Hamiltonian \e{unitarizingham1} becomes
\beq
\cH_{\Phi}~=~\cH+\pi^{}_{A_{0}}\chi^{A_{0}}
+\cT_{A_{0}}\{\cB^{A_{0}}, \cX_{B_{0}} \} \lambda^{B_{0}}
+\cX_{A_{0}}\cP^{A_{0}}
+\bcB_{A_{0}}\{\chi^{A_{0}}, \cT_{B_{0}} \} \cB^{B_{0}}
+{\rm higher~order~terms}~.
\label{unitaringham2}
\eeq
The second (resp.\ third) term on the \rhs is the delta-function term for the
gauge-fixing (resp.\ gauge-generating) constraints \mb{\chi^{A_{0}}} (resp.\ 
\mb{\cT_{A_{0}}}), where the integration is over the \mb{\pi^{}_{A_{0}}}
(resp.\ \mb{\lambda^{A_{0}}}) variables. The fourth term is the delta-function
term for the next generation of superpartners \mb{\cX_{A_{0}}} to the
gauge-generating constraints, where the integration is now over the
\mb{\cP^{A_{0}}} variables. (Recall that \mb{\cX_{A_{0}}} are the new ghost
momenta \mb{\bcP_{A_{0}}} in disguise, \cf
\eqs{bcpindisguise1}{bcpindisguise2}. This fact implies that the first kinetic
term \e{kineticterms} drops out. The last two kinetic terms \e{kineticterms} 
are suppressed by a factor of \mb{\epsilon}, \cf \eq{unitarylimit}.) And 
finally, the fifth term in \eq{unitaringham2} is the Faddeev-Popov determinant
term.

\section{Abelian Case} 
\label{secabel}

\noi
In this Section we provide details for the construction in the Abelian
reducible case. Recall that one may in principle Abelianize {\em any} BRST
operator \mb{\omegazero} by unitary similarity transformations (although
locality and symmetries, such as \eg Lorentz symmetry, may be lost in the
process). Therefore, from a theoretical perspective, it is enough to consider
the Abelian case.

\subsection{Old BRST Operator \mb{\omegazero}}

\begin{table}[b] 
\caption{The rectangular matrix \mb{\Z_{\alpha_{s+1}}{}^{\beta_{s}}} subjected
to the ultimate Abelianization consists entirely of blocks of zero-matrices and
unit-matrices.}
\label{ztable}
\begin{center}
\begin{tabular}{|c||ccccccccc|} \hline
$\alpha~\backslash~\beta $&\multicolumn{2}{c}{$\beta_{0}$}&
\multicolumn{2}{c}{$\beta_{1}$}&$\cdots$&
\multicolumn{2}{c}{$\beta_{L-2}$}&\multicolumn{2}{c|}{$\beta_{L-1}$}
\\ \hline\hline
\lower 1.5ex \hbox{$\alpha_{1}$}&${\bf 0}$&${\bf 1}$&&&&&&&  \\ 
&${\bf 0}$&${\bf 0}$&&&&&&& \\
\lower 1.5ex \hbox{$\alpha_{2}$}&&&${\bf 0}$&${\bf 1}$&&&&& \\ 
&&&${\bf 0}$&${\bf 0}$&&&&& \\
$\vdots$&&&&&$\ddots$&&&& \\ 
\lower 1.5ex \hbox{$\alpha_{L-1}$}&&&&&&${\bf 0}$&${\bf 1}$&& \\ 
&&&&&&${\bf 0}$&${\bf 0}$&& \\ 
$\alpha_{L}$&&&&&&&&${\bf 0}$&${\bf 1}$ \\ \hline
\end{tabular}
\end{center}
\end{table}

\noi
The reducible Abelian Ansatz for \mb{\omegazero} is defined as 
\beq
\omegazero~=~\C^{\alpha}\T_{\alpha}
~=~\C^{\alpha_{0}}\T_{\alpha_{0}}+
\sum_{s=0}^{L-1}\C^{\alpha_{s+1}}
\Z_{\alpha_{s+1}}{}^{\beta_{s}}~\bP_{\beta_{s}}
(-1)^{\eps_{\beta_{s}}+s}~.
\label{abelansatzomegazero}
\eeq
The \mb{\omegazero} nilpotency \e{oldbrstcharge} is equivalent to the following
set of equations:
\bea
\Z_{\alpha_{1}}{}^{\beta_{0}}~\T_{\beta_{0}}&=&0~,\label{abelreductees} \\
\Z_{\alpha_{s+2}}{}^{\beta_{s+1}}~\Z_{\beta_{s+1}}{}^{\gamma_{s}}
&=&0~,\label{abelzzrelation} \\
\{\T_{\alpha_{0}},\T_{\beta_{0}}\}&=&0~,\label{abelinvolution} \\
\{\T_{\alpha_{0}}, \Z_{\beta_{s+1}}{}^{\gamma_{s}}\}&=&0~,\label{abelteez} \\
(-1)^{(\eps_{\beta_{s+1}}+s)(\eps_{\gamma_{s}}+s+1)}
\{\Z_{\alpha_{s+1}}{}^{\gamma_{s}},\Z_{\beta_{s+1}}{}^{\delta_{s}}\}
&=&(-1)^{(\eps_{\gamma_{s}}+s)(\eps_{\delta_{s}}+s)}
(\gamma_{s}\leftrightarrow \delta_{s})~,\label{abelzzsomemore} \\
\{\Z_{\alpha_{r+1}}{}^{\gamma_{r}},\Z_{\beta_{s+1}}{}^{\delta_{s}}\}&=&0
~,~~~~~~~~~~~r\neq s~. \label{abelzzotherwise}
\eea
The \mb{\omegazero} nilpotency \e{oldbrstcharge} does {\em not} guarantee by 
itself that all the structure functions \mb{\Z_{\alpha_{s+1}}{}^{\beta_{s}}} 
commute, \cf \eq{abelzzsomemore}, \ie there could be a small non-Abelian 
remnant left in the Abelian Ansatz \e{abelansatzomegazero}, despite the name.
Nevertheless, it is always possible to ensure that all the structure functions 
\mb{\Z_{\beta_{s+1}}{}^{\gamma_{s}}} commute
\beq
\{\Z_{\alpha_{r+1}}{}^{\gamma_{r}},\Z_{\beta_{s+1}}{}^{\delta_{s}}\}~=~0~,
\label{finalabelianization}
\eeq
\wtho a rotation that preserves the Ansatz \e{abelansatzomegazero}, \cf 
Table~\ref{ztable}.

\subsection{Anti-BRST Operator \mb{\bomegazero}}

\noi
Similarly, the anti-BRST operator \mb{\bomegazero} is in the reducible Abelian
case given as
\beq
\bomegazero~=~\bT^{\alpha}\bP_{\alpha}(-1)^{\eps_{\alpha}}
~=~\bT^{\alpha_{0}}\bP_{\alpha_{0}}(-1)^{\eps_{\alpha_{0}}}
+\sum_{s=1}^{L}\C^{\alpha_{s-1}}\bZ_{\alpha_{s-1}}{}^{\beta_{s}}~
\bP_{\beta_{s}}(-1)^{\eps_{\beta_{s}}+s}~.
\eeq
The \mb{\bomegazero} nilpotency \e{boldbrstcharge} is equivalent to the
following set of equations:
\bea
\bT^{\alpha_{0}}~\bZ_{\alpha_{0}}{}^{\beta_{1}}&=&0~,\label{babelreductees} \\
\bZ_{\alpha_{s}}{}^{\beta_{s+1}}\bZ_{\beta_{s+1}}{}^{\gamma_{s+2}}
&=&0~,\label{babelzzrelation} \\
\{\bT^{\alpha_{0}},\bT^{\beta_{0}}\}&=&0~,\label{babelinvolution} \\
\{\bT^{\alpha_{0}}, \bZ_{\beta_{s}}{}^{\gamma_{s+1}}\}&=&0~,\label{babelteez}\\
(-1)^{(\eps_{\beta_{s}}+s+1)(\eps_{\gamma_{s+1}}+s)}
\{\bZ_{\alpha_{s}}{}^{\gamma_{s+1}},\bZ_{\beta_{s}}{}^{\delta_{s+1}}\}
&=&(-1)^{(\eps_{\alpha_{s}}+s)(\eps_{\beta_{s}}+s)}
(\alpha_{s}\leftrightarrow \beta_{s})~,\label{babelzzsomemore} \\
\{\bZ_{\alpha_{r}}{}^{\gamma_{r+1}},\bZ_{\beta_{s}}{}^{\delta_{s+1}}\}&=&0
~,~~~~~~~~~~~r\neq s~. \label{babelzzotherwise}
\eea
The compatibility \e{boldbrstcharge} of the BRST and the anti-BRST operator,
\ie the fact that they commute, yields
\bea
\bT^{\alpha_{0}}~\T_{\alpha_{0}}&=&0~, \label{abeltees1} \\
\Z_{\alpha_{s}}{}^{\beta_{s-1}}\bZ_{\beta_{s-1}}{}^{\gamma_{s}}
+\bZ_{\alpha_{s}}{}^{\beta_{s+1}}\Z_{\beta_{s+1}}{}^{\gamma_{s}}
&=&0~,\label{abelzzbarrelation} \\
\{\T_{\alpha_{0}},\bT^{\gamma_{0}}\}
+\bZ_{\alpha_{0}}{}^{\beta_{1}}\Z_{\beta_{1}}{}^{\gamma_{0}}
&=&0~,\label{abeltees2} \\
\{\T_{\alpha_{0}}, \bZ_{\beta_{s}}{}^{\gamma_{s+1}}\}&=&0~,\label{abelteez1}\\
\{\Z_{\alpha_{s+1}}{}^{\beta_{s}},\bT^{\gamma_{0}}\}&=&0~,\label{abelteez2}\\
\{\Z_{\alpha_{r+1}}{}^{\gamma_{r}},\bZ_{\beta_{s}}{}^{\delta_{s+1}}\}&=&0~.
\label{abelzzbarotherwise}
\eea
In view of the non-Abelian remnant in \eqs{abelzzsomemore}{babelzzsomemore},
it is a small miracle that the structure functions
\mb{\Z_{\alpha_{r+1}}{}^{\gamma_{r}}} and \mb{\bZ_{\beta_{s}}{}^{\delta_{s+1}}}
commute, \cf \eq{abelzzbarotherwise}. 
The complete \mb{\bT^{\alpha_{0}}} solution to \eq{abeltees1} alone, is
\beq
\bT^{\alpha_{0}}~=~A^{\alpha_{0}\beta_{0}}~\T_{\beta_{0}}
+A^{\beta_{1}}\Z_{\beta_{1}}{}^{\alpha_{0}}~,
\label{abelcompletesolutees1}
\eeq
where \mb{A^{\alpha_{0}\beta_{0}}=
-(-1)^{\eps_{\alpha_{0}}\eps_{\beta_{0}}}A^{\beta_{0}\alpha_{0}}}
and \mb{A^{\beta_{1}}} are arbitrary operators, \cf \eq{completesolutees1}.
Moreover, one may show that there exists an Abelian maximal rank solution
for \mb{\bT^{\alpha_{0}}} and \mb{\bZ_{\beta_{s}}{}^{\gamma_{s+1}}} to all of 
the eqs.\ \e{babelreductees}-\e{abelzzbarotherwise} if all the integers
\mb{m_{s}} are even, where \mb{s\in\{0,\ldots,L\}}.

\subsection{Improved BRST Operators \mb{\omegadelta} and \mb{\bomegadelta}}

\noi
The new ghost sector is Abelianized with \mb{(\cB,\bcP)}-dependent unitary
transformations. The Abelian Ans\"atze for the improved BRST and anti-BRST
operators are
\bea
\omegadelta&=&
\omegazero+\sum_{s=0}^{L}\cB_{1}^{\alpha_{s}}\bcP^{0}_{\alpha_{s}}
~=~\omegazero-\sum_{s=0}^{L}(-1)^{\eps_{\alpha_{s}}+s}
\Pi_{*}^{\alpha_{s}}\Pi_{\alpha_{s}}~, \label{omegadeltaabel} \\
\bomegadelta&=&
\bomegazero-\sum_{s=0}^{L}(s\!+\!1)
\cB_{0}^{\alpha_{s}}\bcP^{1}_{\alpha_{s}}
~=~\bomegazero-\sum_{s=0}^{L}(s\!+\!1)
\B^{\alpha_{s}}\B^{*}_{\alpha_{s}}~,\label{bomegadeltaabel}
\eea
respectively. The two improved charges \mb{\omegadelta} and \mb{\bomegadelta}
are nilpotent, and their normalized commutator yields the new ghost operator 
\mb{\gbb},
\beq
 \{\omegadelta,\bomegadelta\}~=~\sum_{s=0}^{L}(s\!+\!1) 
(\B^{*}_{\alpha_{s}}\Pi_{*}^{\alpha_{s}}(-1)^{\eps_{\alpha_{s}}+s}
-\Pi_{\alpha_{s}}\B^{\alpha_{s}})
~=~ -\sum_{s=0}^{L} (s\!+\!1) \bcP_{A_{s}}\cB^{A_{s}}~=~\gbb~,
\eeq
\cf \eqs{geebb}{ooghomotopy}. The antibracket
\beq
(\B^{\alpha},\B^{*}_{\beta})_{\omegadelta}~=~\delta^{\alpha}_{\beta}
\label{abelomegadeltarankcond}
\eeq
is non-degenerated in the \mb{\B\B^{*}}-sector, as it should be, 
\cf \eq{omegadeltarankcond2}.

\subsection{The \mb{\cS_{1}} Operator}

\noi
The Abelian Ansatz for \mb{\cS_{1}} is
\bea
\cS_{1}&=&\cB^{A}\tcT_{A}
~=~-\Pi_{*}^{\alpha_{0}}\tcY_{\alpha_{0}}
+\sum_{s=0}^{L}\B^{\alpha_{s}}\tcX_{\alpha_{s}} \cr
&=&\Pi_{*}^{\alpha_{0}}\Y_{\alpha_{0}}{}^{\beta_{1}}
\bP_{\beta_{1}}(-1)^{\eps_{\beta_{1}}+1}
+\bP_{\alpha_{0}}\B^{\alpha_{0}}
+\sum_{s=1}^{L}\B^{\alpha_{s}}\Z_{\alpha_{s}}{}^{\beta_{s-1}}
\B^{*}_{\beta_{s-1}}~,
\label{soneexpanabel}
\eea
where the constraints
\mb{\tcT_{A}\equiv\{\tcX_{\alpha};(-1)^{\eps_{\alpha}}\tcY_{\alpha}\}}
are on the form
\beq
\begin{array}{rclcrcll}
\tcX_{\alpha_{0}}&=&(-1)^{\eps_{\alpha_{0}}+1}\bP_{\alpha_{0}}~,&~~~~~~&
\tcY_{\alpha_{0}}&=&\Y_{\alpha_{0}}{}^{\beta_{1}}\bP_{\beta_{1}}
(-1)^{\eps_{\beta_{1}}}~,& \\ 
\tcX_{\alpha_{s}}&=&\Z_{\alpha_{s}}{}^{\beta_{s-1}}\B^{*}_{\beta_{s-1}}~,&&
\tcY_{\alpha_{s}}&=&0~, &s\in\{1,\ldots,L\}~.
\end{array} \label{abeltildeconstr}
\eeq
The higher-stage tilde constraints can be rephrased as
\mb{\tcT_{A_{s+1}}\equiv
\tcZ_{A_{s+1}}{}^{B_{s}}\bcP_{B_{s}}(-1)^{\eps_{B_{s}}+s}}, 
where the tilde structure functions \mb{\tcZ_{A_{s+1}}{}^{B_{s}}} are given
by the old structure functions \mb{\Z_{\alpha_{s+1}}{}^{\beta_{s}}},
\beq
\{\tcT_{A_{s+1}},\cB^{B_{s}}\}~=~\tcZ_{A_{s+1}}{}^{B_{s}}
~=~\twobytwo{0}{\Z_{\alpha_{s+1}}{}^{\beta_{s}}
(-1)^{\eps_{\beta_{s}}+s+1}}{0}{0}
~,~~~~~~~s\in\{0,\ldots,L\!-\!1\}~,
\eeq
and are therefore nilpotent. The rectangular matrix
\mb{\Y_{\alpha_{0}}{}^{\beta_{1}}=\Y_{\alpha_{0}}{}^{\beta_{1}}(q,p)} in
\eq{abeltildeconstr} is the first in a sequence of rectangular matrices
\mb{\Y_{\alpha_{s-1}}{}^{\beta_{s}}=\Y_{\alpha_{s-1}}{}^{\beta_{s}}(q,p)},
\mb{s\in\{1,\ldots,L\}}, which are chosen as a right inverse for the
\mb{\Z_{\alpha_{s}}{}^{\beta_{s-1}}} matrix,
\beq
 \Z_{\alpha_{s}}{}^{\beta_{s-1}}~\Y_{\beta_{s-1}}{}^{\gamma_{s}}
~=~\delta_{\alpha_{s}}^{\gamma_{s}}
-\Y_{\alpha_{s}}{}^{\beta_{s+1}}~\Z_{\beta_{s+1}}{}^{\gamma_{s}}~,
~~~~~~~~s\in\{1,\ldots,L\!-\!1\}~, \label{why0}
\eeq
modulo matrices \mb{\Y_{\alpha_{s}}{}^{\beta_{s+1}}} and 
\mb{\Z_{\beta_{s+1}}{}^{\gamma_{s}}} associated with the next reducibility
stage. One also assumes that the matrices
\mb{\Y_{\alpha_{s-1}}{}^{\beta_{s}}} commute with 1) the constraints
\mb{\T_{\alpha_{0}}}, 2) the structure functions 
\mb{\Z_{\alpha_{s+1}}{}^{\beta_{s}}} and 3) among themselves, to avoid
higher-order terms in expansions later-on. (This is legitimate because after
all, one could just have chosen all the structure functions as constants.) 
\bea
\{\T_{\alpha_{0}}, \Y_{\beta_{s-1}}{}^{\gamma_{s}} \}&=&0~,\label{why1} \\
\{\Z_{\alpha_{r}}{}^{\beta_{r-1}}, \Y_{\gamma_{s-1}}{}^{\delta_{s}} \}&=&0~,
\label{why2} \\
\{\Y_{\alpha_{r-1}}{}^{\beta_{r}}, \Y_{\gamma_{s-1}}{}^{\delta_{s}} \}&=&0
~,~~~~~~~~r,s\in\{1,\ldots,L\}~. \label{why3}
\eea
The first two conditions \es{why1}{why2} are equivalent to
\beq
\{\omegazero,\Y_{\alpha_{s-1}}{}^{\beta_{s}}\}~=~0
~,~~~~~~~~s\in\{1,\ldots,L\}~.
\label{why4}
\eeq
Notice that the tilde constraints \mb{\tcT_{A}} in \eq{abeltildeconstr} are
Abelian, \ie they commute,
\beq
\{\tcT_{A},\tcT_{B}\}~=~0~.
\label{abeltildecommute}
\eeq
The involution \e{abeltildecommute} is not a consequence of any of the
nilpotency relations that we have encountered so far. (However, see
Appendix~\ref{apptomegaone}.) The tilde constraints \mb{\tcT_{A}} are also
Abelian in the antibracket sense, \ie
\beq
(\tcT_{A},\tcT_{B})^{}_{\omegazero}~=~0~,~~~~~~~~~~
(\tcT_{A},\tcT_{B})^{}_{\omegadelta}~=~0~.
\label{tildeantibracketcommute}
\eeq
The antibracket involution \e{tildeantibracketcommute} is closely related to
the master \eq{s2exists}, \cf \eq{s1s1comhighbrow} below.

\subsection{New BRST Operator \mb{\omegaone}}

\noi
The new BRST operator \mb{\omegaone} may be calculated as the commutator of
\mb{\omegadelta} and \mb{\cS_{1}}, \cf \eq{s1ohs1}. One finds
\bea
\omegaone&=&\{\omegadelta,\cS_{1}\}
~=~\{\omegadelta,\cB^{A}\}\tcT_{A}-\cB^{A}\{\tcT_{A},\omegadelta\}  \cr
&=&\Pi_{*}^{\alpha_{0}}\{\tcY_{\alpha_{0}},\omegadelta\}
-\sum_{s=0}^{L}\left(\B^{\alpha_{s}}\{\tcX_{\alpha_{s}},\omegadelta\}
+\Pi_{*}^{\alpha_{s}}\tcX_{\alpha_{s}}\right)\cr
&=&\B^{\alpha_{0}}\T_{\alpha_{0}}
+\Pi_{*}^{\alpha_{0}}(\delta_{\alpha_{0}}^{\gamma_{0}}
-\Y_{\alpha_{0}}{}^{\beta_{1}}\Z_{\beta_{1}}{}^{\gamma_{0}})
\bP_{\gamma_{0}}(-1)^{\eps_{\gamma_{0}}} \cr
&&+\sum_{s=0}^{L-1}\left(\B^{\alpha_{s+1}}
\Z_{\alpha_{s+1}}{}^{\beta_{s}}\Pi_{\beta_{s}}
(-1)^{\eps_{\beta_{s}}+s}
-\Pi_{*}^{\alpha_{s+1}}\Z_{\alpha_{s+1}}{}^{\beta_{s}}
\B^{*}_{\beta_{s}}\right)~,
\label{omegaoneabel1}
\eea
where use is made of \eqs{finalabelianization}{why4} among others. Evidently
the \mb{\omegaone} operator has the Abelian form
\beq
\omegaone~=~\cB^{A}\cT_{A}
~=~\sum_{s=0}^{L}(\B^{\alpha_{s}}\cT_{\alpha_{s}}
-\Pi_{*}^{\alpha_{s}}\cX_{\alpha_{s}})~,
\label{omegaoneabel2}
\eeq
with the new constraints  
\mb{\cT_{A}\equiv\{\cT_{\alpha};(-1)^{\eps_{\alpha}}\cX_{\alpha}\}} given by
\beq
\begin{array}{rclcrcl}
\cT_{\alpha_{0}}&=&-\{\tcX_{\alpha_{0}},\omegazero\}
~=~\T_{\alpha_{0}}~,&~~~~&
\cX_{\alpha_{0}}&=&\tcX_{\alpha_{0}}-\{\tcY_{\alpha_{0}},\omegazero\} \\
&&&&&=&(\delta_{\alpha_{0}}^{\gamma_{0}}
-\Y_{\alpha_{0}}{}^{\beta_{1}}\Z_{\beta_{1}}{}^{\gamma_{0}})
\bP_{\gamma_{0}}(-1)^{\eps_{\gamma_{0}}+1}~, \\
\cT_{\alpha_{s}}&=&-\{\tcX_{\alpha_{s}},\omegadelta\}&&
\cX_{\alpha_{s}}&=&\tcX_{\alpha_{s}}
~=~\Z_{\alpha_{s}}{}^{\beta_{s-1}}\B^{*}_{\beta_{s-1}}~, \\
&=&\Z_{\alpha_{s}}{}^{\beta_{s-1}}~\Pi_{\beta_{s-1}}
(-1)^{\eps_{\beta_{s-1}}+s-1}~,&&&&s\in\{1,\ldots,L\}~,
\end{array}
\label{abelnewconstr}
\eeq
where in general only the new constraints \mb{\cT_{\alpha}} in the first
column of \eq{abelnewconstr} are BRST-invariant.
The new constraints \mb{\cT_{\alpha_{0}}} are just the old constraints, since
they are already BRST-invariant in the Abelian case.
The r\^ole of the \mb{\Y_{\alpha_{0}}{}^{\beta_{1}}} matrix is to
appropriately reduce the number of independent constraints among the
zeroth-stage superpartners \mb{\cX_{\alpha_{0}}}.
We also mention that the new higher-stage constraints can be rephrased as
\mb{\cT_{A_{s+1}}\equiv
\cZ_{A_{s+1}}{}^{B_{s}}\bcP_{B_{s}}(-1)^{\eps_{B_{s}}+s}}, 
where the new structure functions \mb{\cZ_{A_{s+1}}{}^{B_{s}}} are two 
copies of the old structure functions \mb{\Z_{\alpha_{s+1}}{}^{\beta_{s}}},
\beq
\{\cT_{A_{s+1}},\cB^{B_{s}}\}~=~\cZ_{A_{s+1}}{}^{B_{s}}
~=~\twobytwo{\Z_{\alpha_{s+1}}{}^{\beta_{s}}}{0}{0}
{(-1)^{\eps_{\alpha_{s+1}}+\eps_{\beta_{s}}}~
\Z_{\alpha_{s+1}}{}^{\beta_{s}}}~,~~~~~~~s\in\{0,\ldots,L\!-\!1\}~.
\eeq
It is easy to check that the new constraints \e{abelnewconstr} do commute with
the tilde constraints \e{abeltildeconstr},
\beq
\{\tcT_{A},\cT_{B}\}~=~0~,
\label{abelmixedconstrcommute}
\eeq
do commute among themselves,
\beq
\{\cT_{A},\cT_{B}\}~=~0~,
\label{abelnewconstrcommute}
\eeq
and that the new BRST operator \mb{\omegaone=\cB^{A}\cT_{A}} is nilpotent.

\subsection{The \mb{\cS_{2}} Operator} 

\noi
The \mb{\{\cS_{1},\omegaone\}} commutator is \mb{\omegadelta}-exact
\beq
\{\cS_{1},\omegaone\}
~=~(\cS_{1},\cS_{1})_{\omegadelta}
~=~\{\cS_{2},\omegadelta\}~,
\label{s1s1s2}
\eeq
according to the general theory, \cf \eqs{s1s1antibracket}{s2exists}. When the
commutator of \mb{\cS_{1}} and \mb{\omegaone} is calculated using the Abelian
Ans\"atze \es{soneexpanabel}{omegaoneabel1}, one indeed finds that it is
\mb{\omegadelta}-exact:
\bea
\{\cS_{1},\omegaone\}&=&\B^{\alpha_{1}}\Z_{\alpha_{1}}{}^{\beta_{0}}
(2\delta_{\beta_{0}}^{\delta_{0}}
-\Y_{\beta_{0}}{}^{\gamma_{1}}\Z_{\gamma_{1}}{}^{\delta_{0}})
\bP_{\delta_{0}}(-1)^{\eps_{\delta_{0}}}
+\Pi_{*}^{\alpha_{1}}\Z_{\alpha_{1}}{}^{\beta_{0}}~
\Y_{\beta_{0}}{}^{\gamma_{1}}\bP_{\gamma_{1}}(-1)^{\eps_{\gamma_{1}}+1} \cr
&=&\B^{\alpha_{1}}\Z_{\alpha_{1}}{}^{\beta_{0}}
~\Y_{\beta_{0}}{}^{\gamma_{1}}\Z_{\gamma_{1}}{}^{\delta_{0}}
\bP_{\delta_{0}}(-1)^{\eps_{\delta_{0}}}
+\Pi_{*}^{\alpha_{1}}\Z_{\alpha_{1}}{}^{\beta_{0}}~
\Y_{\beta_{0}}{}^{\gamma_{1}}
\bP_{\gamma_{1}}(-1)^{\eps_{\gamma_{1}}+1}\cr
&=&\{\omegadelta,~\B^{\alpha_{1}}\Z_{\alpha_{1}}{}^{\beta_{0}}
~\Y_{\beta_{0}}{}^{\gamma_{1}}
\bP_{\gamma_{1}}(-1)^{\eps_{\gamma_{1}}}\}~. \label{s1s1com}
\eea
Here use is made of \eq{why0} in the second equality. In the irreducible case
\mb{L\!=\!0}, one can just let \mb{\cS_{2}\!=\!0} be zero, but this is no
longer true in the reducible case \mb{L\!>\!0}. (However, for an alternative
solution with \mb{\cS_{2}\!=\!0} even for the reducible case \mb{L\!>\!0}, 
see Subsection~\ref{secaas} below.) In the general Abelian case, it is
consistent to define \mb{\cS_{2}} as
\bea
\cS_{2}&=&\cB^{A}\brcT_{A}~=~\cB^{A}\{\tcT_{A},\cB^{B}\}\tcT_{B}
~=~\sum_{s=0}^{L-1}\cB^{A_{s+1}}\tcZ_{A_{s+1}}{}^{B_{s}}~\tcT_{B_{s}} \cr
&=&-\B^{\alpha_{1}}\Z_{\alpha_{1}}{}^{\beta_{0}}~
\Y_{\beta_{0}}{}^{\gamma_{1}}\bP_{\gamma_{1}}(-1)^{\eps_{\gamma_{1}}} 
~=~-\B^{\alpha_{1}}(\delta_{\alpha_{1}}^{\gamma_{1}}
-\Y_{\alpha_{1}}{}^{\beta_{2}}\Z_{\beta_{2}}{}^{\gamma_{1}})
\bP_{\gamma_{1}}(-1)^{\eps_{\gamma_{1}}}~.
\label{s2abel}
\eea
In particular, the \mb{\brcT_{A}} constraints are linear combinations of the 
tilde constraints \mb{\tcT_{A}},
\beq
 \brcT_{A}~=~\{\tcT_{A},\cB^{B}\}\tcT_{B}~,~~~~~~~~~~
\brcT_{A_{s+1}}~=~\tcZ_{A_{s+1}}{}^{B_{s}}~\tcT_{B_{s}}~.
\label{brctfromtct}
\eeq
It is easy to check that they commute
\beq
\{\brcT_{A},\brcT_{B}\}~=~0~.
\label{brevecommute}
\eeq
One may wonder if there is a broader derivation of the fact that the Abelian
Ansatz \mb{\cS_{1}=\cB^{A}\tcT_{A}} fulfills the master \eq{s2exists}? It turns
out that the question hangs on a somewhat artificially looking formula,
\beq
 (\tcT_{A},\cB^{B})_{\omegadelta}\tcT_{B}~=~
\{\tcT_{A},\cB^{B}\}\{\tcT_{B},\omegadelta\}~,
\label{funnylooking}
\eeq
which is essentially \eq{why0} in disguise. One calculates
\bea
(\cS_{1},\cS_{1})_{\omegadelta}
&=&\{\{\cB^{A}\tcT_{A},\omegadelta\},\cB^{B}\tcT_{B}\} \cr
&=&2\cB^{A}(\tcT_{A},\cB^{B})_{\omegadelta}\tcT_{B}
-\cB^{A}\{\tcT_{A},\cB^{B}\}\{\tcT_{B},\omegadelta\}
-\{\omegadelta,\cB^{A}\}\{\tcT_{A},\cB^{B}\}\tcT_{B} \cr
&=&\cB^{A}\{\tcT_{A},\cB^{B}\}\{\tcT_{B},\omegadelta\}
-\{\omegadelta,\cB^{A}\}\{\tcT_{A},\cB^{B}\}\tcT_{B} \cr
&=&\{\cB^{A}\{\tcT_{A},\cB^{B}\}\tcT_{B},\omegadelta\}
~=~\{\cS_{2},\omegadelta\}~,
\label{s1s1comhighbrow}
\eea
where use has been made of eqs.\ \e{abeltildecommute},
\e{tildeantibracketcommute}, \e{funnylooking} and
\beq
 \{\{\tcT_{A},\cB^{B}\},\omegadelta\}~=~0~.
\eeq
Not surprisingly, there is a complete one-to-one correspondence between this
derivation \e{s1s1comhighbrow} and the first derivation \e{s1s1com}.

\subsection{The Higher \mb{\cS_{k}} Operators} 

\noi
There are no further non-trivial antibrackets,
\bea
(\cS_{1},\cS_{2})_{\omegadelta}&=&\Hf \{\cS_{2},\omegaone\}+\Hf 
\{\cS_{1},\{\omegaone,\cS_{1}\}\}~=~0~, \\
(\cS_{2},\cS_{2})_{\omegadelta}&=&\{\cS_{2},\{\omegaone,\cS_{1}\}\}~=~0~,
\eea
\cf \eq{s1s1s2} and the explicit expressions \e{soneexpanabel}, \e{s1s1com}
and \e{s2abel} for \mb{\cS_{1}}, \mb{\{\cS_{1},\omegaone\}} and \mb{\cS_{2}}.
Therefore it is consistent to put all the higher \mb{\cS_{k}} operators to
zero,
\beq
\cS_{k}~=~0~,~~~~~~~~~ k\in\{3,4,5,\ldots\}~,
\eeq
\cf \eq{ometower}.

\subsection{Another Abelian Solution} 
\label{secaas}

\noi
We now apply the Abelian Ansatz \e{soneexpanabel} to the primed solution
\mb{\cS^{\prime}} from Subsection~\ref{secinvrel}. Recall that
\mb{\cS^{\prime}} is mediated by an \mb{\bomegadelta}-exact Fermionic generator
of the form
\beq
\Psi_{1}~=~\{\bomegadelta, \cS_{1}\}
~=~\B^{\alpha_{0}}
(\Y_{\alpha_{0}}{}^{\beta_{1}}-\bZ_{\alpha_{0}}{}^{\beta_{1}})
\bP_{\beta_{1}}(-1)^{\eps_{\beta_{1}}}
-\Pi_{*}^{\alpha_{0}}\Y_{\alpha_{0}}{}^{\beta_{1}}
\bZ_{\beta_{1}}{}^{\gamma_{2}}\bP_{\gamma_{2}}(-1)^{\eps_{\gamma_{2}}}~.
\label{abelinvpsione}
\eeq
The change in \mb{\cS_{1}} is \mb{\omegadelta}-exact, 
\bea
\cS_{1}-\cS^{\prime}_{1}~=~\{\omegadelta, \Psi_{1}\}
&=&\Pi_{*}^{\alpha_{0}}
\left(\Y_{\alpha_{0}}{}^{\delta_{1}}
-(\delta_{\alpha_{0}}^{\gamma_{0}}-
\Y_{\alpha_{0}}{}^{\beta_{1}}\Z_{\beta_{1}}{}^{\gamma_{0}})
\bZ_{\gamma_{0}}{}^{\delta_{1}}\right)
\bP_{\delta_{1}}(-1)^{\eps_{\delta_{1}}+1} \cr
&&-\B^{\alpha_{0}}
(\Y_{\alpha_{0}}{}^{\beta_{1}}-\bZ_{\alpha_{0}}{}^{\beta_{1}})
\Z_{\beta_{1}}{}^{\gamma_{0}}\bP_{\gamma_{0}}(-1)^{\eps_{\gamma_{0}}}~,
\eea
\cf \eqs{maximalarbitrariness1}{abelzzbarrelation}. Then the primed solution
\mb{\cS^{\prime}_{1}} reads
\bea
\cS^{\prime}_{1}&=&\Pi_{*}^{\alpha_{0}}
(\delta_{\alpha_{0}}^{\gamma_{0}}-
\Y_{\alpha_{0}}{}^{\beta_{1}}\Z_{\beta_{1}}{}^{\gamma_{0}})
\bZ_{\gamma_{0}}{}^{\delta_{1}}\bP_{\delta_{1}}(-1)^{\eps_{\delta_{1}}+1} \cr
&&-\B^{\alpha_{0}}\left( \delta_{\alpha_{0}}^{\gamma_{0}}
-(\Y_{\alpha_{0}}{}^{\beta_{1}}-\bZ_{\alpha_{0}}{}^{\beta_{1}})
\Z_{\beta_{1}}{}^{\gamma_{0}}\right)
\bP_{\gamma_{0}}(-1)^{\eps_{\gamma_{0}}}
+\sum_{s=1}^{L}\B^{\alpha_{s}}\Z_{\alpha_{s}}{}^{\beta_{s-1}}
\B^{*}_{\beta_{s-1}}~.
\label{zbsoneexpanabelalternative}
\eea
Derived from a canonical transformation \e{maximalarbitrariness1}, the primed 
solution \e{zbsoneexpanabelalternative} is guaranteed to meet the correct
boundary condition \e{s1ohs1}. Moreover, the primed solution has the remarkable
property that the \mb{\cS^{\prime}_{2}} operator vanishes identically,
\beq
 \cS^{\prime}_{2}~=~0~,
\eeq  
if one chooses \mb{\Psi_{2}\!=\!0}. This is because the change in \mb{\cS_{2}}
is given by
\beq
\cS^{\prime}_{2}-\cS_{2}~=~\{\{\omegadelta, \Psi_{1}\},\cS_{1}\}
~=~\B^{\alpha_{1}}\Z_{\alpha_{1}}{}^{\beta_{0}}~
\Y_{\beta_{0}}{}^{\gamma_{1}}\bP_{\gamma_{1}}(-1)^{\eps_{\gamma_{1}}}
~=~-\cS_{2}~,
\eeq
\cf eqs.\ \e{maximalarbitrariness2}, \es{why0}{s2abel}. Hence it is consistent
to choose all the higher operators \mb{S^{\prime}_{k\geq 2}=0} equal to zero.
To simplify, let us specialize to the case of trivial anti-BRST symmetry
\mb{\bomegazero=0}. In this case, the structure functions
\mb{\bZ_{\alpha_{0}}{}^{\beta_{1}}=0} vanish, and the \mb{\cS^{\prime}_{1}}
operator \e{zbsoneexpanabelalternative} becomes
\beq
\cS^{\prime}_{1}~=~-\B^{\alpha_{0}}( \delta_{\alpha_{0}}^{\gamma_{0}}
-\Y_{\alpha_{0}}{}^{\beta_{1}}\Z_{\beta_{1}}{}^{\gamma_{0}})
\bP_{\gamma_{0}}(-1)^{\eps_{\gamma_{0}}}
+\sum_{s=1}^{L}\B^{\alpha_{s}}\Z_{\alpha_{s}}{}^{\beta_{s-1}}
\B^{*}_{\beta_{s-1}}~.
\label{soneexpanabelalternative}
\eeq
Clearly the primed solution \e{soneexpanabelalternative} (or its alter ego 
\e{zbsoneexpanabelalternative}) is too complicated to serve as a first
principle. {}For instance, the projection inside the
\mb{\B^{\alpha_{0}}\bP_{\gamma_{0}}} term looks rather artificial if
postulated from scratch. Also the Hessian of the primed solution
\e{soneexpanabelalternative} has a smaller rank than its unprimed
counterpart, and hence it represents a solution that is not proper.

\section{Algebras of Constraints}
\label{secalgconstr}

\noi
We now return to the fully non-Abelian case. An interesting question is what
type of algebra do the new constraints \mb{\cT_{A}} and tilde constraints 
\mb{\tcT_{A}} in general obey? We will address this topic in this Section.

\subsection{Unitary Transformations}
\label{secut}

\noi
To study the question of constraint algebras, it is useful to observe how the
new structure functions behave under unitary transformations 
\beq
A~=~e^{-\frac{i}{\hbar}G}A^{(0)}e^{\frac{i}{\hbar}G}
~=~e^{\{G,\cdot\}}A^{(0)}
\label{ut}
\eeq
in the new ghost sector,
where \mb{A^{(0)}} (resp.\ \mb{A}) denotes any operator before (resp.\ after)
a unitary transformation. Here \mb{G\!=\!G^{\dagger}} is a finite Hermitian 
generator of the transformation,
\bea
G&=&G_{0}+\sum_{s=0}^{L}\cB^{A_{s}}G_{A_{s}}{}^{B_{s}}~
\bcP_{B_{s}}(-1)^{\eps_{B_{s}}+s} \cr
&&+\Hf\!\!\!\!\!\!\!\!\sum_{\footnotesize
\begin{array}{c}r,s\geq 0\cr r\!+\!s\!+\!1\leq L\end{array}}
\!\!\!\!\!\!\!\!\cB^{B_{s}}\cB^{A_{r}}~
G_{A_{r}B_{s}}{}^{C_{r+s+1}}~\bcP_{C_{r+s+1}}
(-1)^{\eps_{B_{s}}+\eps_{C_{r+s+1}}+r+1} \cr
&&+\Hf\!\!\!\!\!\!\!\!\sum_{\footnotesize
\begin{array}{c}r,s\geq 0\cr r\!+\!s\!+\!1\leq L\end{array}}
\!\!\!\!\!\!\!\!\cB^{C_{r+s+1}}~
G_{C_{r+s+1}}{}^{B_{s}A_{r}}~\bcP_{A_{r}}\bcP_{B_{s}}(-1)^{\eps_{A_{r}}+r}
+{\cal O}(\cB^{3}\bcP,\cB^{2}\bcP^{2},\cB\bcP^{3})~, 
\label{gexpan}
\eea
and where \mb{G_{A\ldots}{}^{B\ldots}\!=\!G_{A\ldots}{}^{B\ldots}(q,p;\C,\bP)}
depends on the old phase space variables. See \Ref{BatTyu05} for a related
discussion. The quantum numbers for \mb{G} are
\beq
\eps(G)~=~0~,~~~~~~~~~~~~~~~~~\gh(G)~=~0~=~\ngh(G)~.
\eeq
Hermiticity of \mb{G\!=\!G^{\dagger}} imposes non-trivial conditions on the
\mb{G_{A\ldots}{}^{B\ldots}} structure functions \cite{BatFra83b}.
The \mb{G_{0}} generates canonical transformations and rotations within the old
sector, and it plays only a relatively minor r\^ole in the new ghost sector, 
so we shall assume that \mb{G_{0}\!=\!0} is zero in this Section to keep the 
formulas as simple as possible. Then the BRST operator
\beq
\omegazero~=~\omegazero^{(0)}
\eeq
is invariant under such unitary transformations \e{ut}. Similarly, the
constraints \mb{\cT_{A_{0}}}, \mb{\tcT_{A_{0}}} and \mb{\brcT_{A_{0}}}
transform covariantly,
\bea
\cT_{A_{0}}&=&\Lambda_{A_{0}}{}^{B_{0}}~\cT_{B_{0}}^{(0)}~,
\label{ctcovtransf} \\
 \tcT_{A_{0}}&=&\Lambda_{A_{0}}{}^{B_{0}}~\tcT_{B_{0}}^{(0)}~, 
\label{tctcovtransf} \\
 \brcT_{A_{1}}&=&\Lambda_{A_{1}}{}^{B_{1}}~\brcT_{B_{1}}^{(0)}~,
\label{brctcovtransf}
\eea
where \mb{\Lambda_{A}{}^{B}} denotes the exponential of the matrix
\mb{G_{A}{}^{B}}, \ie
\beq
\Lambda_{A}{}^{B}~\equiv~\delta_{A}^{B}
+G_{A}{}^{B}+\Hf G_{A}{}^{C}G_{C}{}^{B}+
\frac{1}{6}G_{A}{}^{C}G_{C}{}^{D}G_{D}{}^{B}
+ \ldots~.
\eeq
One deduces via Abelianization that there exists a set of new breve structure
functions \mb{\brcZ_{A_{1}}{}^{B_{0}}} such that
\beq
   \brcT_{A_{1}}~=~\brcZ_{A_{1}}{}^{B_{0}}~\tcT_{B_{0}}~=~{\cal O}(\tcT)~,
\eeq
because this holds in the Abelian case, \cf \eq{brctfromtct}, and because both
types of constraints \mb{\tcT_{A_{0}}} and \mb{\brcT_{A_{0}}} transform
covariantly, \cf \eqs{tctcovtransf}{brctcovtransf}.
({}For the purely rotational case, where the generator
\mb{G=\cB^{A}G_{A}{}^{B}\bcP_{B}(-1)^{\eps_{B}}} is linear in both \mb{\cB^{A}}
and \mb{\bcP_{A}}, the breve structure functions
\mb{\brcZ_{A_{1}}{}^{B_{0}}=\tcZ_{A_{1}}{}^{B_{0}}} match their tilde
counterparts as we shall soon see, \cf \eq{tcztransf} below.)
In detail, let \mb{(\Lambda^{\lambda})_{A}{}^{B}} denote the \mb{\lambda}th
power of the \mb{\Lambda_{A}{}^{B}} matrix, \ie the exponential of the matrix
\mb{\lambda G_{A}{}^{B}}. Then the structure functions
\mb{\cZ_{A_{s+1}}{}^{B_{s}}}, \mb{\tcZ_{A_{s+1}}{}^{B_{s}}} and
\mb{\cV_{A_{s}}{}^{B_{s}}} transform as
\bea
\cZ_{A_{s+1}}{}^{B_{s}}&=&
-\int_{0}^{1} \!\!\! d\lambda~(\Lambda^{1-\lambda})_{A_{s+1}}{}^{C_{s+1}}~
G_{C_{s+1}}{}^{D_{s}E_{0}}~(\Lambda^{\lambda})_{E_{0}}{}^{F_{0}}~
\cT_{F_{0}}^{(0)}~(\Lambda^{\lambda-1})_{D_{s}}{}^{B_{s}}\cr
&&+\Lambda_{A_{s+1}}{}^{C_{s+1}}~
\cZ_{C_{s+1}}^{(0)}{}^{D_{s}}~(\Lambda^{-1})_{D_{s}}{}^{B_{s}} 
+\left(\rule[-1ex]{0ex}{3ex}
{\cal O}(\hbar)~{\rm terms,~if}~R\geq 2 \right)~, \\
\tcZ_{A_{s+1}}{}^{B_{s}}&=&
\int_{0}^{1} \!\!\! d\lambda~(\Lambda^{1-\lambda})_{A_{s+1}}{}^{C_{s+1}}~
G_{C_{s+1}}{}^{D_{s}E_{0}}~(\Lambda^{\lambda})_{E_{0}}{}^{F_{0}}~
\tcT_{F_{0}}^{(0)}~(\Lambda^{\lambda-1})_{D_{s}}{}^{B_{s}}
(-1)^{\eps_{D_{s}}+s} \cr
&&+\Lambda_{A_{s+1}}{}^{C_{s+1}}~
\tcZ_{C_{s+1}}^{(0)}{}^{D_{s}}~(\Lambda^{-1})_{D_{s}}{}^{B_{s}}
+\left(\rule[-1ex]{0ex}{3ex}
{\cal O}(\hbar)~{\rm terms,~if}~R\geq 2 \right)~, \label{tcztransf} \\
\cV_{A_{s}}{}^{D_{s}}&=&\Lambda_{A_{s}}{}^{B_{s}}\cV_{B_{s}}^{(0)}{}^{C_{s}}~
(\Lambda^{-1})_{C_{s}}{}^{D_{s}}
+(-1)^{\eps_{A_{s}}+s}\{\omegazero,\Lambda_{A_{s}}{}^{B_{s}}\}
(\Lambda^{-1})_{B_{s}}{}^{D_{s}}\cr
&&+\left(\rule[-1ex]{0ex}{3ex}
{\cal O}(\hbar)~{\rm terms,~if}~R\geq 2~{\rm and}~s\geq 1 \right)~, 
\label{vtransf}
\eea
Note that \eq{vtransf} resembles the transformation law for a connection
one-form, if one identifies \mb{\cV_{A_{s}}{}^{B_{s}}} with the connection
one-form and the BRST transformation \mb{\{\omegazero,\cdot\}} with the
de Rham exterior derivative. In this interpretation, the \rhs of 
\eq{brstinvdoubleconstr} behaves as a covariant derivative, so that the new
constraints \mb{\cT_{A_{0}}} on the corresponding \lhs can transform
covariantly, \cf \eq{ctcovtransf}. The structure functions
\mb{\brcU_{A_{0}B_{0}}} inside \mb{\cS_{2}} transform as
\beq
\brcU_{A_{0}B_{0}}
~=~\brLambda_{A_{0}B_{0}}{}^{C_{0}D_{0}}~\brcU_{C_{0}D_{0}}^{(0)}
+\int_{0}^{1} \!\!\! d\lambda~
(\brLambda^{1-\lambda})_{A_{0}B_{0}}{}^{C_{0}D_{0}}~G_{C_{0}D_{0}}{}^{E_{1}}~
(\Lambda^{\lambda})_{E_{1}}{}^{F_{1}}~\brcT_{F_{1}}^{(0)}~.
\eeq
Here \mb{\brLambda_{A_{0}B_{0}}{}^{C_{0}D_{0}}} denotes the exponential of the
matrix
\beq
\brG_{A_{0}B_{0}}{}^{C_{0}D_{0}}
~=~\Hf \left( G_{A_{0}}{}^{C_{0}}~\delta_{B_{0}}^{D_{0}}
-(-1)^{\eps_{A_{0}}\eps_{B_{0}}}(A_{0}\leftrightarrow B_{0})\right)
-(-1)^{\eps_{C_{0}}\eps_{D_{0}}}(C_{0}\leftrightarrow D_{0})~.\label{brevegee}
\eeq
The antisymmetrization in the above \eq{brevegee} ensures the antisymmetry of
the structure functions
\mb{\brcU_{A_{0}B_{0}}=-(-1)^{\eps_{A_{0}}\eps_{B_{0}}}\brcU_{B_{0}A_{0}}}.
Since \mb{\brcU_{A_{0}B_{0}}^{(0)}=0} in the Abelian case, there exist breve 
structure functions \mb{\brcU_{A_{0}B_{0}}{}^{C_{1}}}, such that
\beq
\brcU_{A_{0}B_{0}}~=~\brcU_{A_{0}B_{0}}{}^{C_{1}}~\brcT_{C_{1}}
~=~{\cal O}(\brcT)~.
\eeq
In the purely rotational case, the breve structure functions 
\mb{\brcU_{A_{0}B_{0}}=0} vanish.

\subsection{Algebra of Constraints}

\noi
The \mb{\cT_{A_{0}}} and \mb{\tcT_{A_{0}}} constraints are in weak involution,
\bea
\{\cT_{A_{0}},\cT_{B_{0}}\}&=&\cU_{A_{0}B_{0}}{}^{C_{0}}\cT_{C_{0}} ~, 
\label{newweaklycommute} \\
\{\tcT_{A_{0}},\cT_{B_{0}}\}&=&\cE_{A_{0}B_{0}}{}^{C_{0}}\tcT_{C_{0}}
+\tcE_{A_{0}B_{0}}{}^{C_{0}}\cT_{C_{0}}~,\label{mixedweaklycommute} \\
\{\tcT_{A_{0}},\tcT_{B_{0}}\}&=&\tcF_{A_{0}B_{0}}{}^{C_{0}}\tcT_{C_{0}}~,
\label{tildeweaklycommute}
\eea
where \mb{\cU_{A_{0}B_{0}}{}^{C_{0}}} is part of \mb{\omegaone}, \cf 
\eq{omegaoneexpan}, and where \mb{\cE_{A_{0}B_{0}}{}^{C_{0}}}, 
\mb{\tcE_{A_{0}B_{0}}{}^{C_{0}}} and \mb{\tcF_{A_{0}B_{0}}{}^{C_{0}}}
are some new structure functions. The first involution \e{newweaklycommute}
follows from the \mb{\omegaone} nilpotency \e{newbrstcharge}.
The involutions \es{mixedweaklycommute}{tildeweaklycommute} are not
consequences of any of the nilpotency relations that we have encountered so
far (However, see Appendix~\ref{apptomegaone}), but follows for instance
because of Abelianization. Namely, recall that there exist unitarily
equivalent Abelian constraints in strong involution,
\bea
\{\cT_{A_{0}}^{(0)},\cT_{B_{0}}^{(0)}\}&=&0~,\\
\{\tcT_{A_{0}}^{(0)},\cT_{B_{0}}^{(0)}\}&=&0~,\\
\{\tcT_{A_{0}}^{(0)},\tcT_{B_{0}}^{(0)}\}&=&0~,
\eea
\cf eqs.\ \e{abeltildecommute}, 
\es{abelmixedconstrcommute}{abelnewconstrcommute}, which imply that the
\mb{\cT_{A_{0}}} and \mb{\tcT_{A_{0}}} constraints are in general in weak
involution as displayed in eqs.\ \e{newweaklycommute},
\es{mixedweaklycommute}{tildeweaklycommute}.
In the purely rotational case, the structure functions read
\bea
\cU_{A_{0}B_{0}}{}^{E_{0}}
&=&\left(\Lambda_{A_{0}}{}^{C_{0}}
\{\cT_{C_{0}}^{(0)},\Lambda_{B_{0}}{}^{D_{0}}\}
+\Hf(-1)^{\eps_{B_{0}}\eps_{D_{0}}}
\{\Lambda_{A_{0}}{}^{D_{0}},\Lambda_{B_{0}}{}^{C_{0}}\}
\cT_{C_{0}}^{(0)}\right)(\Lambda^{-1})_{D_{0}}{}^{E_{0}} \cr
&&-(-1)^{\eps_{A_{0}}\eps_{B_{0}}}(A_{0}\leftrightarrow B_{0})~, \\
\tcF_{A_{0}B_{0}}{}^{E_{0}}
&=&\left(\Lambda_{A_{0}}{}^{C_{0}}
\{\tcT_{C_{0}}^{(0)},\Lambda_{B_{0}}{}^{D_{0}}\}
+\Hf(-1)^{(\eps_{B_{0}}+1)(\eps_{D_{0}}+1)}
\{\Lambda_{A_{0}}{}^{D_{0}},\Lambda_{B_{0}}{}^{C_{0}}\}
\tcT_{C_{0}}^{(0)}\right)(\Lambda^{-1})_{D_{0}}{}^{E_{0}} \cr
&&-(-1)^{(\eps_{A_{0}}+1)(\eps_{B_{0}}+1)}(A_{0}\leftrightarrow B_{0})~, \\
\cE_{A_{0}B_{0}}{}^{E_{0}}
&=&\left(-(-1)^{(\eps_{A_{0}}+1)\eps_{B_{0}}}\Lambda_{B_{0}}{}^{C_{0}}
\{\cT_{C_{0}}^{(0)},\Lambda_{A_{0}}{}^{D_{0}}\} \right. \cr
&& +\Hf\left. (-1)^{\eps_{B_{0}}(\eps_{D_{0}}+1)}
\{\Lambda_{A_{0}}{}^{D_{0}},\Lambda_{B_{0}}{}^{C_{0}}\}
\cT_{C_{0}}^{(0)}\right)(\Lambda^{-1})_{D_{0}}{}^{E_{0}}~, \\
\tcE_{A_{0}B_{0}}{}^{E_{0}}
&=&\left(\Lambda_{A_{0}}{}^{C_{0}}
\{\tcT_{C_{0}}^{(0)},\Lambda_{B_{0}}{}^{D_{0}}\} \right. \cr
&& -\Hf\left. (-1)^{(\eps_{A_{0}}+1)(\eps_{B_{0}}+\eps_{D_{0}})}
\{\Lambda_{B_{0}}{}^{D_{0}},\Lambda_{A_{0}}{}^{C_{0}}\}
\tcT_{C_{0}}^{(0)}\right)(\Lambda^{-1})_{D_{0}}{}^{E_{0}}~.
\eea

\subsection{Antibracket Algebra of Constraints}
\label{secantialg}

\noi
The complete set of \mb{\cT_{A_{0}}} and \mb{\tcT_{A_{0}}} constraints form a 
closed antibracket algebra,
\bea
(\cT_{A_{0}},\cT_{B_{0}})^{}_{\omegazero}
&=&\cQ_{A_{0}B_{0}}{}^{C_{0}}\cT_{C_{0}} ~, 
\label{newweaklyantibracketcommute} \\
(\tcT_{A_{0}},\cT_{B_{0}})^{}_{\omegazero}
&=&\cR_{A_{0}B_{0}}{}^{C_{0}}\tcT_{C_{0}}
+\tcR_{A_{0}B_{0}}{}^{C_{0}}\cT_{C_{0}}~,
\label{mixedweaklyantibracketcommute} \\
(\tcT_{A_{0}},\tcT_{B_{0}})^{}_{\omegazero}
&=&\cS_{A_{0}B_{0}}{}^{C_{0}}\tcT_{C_{0}}+
\tcS_{A_{0}B_{0}}{}^{C_{0}}\cT_{C_{0}}~,
\label{tildeweaklyantibracketcommute}
\eea
where the structure functions \mb{\cQ_{A_{0}B_{0}}{}^{C_{0}}},
\mb{\cR_{A_{0}B_{0}}{}^{C_{0}}}, \mb{\tcR_{A_{0}B_{0}}{}^{C_{0}}},
\mb{\cS_{A_{0}B_{0}}{}^{C_{0}}} and \mb{\tcS_{A_{0}B_{0}}{}^{C_{0}}} are given
by
\bea
\rule[-1ex]{0ex}{4ex} 2\cQ_{A_{0}B_{0}}{}^{C_{0}}
&=&\{\cT_{A_{0}},\cV_{B_{0}}{}^{C_{0}}\}(-1)^{\eps_{B_{0}}}
-\cV_{A_{0}}{}^{D_{0}}~\cU_{D_{0}B_{0}}{}^{C_{0}}
-(-1)^{(\eps_{A_{0}}+1)(\eps_{B_{0}}+1)}(A_{0}\leftrightarrow B_{0})~, \\
\rule[-1ex]{0ex}{4ex} 2\cR_{A_{0}B_{0}}{}^{C_{0}}
&=&-\{\cT_{B_{0}},\cV_{A_{0}}{}^{C_{0}}\}(-1)^{\eps_{A_{0}}\eps_{B_{0}}}
+\cV_{A_{0}}{}^{D_{0}}\cE_{D_{0}B_{0}}{}^{C_{0}} \cr
\rule[-2ex]{0ex}{4ex} &&+\cV_{B_{0}}{}^{D_{0}}\cE_{A_{0}D_{0}}{}^{C_{0}}
(-1)^{(\eps_{A_{0}}+1)(\eps_{B_{0}}+\eps_{D_{0}}+1)+\eps_{B_{0}}}~, \\
\rule[-1ex]{0ex}{4ex} 2\tcR_{A_{0}B_{0}}{}^{C_{0}}
&=&\{\tcT_{A_{0}},\cV_{B_{0}}{}^{C_{0}}\}(-1)^{\eps_{B_{0}}}
+\cV_{A_{0}}{}^{D_{0}}\tcE_{D_{0}B_{0}}{}^{C_{0}} \cr
\rule[-2ex]{0ex}{4ex} &&+\cV_{B_{0}}{}^{D_{0}}\tcE_{A_{0}D_{0}}{}^{C_{0}}
(-1)^{(\eps_{A_{0}}+1)(\eps_{B_{0}}+\eps_{D_{0}}+1)+\eps_{B_{0}}}
-\cU_{A_{0}B_{0}}{}^{C_{0}}~, \\
\rule[-1ex]{0ex}{4ex} 2\cS_{A_{0}B_{0}}{}^{C_{0}}
&=&\cV_{A_{0}}{}^{D_{0}}\tcF_{D_{0}B_{0}}{}^{C_{0}}
+\left(\{\tcT_{A_{0}},\cV_{B_{0}}{}^{C_{0}}\}
-\cE_{A_{0}B_{0}}{}^{C_{0}}\right)(-1)^{\eps_{B_{0}}} \cr
\rule[-2ex]{0ex}{4ex} 
&&-(-1)^{\eps_{A_{0}}\eps_{B_{0}}}(A_{0}\leftrightarrow B_{0})~, \\
\rule[-1ex]{0ex}{4ex} 2\tcS_{A_{0}B_{0}}{}^{C_{0}}
&=&-\tcE_{A_{0}B_{0}}{}^{C_{0}}(-1)^{\eps_{B_{0}}} 
-(-1)^{\eps_{A_{0}}\eps_{B_{0}}}(A_{0}\leftrightarrow B_{0})~, 
\eea
\cf eqs.\ \e{operatorantibracket}, \e{brstinvnewconstr},
\e{brstinvdoubleconstr}, \e{newweaklycommute}, 
\es{mixedweaklycommute}{tildeweaklycommute}. 

\noi
It is an important observation that the new \mb{\cT_{A_{0}}} constraints by
themselves form a closed antibracket algebra \e{newweaklyantibracketcommute}.
On the other hand, when considering the tilde constraints \mb{\tcT_{A_{0}}} by
themselves, they are {\em not} necessarily in involution \wrt the antibracket
\mb{(\cdot,\cdot)_{\omegazero}^{}} --- not even at the classical level.
This is despite the fact that the Abelian constraints \mb{\tcT_{A_{0}}^{(0)}}
satisfy \mb{(\tcT_{A_{0}}^{(0)},\tcT_{B_{0}}^{(0)})_{\omegazero}^{}=0}, \cf 
\eq{tildeantibracketcommute}. Classically and on-shell \wrt the constraints 
\mb{\tcT_{A_{0}}}, the antibracket
\mb{(\tcT_{A_{0}},\tcT_{B_{0}})_{\omegazero}^{}} contains non-vanishing
contributions 
\bea
(\tcT_{A_{0}},\tcT_{B_{0}})_{\omegazero}^{}&\longrightarrow &
\Hf \Lambda_{A_{0}}{}^{C_{0}}\{\tcT_{C_{0}}^{(0)},
\Lambda_{B_{0}}{}^{D_{0}}\}_{PB}^{}
\{\tcT_{D_{0}}^{(0)},\omegazero\}_{PB}^{}(-1)^{\eps_{B_{0}}} \cr 
&&+{\cal O}(\tcT)-(-1)^{\eps_{A_{0}}\eps_{B_{0}}}(A_{0}\leftrightarrow B_{0})
~~~~~~~~~~~~~~~~~~~~{\rm for}~~~~~\hbar\to 0~. 
\label{tildetantibracket}
\eea
There are practically no conditions on the rotation matrix
\mb{\Lambda_{A_{0}}{}^{B_{0}}\!=\!\Lambda_{A_{0}}{}^{B_{0}}(q,p;\C,\bP)},
and hence the \rhs of \eq{tildetantibracket} does not always vanish.
The crucial difference between \eq{tildetantibracket} and the above commutator
involutions \e{newweaklycommute}, \es{mixedweaklycommute}{tildeweaklycommute}, 
is, that the operator antibracket \e{operatorantibracket} does not satisfy the
pertinent Leibniz rule, while the commutator does. One can easily check \wtho
\eq{brstinvdoubleconstr} that the emerging extra contributions are proportional
to the new \mb{\cT_{A_{0}}} constraints, \cf last term on the \rhs of 
\eq{tildeweaklyantibracketcommute}.

\section{Conclusion}
\label{secconcl}

\noi
We have extended the construction of BRST-invariant constraints in
\Ref{BatTyu03} to include reducible gauge algebras. We have also stressed a
deep relationship with BRST/anti-BRST symmetric models. Here the two
nilpotent, Grassmann-odd, mutually anti-commuting BRST operators come from a
deformed version \mb{\omegadelta} of the ordinary BRST operator
\mb{\omegazero}, and a new BRST operator \mb{\omegaone=\cB^{A}\cT_{A}+\ldots},
which encodes the new constraints \mb{\cT_{A}}. Note however, that
all three charges \mb{\omegazero}, \mb{\omegadelta} and \mb{\omegaone}
have ordinary ghost number \mb{+1}, and only the latter
operator is charges \wrt the new ghost number, \mb{\ngh(\omegaone)=1}, 
which is different from the usual BRST/anti-BRST formulation. 
Nevertheless, one finds many similarities at the algebraic level. Some of them
is exposed in Table~\ref{analogytable}. 
In particular, we have constructed a unitarizing Hamiltonian that respects the
two BRST operators \mb{\omegadelta} and \mb{\omegaone} \wtho a Gauge Boson.

\noi
The \mb{\cS\!\equiv\!\sum_{k=0}^{\infty}\cS_{k}} operator, which satisfies an
operator master \eq{ome}, plays a prominent r\^ole in the construction. {}For
instance, the operator \mb{\cS_{1}=\cB^{A}\tcT_{A}+\ldots} contains the tilde 
constraints \mb{\tcT_{A}}, which decent through \eq{brstinvdoubleconstrintro}
to the BRST-invariant constraints \mb{\cT_{A}}.
The existence of the \mb{\cS} operator is deduced from cohomological
considerations of the pertinent BRST operators.
In Appendix~\ref{appome} we have considered various candidates to the operator
master equation. In particular, we have analyzed the simplest cases, which are 
likely to become important for practical calculations.

\noi
We have also investigated the algebras of new constraints \mb{\cT_{A}} and
tilde constraints \mb{\tcT_{A}}, \cf Section~\ref{secalgconstr}. It is found
that the corresponding commutator algebras are closed, but only the former 
follows (with the machinery introduced in the main text excluding 
Appendix~\ref{apptomegaone}) from a BRST nilpotency relation. The full
antibracket algebra of \mb{\cT_{A}} and \mb{\tcT_{A}} constraints is also
closed. So is the antibracket algebra of \mb{\cT_{A}} constraints. However, the
antibracket algebra of tilde constraints \mb{\tcT_{A}} is, on the other hand,
in general an open algebra.

\vspace{0.8cm}

\noi
{\sc Acknowledgement:}~I.A.B.\ thanks R.~Marnelius for numerous discussions.
I.A.B.\ also thanks R.~von~Unge and the Masaryk University for the warm
hospitality extended to him in Brno. The work of I.A.B.\ and K.B.\ is
supported by the Ministry of Education of the Czech Republic under the project
MSM 0021622409. The work of I.A.B.\ is also supported by grants
RFBR 05-01-00996, RFBR 05-02-17217 and LSS-4401.2006.2.

\appendix

\section{Critical Open Bosonic String Theory}
\label{appbosonstring}

\noi
{}For completeness, we here list the standard definitions that go into
critical open Bosonic string theory. Let \mb{\eta_{\mu\nu}\!=\!\eta_{\nu\mu}}
be a constant invertible target space metric. There are Bosonic matter fields 
\mb{\alpha_{m}^{\mu}} and Fermionic ghost variables \mb{c_{m}} and \mb{b_{m}},
where \mb{\mu\in\{1,\ldots,D\}} and \mb{m\in\{\ldots,-2,-1,0,1,2\ldots\}}
is an integer. The canonical commutation relations are
\beq
[\alpha_{m}^{\mu},\alpha_{n}^{\nu}]
~=~\hbar\kappa^{|m|}m\delta_{n+m}^{0}\eta^{\mu\nu}
~,~~~~~~~~~~~[c_{m},b_{n}]~=~\hbar\kappa^{|m|} \delta_{n+m}^{0}~,
\label{bosonstringccr}
\eeq
and zero in all the remaining sectors.
Here \mb{\kappa\to 1} is a regularization parameter, which at the end of the
calculations should be set to \mb{1}. {}From a world-sheet perspective,
the \mb{\kappa}-regularization smears the singularity of the operator product
expansion (\ie the OPE relation). The normal ordering ``\mb{:~:}'', \aka Wick
ordering, is
\bea
 :\alpha_{m}^{\mu}\alpha_{n}^{\nu}:
&=&\theta(n\!-\!m)\alpha_{m}^{\mu}\alpha_{n}^{\nu}
+\theta(m\!-\!n)\alpha_{n}^{\nu}\alpha_{m}^{\mu}~,
\label{bosonstringnormal1} \\
 -:b_{n}c_{m}: ~=~:c_{m}b_{n}: 
&=&\theta(n\!-\!m)c_{m}b_{n}-\theta(m\!-\!n)b_{n}c_{m}~,
\label{bosonstringnormal2}
\eea
where \mb{\theta} denotes the Heaviside step function with
\mb{\theta(0)=\Hf}.

\noi
Here we outline a simple (and we think compelling) proof of the conformal
anomaly, which does not rely on zeta function regularization or a choice of
vacuum. (In reality, one should make sure that the choice of normal ordering
prescription can be accompanied with a compatible choice of bra and ket vacuum
states. This of course is the case.) One first rewrites the commutators of
Virasoro constraints \es{tvirasoroconstraints}{ctvirasoroconstraints} in terms
of anti-supercommutators of the elementary modes by straightforward algebraic
manipulation, which uses the commutation relations \e{bosonstringccr},
\bea
[\T_{m},\T_{n}]&=&
\frac{\hbar}{2}\eta_{\mu\nu}\sum_{i}i\kappa^{|i|}
[\alpha_{m-i}^{\mu},\alpha_{n+i}^{\nu}]_\plus ~, 
\label{ttcomrega}\\
{}[\T_{m}^{(c)},\T_{n}^{(c)}]
&=&\frac{\hbar}{2}\sum_{i}(m\!+\!i)(2n\!+\!i)\kappa^{|i|}
[b_{m-i},c_{n+i}]_\plus -(m\leftrightarrow n)~.
\label{ttcomregc}
\eea
When one normal-orders the anti-supercommutators, \cf \eq{supercom}, one gets
two terms: one quadratic and one constant in the elementary modes, 
\bea
[\alpha_{m}^{\mu},\alpha_{n}^{\nu}]_\plus&=&
2:\alpha_{m}^{\mu}\alpha_{n}^{\mu}:
+\hbar\kappa^{|m|}|m|\delta_{n+m}^{0}\eta^{\mu\nu}~, \\
{}[b_{m},c_{n}]_\plus&=&
2:b_{m}c_{n}:+\hbar\kappa^{|m|}\sgn(m)\delta_{n+m}^{0}~.
\eea
Here \mb{\sgn(m)} denotes the sign of a number \mb{m} with \mb{\sgn(0)=0}.
In the \mb{\kappa\to 1} limit, the quadratic pieces in
\eqs{ttcomrega}{ttcomregc} become
\mb{\hbar(m\!-\!n)(\T_{m+n}\!+\!\hbar a\delta_{m+n}^{0})} and
\mb{\hbar(m\!-\!n)\T_{m+n}^{(c)}}, respectively.
On the other hand, for any \mb{\kappa} with \mb{|\kappa|<1}, the constant
pieces in \eqs{ttcomrega}{ttcomregc} are of the form
\mb{\hbar^{2}A_{m}^{(\alpha)}\delta_{m+n}^{0}} and
\mb{\hbar^{2}A_{m}^{(c)}\delta_{m+n}^{0}}, respectively, where
\bea
A_{m}^{(\alpha)}&=&D\!\!\!\!\!\!\!\!
\sum_{\footnotesize\begin{array}{c}i,j\cr i+j=m \end{array}}
\!\!\!\!\!\!\!\! i|j|\kappa^{|i|+|j|}
~=~ \frac{Dm(m^2-1)}{12}\kappa^{|m|}~, \label{confanomalya} \\
A_{m}^{(c)}&=&-\!\!\!\!\!\!\!\!
\sum_{\footnotesize\begin{array}{c}i,j\cr i+j=m \end{array}}
\!\!\!\!\!\!\!\! (m\!+\!i)(m\!+\!j)\sgn(j)\kappa^{|i|+|j|}
~=~ \frac{m(1-13m^2)}{6}\kappa^{|m|}~.
\label{confanomalyc}
\eea
A few remarks are in order.
In the restricted double summations \es{confanomalya}{confanomalyc}, which are
absolutely and unconditionally convergent for \mb{|\kappa|<1}, note that the
\mb{(i,j)}'th term is antisymmetric under an \mb{(i\leftrightarrow j)} exchange
if the summation variables \mb{i} and \mb{j} have opposite signs. Therefore one
only has to consider \mb{i}'s and \mb{j}'s with weakly the same sign. (The word
{\em weakly} refers to that \mb{i} or \mb{j} could be \mb{0}.) Since at the
same time the sum \mb{i\!+\!j=m} of \mb{i} and \mb{j} is held fixed, the 
restricted \mb{(i,j)} double sum contains only finitely many terms, which may
be readily summed to give the familiar expression for the conformal anomaly,
\cf \eqs{confanomalya}{confanomalyc}. 
In retrospect, the \mb{\kappa}-regularization has picked a particular
(although very natural) summation ordering for two infinite sums, which are
conditionally convergent.

\noi
The BRST charge \mb{\omegazero} and the ghost  operator \mb{G_{c}} read 
\cite{Hwa83}
\bea
\omegazero&=&\sum_{m} \T_{m} c_{-m}
+\Hf\sum_{m,n}(m\!-\!n) :b_{m+n}c_{-n}c_{-m}:
~=~\sum_{m}:( \T_{m}+\Hf  \T_{m}^{(c)}) c_{-m}: \cr
&=&\sum_{m} \cT_{m} c_{-m}-\Hf\sum_{m,n}(m\!-\!n)~b_{m+n}c_{-n}c_{-m}~,
\label{bosonstringomega} \\
G_{c}&=&\sum_{m}:c_{-m}b_{m}:~.
\eea
The last expression in \eq{bosonstringomega}, which has only normal ordering
inside \mb{\cT_{m}}, is useful when proving the nilpotency relation
\e{bosonstringnilp}. The Hermitian conjugate ``\mb{\dagger}'' is defined in
\eq{bosonstringherm1}, which leads to
\beq
\cT_{m}^{\dagger}~=~\cT_{-m}~,~~~~~\omegazero^{\dagger}~=~\omegazero~,~~~~~
G_{c}^{\dagger}~=~G_{c}~.
\label{bosonstringherm2} 
\eeq

\section{Superfield Formulation}
\label{appsuperfield}

\noi
A peculiar (although absolutely consistent) feature of the new ghosts sector
is that the new ghosts can be organized in superpartners that carry shifted 
ghost numbers. It is tempting to rewrite the superpartner fields 
as \mb{N\!=\!1} superfields by introducing a Fermionic \mb{\theta}-coordinate,
and absorb the ghost number deficit into this \mb{\theta},
\beq
 \gh(\theta)~=~-1~,~~~~~~~~\ngh(\theta)~=~0~,~~~~~~~~\eps(\theta)~=~1~.
\eeq
In detail, one may rewrite the construction as follows 
\beq
\begin{array}{rccclcrcccl}
\cB^{A}&\equiv&\{\cB_{0}^{\alpha};~\cB_{1}^{\alpha}\}
&\equiv&\{\B^{\alpha};~
(-1)^{\eps_{\alpha}+1}\Pi_{*}^{\alpha}\}
&\longrightarrow&
\cB^{\alpha}(\theta)&\equiv&\cB_{0}^{\alpha}+\theta \cB_{1}^{\alpha}
&=&\B^{\alpha}-\Pi_{*}^{\alpha}\theta~, \\
\bcP_{A}&\equiv&\{\bcP^{0}_{\alpha};~\bcP^{1}_{\alpha}\}
&\equiv&\{\Pi_{\alpha};~\B^{*}_{\alpha}\}&\longrightarrow&
\bcP_{\alpha}(\theta)&\equiv&
-\bcP^{0}_{\alpha}\theta+\bcP^{1}_{\alpha}
&=&-\Pi_{\alpha}\theta+\B^{*}_{\alpha}~,\\
\cT_{A}&\equiv&\{\cT^{0}_{\alpha};~\cT^{1}_{\alpha}\}
&\equiv&\{\cT_{\alpha};~
(-1)^{\eps_{\alpha}}\cX_{\alpha}\}
&\longrightarrow&\cT_{\alpha}(\theta)&\equiv&
-\cT^{0}_{\alpha}\theta+\cT^{1}_{\alpha}
&=&-\cT_{\alpha}\theta
+\cX_{\alpha}(-1)^{\eps_{\alpha}}~.
\end{array}
\label{introsusy}
\eeq
As a result, the above superfields carry definite ghost number,
\beq
\gh(\cB^{\alpha_{s}}(\theta))~=~s\!+\!1~=~-\gh(\cT_{\alpha_{s}}(\theta))
~=~-\gh(\bcP_{\alpha_{s}}(\theta))\!-\!1~.
\eeq
A similar trick may be applied to the non-minimal variables 
\e{nonminnewstatistics} and the gauge-fixing condition \e{gaugefixingcond}.
The superfield transcription \e{introsusy} basically amounts to rewrite all
previous capital indices, \mb{A;B;C;\ldots}, in the corresponding superfield
indices,
\mb{\alpha,\theta;\beta,\theta^{\prime};C,\theta^{\prime\prime};\ldots}. A sum
\mb{\sum_{A}} over a repeated dummy index, which one normally does not write
explicitly, now also involve a Berezin integration
\mb{ \sum_{\alpha}\int \! \! d\theta}, and so forth. In detail, we use the
following superconventions,
\beq
 \int \! \! d\theta~\theta~\equiv~1~,~~~\delta(\theta)~\equiv~\theta~.
\eeq
Then the canonical commutation relations \e{ccr} read
\beq
 \{ \cB^{\alpha}(\theta), \bcP_{\beta}(\theta^{\prime})\}
~=~ \delta^{\alpha}_{\beta}\delta(\theta\!-\!\theta^{\prime})
~=~ \{ \bcP_{\beta}(\theta), \cB^{\alpha}(\theta^{\prime})\}~.
\eeq
Likewise, one gets
\beq
\int \! \! d\theta~\cB^{\alpha}(\theta)\cT_{\alpha}(\theta)
~=~\B^{\alpha}\cT_{\alpha}-\Pi_{*}^{\alpha}\cX_{\alpha}~=~\cB^{A}\cT_{A}~,
\eeq

\beq
\omegadelta~=~\omegazero+\int \! \! d\theta~\cB^{\alpha}(\theta)
\cV_{\alpha}{}^{\beta}(\theta,\theta^{\prime})\bcP_{\alpha}(\theta^{\prime})
d\theta^{\prime}+\ldots~,
\eeq

\beq
\omegaone~=~\int \! \! d\theta~
\cB^{\alpha}(\theta)\cT_{\alpha}(\theta)
+\Hf \int \! \! d\theta d\theta^{\prime}~
\cB^{\beta}(\theta^{\prime})\cB^{\alpha}(\theta)
\cU_{\alpha\beta}{}^{\gamma}
(\theta,\theta^{\prime},\theta^{\prime\prime})
\bcP_{\gamma}(\theta^{\prime\prime})d\theta^{\prime\prime}
+\ldots~,
\eeq
and
\beq
\{\cT_{\alpha}(\theta),\cT_{\beta}(\theta^{\prime}) \}
~=~\int \cU_{\alpha\beta}{}^{\gamma}
(\theta,\theta^{\prime},\theta^{\prime\prime})~
d\theta^{\prime\prime}~\cT_{\gamma}(\theta^{\prime\prime})~.
\eeq
While aesthetically nice, the super-transcription unfortunately tend to
increase the formula size, which is why the superfield formulation is not used
in the main part of the paper. We also stress that this superfield formulation
only affects the new ghost sector \mb{(\cB^{A},\bcP_{B})}, while the old phase
variables \mb{(q^{i},p_{j};\C^{\alpha},\bP_{\beta})} remain in a
non-supersymmetric formulation.

\section{Matrix Formulation}
\label{appmatrix}

\begin{table}[b] 
\caption{The operator-valued matrix \mb{\hat{\Omega}_{A}{}^{B}}.} 
\label{omegahattable}
\begin{center}
\begin{tabular}{|c||ccccc|} \hline
$A ~\backslash~ B $&$B_{0}$&$B_{1}$~&$B_{2}$
&$\cdots$& \\ \hline\hline
\rule[-2ex]{0ex}{5ex}$A_{0}$&
$(-1)^{\eps_{A_{0}}}\omegazero\delta_{A_{0}}^{B_{0}}
\!-\!i\hbar\cV_{A_{0}}{}^{B_{0}}$&&&& \\ 
\rule[-2ex]{0ex}{5ex}$A_{1}$&$-i\hbar\cZ_{A_{1}}{}^{B_{0}}$&
$-(-1)^{\eps_{A_{1}}}\omegazero\delta_{A_{1}}^{B_{1}}
\!-\!i\hbar\cV_{A_{1}}{}^{B_{1}}$&&&\\ 
\rule[-2ex]{0ex}{5ex}$A_{2}$&&$-i\hbar\cZ_{A_{2}}{}^{B_{1}}$&
$(-1)^{\eps_{A_{2}}}\omegazero\delta_{A_{2}}^{B_{2}}
\!-\!i\hbar\cV_{A_{2}}{}^{B_{2}}$&&\\ 
\rule[-2ex]{0ex}{5ex}$A_{3}$&&&$-i\hbar\cZ_{A_{3}}{}^{B_{2}}$&$\cdots$&\\ 
\rule[-2ex]{0ex}{5ex}$\vdots$&&&&$\cdots$&\\ \hline
\end{tabular}
\end{center}
\end{table}

\noi
Here we give a matrix formulation of certain aspects of a rank \mb{1} theory,
\ie a theory with no terms of the form \mb{{\cal O}(\cB\bcP^{2})} in the 
power series expansions for \mb{\omegadelta} and \mb{\omegaone}, \cf 
\eqs{omegadeltaexpan}{omegaoneexpan}.
Some of the consequences of the \mb{\omegadelta}-nilpotency 
\e{omegadeltaquantumnumbers} then read
\bea
[\omegazero,\omegazero ]&=&0~, \label{omegazeronilpmatrixsec} \\
{}[\cV_{A_{s}}{}^{C_{s}},\omegazero ](-1)^{\eps_{C_{s}}+s}
&=&i\hbar \cV_{A_{s}}{}^{B_{s}}\cV_{B_{s}}{}^{C_{s}}~, \\
(-1)^{(\eps_{B_{s}}+s+1)(\eps_{C_{s}}+s+1)}
[\cV_{A_{s}}{}^{C_{s}},\cV_{B_{s}}{}^{D_{s}}]
&=&(-1)^{(\eps_{A_{s}}+s)(\eps_{B_{s}}+s)}(A_{s}\leftrightarrow B_{s})~, \\
{}[\cV_{A_{r}}{}^{C_{r}},\cV_{B_{s}}{}^{D_{s}}]&=&0
~,~~~~~~~~~~~r\neq s~.
\eea
where \mb{r,s\in\{0,\ldots,L\}}.
Similarly, some of the consequences of the \mb{\omegadelta}-closeness
\e{omegaonedeltaclosed} for \mb{\omegaone} read
\bea
[\cZ_{A_{s+1}}{}^{C_{s}},\omegazero ](-1)^{\eps_{C_{s}}+s}
&=&i\hbar\cZ_{A_{s+1}}{}^{B_{s}}\cV_{B_{s}}{}^{C_{s}}+
i\hbar\cV_{A_{s+1}}{}^{B_{s+1}}\cZ_{B_{s+1}}{}^{C_{s}}~,\\
(-1)^{(\eps_{B_{s}}+s+1)(\eps_{C_{s}}+s+1)}
[\cZ_{A_{s+1}}{}^{C_{s}},\cV_{B_{s}}{}^{D_{s}}]
&=&(-1)^{(\eps_{C_{s}}+s)(\eps_{D_{s}}+s)}(C_{s}\leftrightarrow D_{s})~, \\
(-1)^{(\eps_{B_{s}}+s+1)(\eps_{C_{s}}+s+1)}
[\cV_{A_{s}}{}^{C_{s}},\cZ_{B_{s}}{}^{D_{s-1}}]
&=&(-1)^{(\eps_{A_{s}}+s)(\eps_{B_{s}}+s)}(A_{s}\leftrightarrow B_{s})
~,~~~~~~s\neq0~, \\
{}[\cZ_{A_{s+1}}{}^{C_{s}},\cV_{B_{r}}{}^{D_{r}}]&=&0
~,~~~~~r-s\neq\left\{\begin{array}{c} 0~,\cr 1~, \end{array} \right.
\eea
where \mb{r\in\{0,\ldots,L\} } and \mb{s\in\{0,\ldots,L\!-\!1\}}.
And finally, some of the consequences of the \mb{\omegaone}-nilpotency
\e{newbrstcharge} read
\bea
\cZ_{A_{s+1}}{}^{B_{s}}\cZ_{B_{s}}{}^{C_{s-1}}&=&0~,~~~~~~s\neq0~, \\
(-1)^{(\eps_{B_{s+1}}+s)(\eps_{C_{s}}+s+1)}
[\cZ_{A_{s+1}}{}^{C_{s}},\cZ_{B_{s+1}}{}^{D_{s}}]
&=&(-1)^{(\eps_{C_{s}}+s)(\eps_{D_{s}}+s)}(C_{s}\leftrightarrow D_{s})~, \\
{}[\cZ_{A_{r+1}}{}^{C_{r}},\cZ_{B_{s+1}}{}^{D_{s}}]&=&0
~,~~~~~~~~~~~r\neq s~, \label{zzotherwisematrixsec}
\eea
where \mb{r,s\in\{0,\ldots,L\!-\!1\}}.
All these relations \e{omegazeronilpmatrixsec}-\e{zzotherwisematrixsec}
can precisely be recast into a nilpotency condition
\beq
[\hat{\Omega},\hat{\Omega}]~=~0 \label{omegahatnilp}
\eeq
for an operator
\bea
\hat{\Omega}~=~\cB^{A}\hat{\Omega}_{A}{}^{B}\bcP_{B}(-1)^{\eps_{B}}
&=&-\omegazero\sum_{s=0}^{L}\cB^{A_{s}}\bcP_{A_{s}}(-1)^{\eps_{A_{s}}+s}
-i\hbar\sum_{s=0}^{L}\cB^{A_{s}}
\cV_{A_{s}}{}^{B_{s}}\bcP_{B_{s}}(-1)^{\eps_{B_{s}}+s}\cr
&&-i\hbar\sum_{s=0}^{L-1}\cB^{A_{s+1}}
\cZ_{A_{s+1}}{}^{B_{s}}\bcP_{B_{s}}(-1)^{\eps_{B_{s}}+s}~,\label{omegahat}
\eea
\cf Table~\ref{omegahattable}. The second and third term on the \rhs of
\eq{omegahat} contain the parts of the \mb{\omegadelta} and \mb{\omegaone}
operator that are linear in both \mb{\cB^{A}} and \mb{\bcP_{A}}. 
The \mb{\hat{\Omega}} operator \e{omegahat} is Grassmann-odd, has ghost number
\mb{\gh(\hat{\Omega})=1}, has indefinite new ghost number
(either \mb{0} or \mb{1}), and is not necessarily Hermitian. In turn,
the \mb{\hat{\Omega}} nilpotency \e{omegahatnilp} is equivalent to
following conditions for the matrix elements 
\mb{\hat{\Omega}_{A}{}^{B}=\hat{\Omega}_{A}{}^{B}(q,p,\C,\bP)},
\bea
 \hat{\Omega}_{A}{}^{B}\hat{\Omega}_{B}{}^{C}&=&0~, \label{omegahatone} \\
(-1)^{(\eps_{B}+1)(\eps_{C}+1)}[\hat{\Omega}_{A}{}^{C},\hat{\Omega}_{B}{}^{D}]
&=& \left\{\begin{array}{c}
(-1)^{\eps_{A}\eps_{B}}(A\leftrightarrow B)~,\cr
(-1)^{\eps_{C}\eps_{D}}(C\leftrightarrow D)~.\end{array} \right.
\label{omegahattwo}
\eea
In particular, one sees that the operator-valued matrix element
\mb{\hat{\Omega}_{A}{}^{B}} are nilpotent in a mixed operator/matrix sense,
\cf \eq{omegahatone}. The two possible right-hand sides of \eq{omegahattwo}
are one and the same condition written twice. 
A similar story is true for \mb{\bomegadelta}, and the
\mb{\omegadelta\leftrightarrow\bomegadelta} interplay yields an interesting
canonical commutation relation in the operator/matrix sense,
\beq
 \hat{\Omega}_{A}{}^{B}\hat{\bar{\Omega}}_{B}{}^{C}
+\hat{\bar{\Omega}}_{A}{}^{B}\hat{\Omega}_{B}{}^{C}
~=~i\hbar \delta_{A}^{C}~,
\eeq
\cf eq.~(5.11) of \Ref{BatTyu03}.

\section{Operator Master Equations}
\label{appome}

\noi
In this Appendix we consider candidates to the operator master equations for
\mb{\eS} and \mb{\cS}, \cf \eqs{omezero}{ome}, respectively. We start with
candidates \mb{\cM\!=\!0} to the master eq.\ for \mb{\cS}, and we assume that
\mb{\cM} takes the form
\beq
\cM~\equiv~\sum_{n=0}^{N}\alpha_{n}\Phi_{\omegadelta}^{n}
(\underbrace{\cS,\cS,\ldots,\cS}_{n~{\rm entries}})
~,~~~~~~~~~~~~\alpha_{N}\neq 0~,
\label{emma}
\eeq
where \mb{\Phi_{\omegadelta}^{n}(\cS,\cS,\ldots,\cS)} are higher antibrackets
\cite{BatMar98c,Ber06}. In general the \mb{n}'th (normalized, operator)
antibracket of \mb{n} operators \mb{A_{1}}, \mb{A_{2}}, \mb{\ldots},
\mb{A_{n}}, is defined as
\beq
 \Phi_{\omegadelta}^{n}(A_{1},A_{2},\ldots,A_{n})
~\equiv~\frac{1}{n!}\sum_{\pi\in S_{n}} (-1)^{\eps_{\pi,A}}
\underbrace{ \{ \{ \ldots \{ \omegadelta, A_{\pi(1)}\}, \ldots \}, 
A_{\pi(n)}\}}_{n~{\rm normalized~commutators}}
~,~~~~~~~~~~~~\Phi_{\omegadelta}^{0}~\equiv~\omegadelta~.
\label{derivedbrackets}
\eeq
The sign-factor \mb{\eps_{\pi,A}} arises from permuting the operators
\mb{A_{1}}, \mb{A_{2}}, \mb{\ldots}, \mb{A_{n}} under the permutation
\mb{\pi\in S_{n}}, see \Ref{Ber06} for details, and see
\Refs{BerDamAlf96}{BatBerDam96} for early field-antifield formulations of 
higher antibrackets.
The Ansatz \e{emma} is partly motivated by the fact that the \mb{\cM} operator
has a well-defined classical limit. This is because the multiple nested
normalized commutators \e{normsupercom} inside the higher antibracket
\e{derivedbrackets} simply reduce to multiple nested Poisson brackets
\mb{\{\cdot,\cdot\}_{PB}} when \mb{\hbar\to 0}.
Let us also assume that \mb{\cS_{0}} is a linear combination of the old and the
new ghost  operator, 
\beq
\cS_{0}~=~\mu G + \nu\gbb~.
\eeq
We are interested in which \mb{\cS_{0}} and \mb{\cM} that would guarantee a
solution for \mb{\cS\equiv\sum_{k=0}^{\infty}\cS_{k}}. In other words, which
coefficients \mb{\mu}, \mb{\nu}, \mb{\alpha_{0}}, \mb{\alpha_{1}},
\mb{\alpha_{2}}, \mb{\ldots}, would provide the existence of \mb{\cS}? Note
that \mb{\cM=\sum_{k=0}^{\infty}\cM_{k}} has quantum numbers
\beq
\eps(\cM)~=~\eps(\cM_{k})~=~1~,~~~~~~~~~\ngh(\cM_{k})~=~k~,~~~~~~~~~
\gh(\cM)~=~\gh(\cM_{k})~=~1~.
\eeq
Let us for the sake of simplicity restrict attention to the second-order case
with \mb{N\!=\!2}, \cf \eq{emma}. We will show in this case that for each
choice of \mb{\cS_{0}} (more precisely, for each choice of \mb{\mu} and
\mb{\nu\neq 0}), there exists a unique \mb{\cM} up to an over-all normalization
factor. The proof goes as follows:
\bea
\cM_{k}&=&\alpha_{2}\sum_{j=0}^{k}\Phi_{\omegadelta}^{2}(\cS_{j},\cS_{k-j})
+\alpha_{1}\Phi_{\omegadelta}^{1}(\cS_{k})
+\delta_{k}^{0} \alpha_{0}\Phi_{\omegadelta}^{0}\label{cmkay1} \\
&=&\left\{\begin{array}{lcl}
\alpha_{2}\sum_{j=1}^{k-1}\Phi_{\omegadelta}^{2}(\cS_{j},\cS_{k-j})
+(\alpha_{1}-\alpha_{2}(2\mu+k\nu))\Phi_{\omegadelta}^{1}(\cS_{k})
&{\rm for}&k\geq 1~,\cr
(\alpha_{1}-\alpha_{2}(2\mu+\nu))\omegaone&{\rm for}& k=1~,\cr
(\mu^{2}\alpha_{2}-\mu \alpha_{1}+\alpha_{0})\omegadelta&{\rm for}& k=0~.
\end{array} \right.
\label{cmkay2}
\eea
The equations \mb{\cM_{0}\!=\!0} and \mb{\cM_{1}\!=\!0} yield two conditions,  
\beq
\frac{\alpha_{0}}{\alpha_{2}}~=~\mu(\mu+\nu)
~,~~~~~~~~~\frac{\alpha_{1}}{\alpha_{2}}~=~2\mu+\nu
~,~~~~~~~~~\alpha_{2}\neq 0~.\label{a01}
\eeq
The equation \mb{\cM_{2}\!=\!0} then becomes
\beq
 -(\cS_{1},\cS_{1})_{\omegadelta}
~\equiv~\Phi_{\omegadelta}^{2}(\cS_{1},\cS_{1})
~=~\nu\Phi_{\omegadelta}^{1}(\cS_{2})~\equiv~\nu \{\omegadelta,\cS_{2}\}~.
\eeq
One knows that the antibracket \mb{(\cS_{1},\cS_{1})_{\omegadelta}\neq 0}
does not always vanish, see for instance \eq{s1s1com}, so one must demand that
\beq
\nu~\neq~0~.
\eeq
One next seeks for an integrability condition \mb{\cI\!=\!0}, where
\beq
\cI~\equiv~\!\!\!\!\!\!\!\!
\sum_{\footnotesize\begin{array}{c}m,n\cr m\!+\!n\leq N \end{array}}
\!\!\!\!\!\!\!\!  \beta_{nm} \Phi_{\omegadelta}^{n}(
\Phi_{\cM}^{m}(\underbrace{\cS,\cS,\ldots,\cS}_{m~{\rm entries}}),
\underbrace{\cS,\cS,\ldots,\cS}_{n\!-\!1~{\rm entries}})
~,~~~~~~~\eps(\cI)~=~0~,~~~~~~~\gh(\cI)~=~2~.
\label{integrabcond}
\eeq
The coefficients \mb{\beta_{nm}} are to be adjusted such that
\eq{integrabcond} is a non-trivial linear combination of generalized Jacobi
identities for the antibracket hierarchy. In the \mb{N\!=\!2} case, the
relevant generalized Jacobi identities are \cite{Ber06}
\bea
\Phi_{\omegadelta}^{1}(\Phi_{\omegadelta}^{0})&=&0~, \\
\Phi_{\omegadelta}^{1}(\Phi_{\omegadelta}^{1}(\cS))&=&0~, \\
\Phi_{\omegadelta}^{2}(\Phi_{\omegadelta}^{0},\cS)&=&0~, \\
\Phi_{\omegadelta}^{1}(\Phi_{\omegadelta}^{2}(\cS,\cS))
+2\Phi_{\omegadelta}^{1}(\Phi_{\omegadelta}^{1}(\cS),\cS)&=&0~, \\
6\Phi_{\omegadelta}^{2}(\Phi_{\omegadelta}^{2}(\cS,\cS),\cS)
-\Phi_{\omegadelta}^{1}(\Phi_{\omegadelta}^{3}(\cS,\cS,\cS))&=&0~.
\eea
They are all consequences of the \mb{\omegadelta} nilpotency
\e{omegadeltaquantumnumbers}. In the \mb{N\!=\!2} case, the integrability
condition \e{integrabcond} reads
\bea
\cI&\equiv&\beta_{10}\Phi_{\omegadelta}^{1}(\Phi_{\cM}^{0})
+\beta_{11}\Phi_{\omegadelta}^{1}(\Phi_{\cM}^{1}(\cS))
+\beta_{20}\Phi_{\omegadelta}^{2}(\Phi_{\cM}^{0},\cS) \cr
&=&\beta_{10}\Phi_{\omegadelta}^{1}
\left(\alpha_{2}\Phi_{\omegadelta}^{2}(\cS,\cS)
+\alpha_{1}\Phi_{\omegadelta}^{1}(\cS)
+\alpha_{0}\Phi_{\omegadelta}^{0}\right) \cr
&&+\beta_{11}\Phi_{\omegadelta}^{1}\left(
\alpha_{2}\Phi_{\omegadelta}^{3}(\cS,\cS,\cS)
+\alpha_{1}\Phi_{\omegadelta}^{2}(\cS,\cS)
+\alpha_{0}\Phi_{\omegadelta}^{1}(\cS)\right) \cr
&&+\beta_{20}\Phi_{\omegadelta}^{2}\left(
\alpha_{2}\Phi_{\omegadelta}^{2}(\cS,\cS)
+\alpha_{1}\Phi_{\omegadelta}^{1}(\cS)
+\alpha_{0}\Phi_{\omegadelta}^{0},~\cS)\right) \cr
&=&(\beta_{10}\alpha_{2}+\beta_{11}\alpha_{1})
\Phi_{\omegadelta}^{1}(\Phi_{\omegadelta}^{2}(\cS,\cS))
+\beta_{11}\alpha_{2}\Phi_{\omegadelta}^{1}
(\Phi_{\omegadelta}^{3}(\cS,\cS,\cS)) \cr
&&+\beta_{20}\alpha_{2}\Phi_{\omegadelta}^{2}
(\Phi_{\omegadelta}^{2}(\cS,\cS),\cS)
+\beta_{20}\alpha_{1}\Phi_{\omegadelta}^{2}
(\Phi_{\omegadelta}^{1}(\cS),\cS)~.
\label{ourintegrabcond}
\eea
The \mb{\beta_{nm}} coefficients are then tuned so that \mb{\cI\!=\!0} is 
automatically satisfied. This yield two conditions:
\beq
\frac{\beta_{20}}{\beta_{11}}~=~-6~,~~~~~~~~~~\frac{\beta_{10}}{\beta_{11}}
~=~-4\frac{\alpha_{1}}{\alpha_{2}}~=~-4(2\mu+\nu)~,~~~~~~~~~\beta_{11}\neq 0~.
\label{b1020}
\eeq
\mb{\cI=\sum_{k=0}^{\infty}\cI_{k}} is itself a sum of terms
\mb{\cI_{k}} with quantum numbers,
\beq
\eps(\cI_{k})~=~0~,~~~\ngh(\cI_{k})~=~k~,~~~\gh(\cI_{k})~=~2~.
\eeq
One can now show the existence of a solution \mb{\cS} to the master equation
\mb{\cM\!=\!0}. Recall that the existence of \mb{\cS_{1}} and \mb{\cS_{2}}
follows from the \mb{\omegadelta}-closeness \e{omegaonedeltaclosed} and the
nilpotency \e{omegaonenilp} of \mb{\omegaone}, \cf Subsections~\ref{secessone}
and \ref{secesstwo}, respectively. One now proceeds by mathematical induction
in the integer \mb{k\geq 3}. Assume that there exist \mb{\cS_{1}},
\mb{\cS_{2}}, \mb{\ldots}, \mb{\cS_{k-1}}, such that
\mb{\cM_{0}\!=\!\cM_{1}\!=\!\cM_{2}\!=\!\ldots\!=\!\cM_{k-1}\!=\!0}. One wants 
to prove that there exists \mb{\cS_{k}}, such that \mb{\cM_{k}\!=\!0}. The
\mb{k}'th integrability condition reads
\bea
\cI_{k}&=&\beta_{10}\{\omegadelta,\cM_{k}\}
+\beta_{11}\{\omegadelta,\{\cM_{k},\cS_{0}\}\}
+\frac{\beta_{20}}{2}\left(\rule[-1ex]{0ex}{3ex} 
\{\{\omegadelta,\cM_{k}\},\cS_{0}\}+ 
\{\{\omegadelta,\cS_{0}\},\cM_{k}\} \right) \cr
&=&2(k\!-\!2)\nu \beta_{11}\{\omegadelta,\cM_{k}\}~,
\label{kthintegrab}
\eea
\cf \eq{ourintegrabcond}. \mb{\cI_{k}\!=\!0} implies that \mb{\cM_{k}} is
\mb{\omegadelta}-closed, and hence \mb{\omegadelta}-exact, \cf statement after
\eq{ooghomotopy}. Therefore the \mb{\Phi^{2}}-term in \eq{cmkay2} must be
\mb{\omegadelta}-exact, and hence there exists an \mb{\cS_{k}} that makes
\mb{\cM_{k}\!=\!0} vanish. The factor \mb{(k\!-\!2)} in \eq{kthintegrab}
indicates that non-trivial information must be added to the construction at the
second induction step, \mb{k\!=\!2}, namely the \mb{\omegaone} nilpotency
\e{omegaonenilp}.

\noi
The master \eq{emma} simplifies further if \mb{\alpha_{0}\!=\!0}, \ie if
\mb{\mu} is either equal to \mb{0} or \mb{-\nu}, \cf \eq{a01}. In the main part
of this paper we choose\footnote{Note that \Ref{BatTyu03} has
\mb{\cS_{0}\!=\!G\!-\!\gbb} corresponding to \mb{(\mu,\nu)=(1,-1)}. The
operator master \eq{ome} remains the same, while there is a change of sign in
\eq{s2exists}; see eqs.\ (5.22) and (5.19) in \Ref{BatTyu03}.}
\mb{\cS_{0}\!=\!\gbb} corresponding to \mb{(\mu,\nu)=(0,1)}.

\noi
Now let us drop the use of calligraphic letters and consider candidates
\mb{\eM\!=\!0} to the master eq.\ for \mb{\eS} in the old sector, \cf 
\eq{omezero}. The analysis is very similar to the above case, so we shall be
brief. Let us assume that \mb{\eS_{0}} is proportional to the old ghost
operator, 
\beq
\eS_{0}~=~\mu G ~.
\eeq
Note that \mb{\eM=\sum_{k=0}^{\infty}\eM_{1-2k}} has quantum numbers
\beq
\eps(\eM)~=~\eps(\eM_{1-2k})~=~1~,~~~~~~~~~~\gh(\eM_{1-2k})~=~1\!-\!2k~.
\eeq
One finds
\bea
\eM_{1-2k}&=&\alpha_{2}\sum_{j=0}^{k}
\Phi_{\omegazero}^{2}(\eS_{-2j},\eS_{2(j-k)})
+\alpha_{1}\Phi_{\omegazero}^{1}(\eS_{-2k})
+\delta_{k}^{0} \alpha_{0}\Phi_{\omegazero}^{0}\label{emkay1} \\
&=&\left\{\begin{array}{lcl}
\alpha_{2}\sum_{j=1}^{k-1}\Phi_{\omegazero}^{2}(\eS_{-2j},\eS_{2(j-k)})
+(\alpha_{1}+2(k\!-\!1)\mu\alpha_{2})\Phi_{\omegazero}^{1}(\eS_{-2k})
&{\rm for}&k\geq 1~,\cr
\alpha_{1}\bomegazero&{\rm for}& k=1~,\cr
(\mu^{2}\alpha_{2}-\mu \alpha_{1}+\alpha_{0})\omegazero&{\rm for}& k=0~.
\end{array} \right.
\label{emkay2}
\eea
The equations \mb{\eM_{1}\!=\!0} and \mb{\eM_{-1}\!=\!0} yield two conditions, 
\beq
\frac{\alpha_{0}}{\alpha_{2}}~=~-\mu^{2}
~,~~~~~~~~~\frac{\alpha_{1}}{\alpha_{2}}~=~0
~,~~~~~~~~~\alpha_{2}\neq 0~.\label{a01zero}
\eeq
The equation \mb{\eM_{-3}\!=\!0} then becomes
\beq
 (\eS_{-2},\eS_{-2})_{\omegazero}
~\equiv~-\Phi_{\omegazero}^{2}(\eS_{-2},\eS_{-2})
~=~2\mu\Phi_{\omegazero}^{1}(\eS_{-4})
~\equiv~2\mu \{\omegazero,\eS_{-4}\}~.
\eeq
One knows that the antibracket \mb{(\eS_{-2},\eS_{-2})_{\omegazero}\neq 0}
does not always vanish, so one must demand that
\beq
\mu~\neq~0~.
\eeq
As before one adjusts the \mb{\beta_{nm}} coefficients so that the
integrability condition \e{ourintegrabcond} is automatically satisfied.
This yield two conditions:
\beq
\frac{\beta_{20}}{\beta_{11}}~=~-6~,~~~~~~~~~~\frac{\beta_{10}}{\beta_{11}}
~=~-4\frac{\alpha_{1}}{\alpha_{2}}~=~0~,~~~~~~~~~\beta_{11}\neq 0~.
\label{b1020zero}
\eeq
\mb{\eI=\sum_{k=0}^{\infty}\eI_{2(1-k)}} is itself a sum of terms
\mb{\eI_{2(1-k)}} with quantum numbers,
\beq
\eps(\eI_{2(1-k)})~=~0~,~~~~~~\gh(\eI_{2(1-k)})~=~2(1\!-\!k)~.
\eeq
The \mb{k}'th integrability condition reads
\bea
\eI_{2(1-k)}&=&\beta_{10}\{\omegazero,\eM_{1-2k}\}
+\beta_{11}\{\omegazero,\{\eM_{1-2k},\eS_{0}\}\}
+\frac{\beta_{20}}{2}\left(\rule[-1ex]{0ex}{3ex}
\{\{\omegazero,\eM_{1-2k}\},\eS_{0}\}+ 
\{\{\omegazero,\eS_{0}\},\eM_{1-2k}\} \right) \cr
&=&4(2\!-\!k)\mu \beta_{11}\{\omegazero,\eM_{1-2k}\}~.
\label{kthintegrabzero}
\eea
In the main text we have chosen \mb{\eS_{0}\!=\!G} corresponding to
\mb{\mu\!=\!1}.

\section{A New Tilde BRST Operator \mb{\tomegaone}}
\label{apptomegaone}

\noi 
In this Appendix we briefly outline how the involution relations
\es{mixedweaklycommute}{tildeweaklycommute}
could be reformulated as nilpotency requirement for a new tilde BRST 
operator \mb{\tomegaone},
\beq
\{\tomegaone,\tomegaone \}~=~ 0~,~~~~~~\label{tomegaonenilp} 
\eps(\tomegaone)~=~1~,~~~~~~\gh(\tomegaone)~=~1~,~~~~~~\ngh(\tomegaone)~=~1~.
\label{newtbrstcharge}
\eeq
The tilde BRST operator \mb{\tomegaone} is a deformation of \mb{\omegaone},
\bea
\tomegaone&=&\omegaone+\sum_{s=0}^{L}\tcB^{A_{s}}\tcT_{A_{s}}
+\Hf\!\!\!\!\!\!\!\!\sum_{\footnotesize
\begin{array}{c}r,s\geq 0\cr r\!+\!s\leq L\end{array}}
\!\!\!\!\!\!\!\!\tcB^{B_{s}}\tcB^{A_{r}}~
\tcF_{A_{r}B_{s}}{}^{C_{r+s}}~\tbcP_{C_{r+s}}
(-1)^{\eps_{B_{s}}+\eps_{C_{r+s}}+r} \cr 
&&+\!\!\!\!\!\!\!\!\sum_{\footnotesize
\begin{array}{c}r,s\geq 0\cr r\!+\!s\leq L\end{array}}
\!\!\!\!\!\!\!\!\cB^{B_{s}}\tcB^{A_{r}}\left(
\tcE_{A_{r}B_{s}}{}^{C_{r+s}}~\bcP_{C_{r+s}}
-\cE_{A_{r}B_{s}}{}^{C_{r+s}}~\tbcP_{C_{r+s}}\right)
(-1)^{\eps_{B_{s}}+\eps_{C_{r+s}}+r}+\ldots~.
\eea
Here we have introduced a new set of tilde ghosts 
\mb{\tcB^{A}\!\equiv\!\{\tcB_{0}^{\alpha},\tcB_{1}^{\alpha}\}} and
\mb{\tbcP_{A}\!\equiv\!\{\tbcP^{0}_{\alpha},\tbcP^{1}_{\alpha}\}}
with canonical commutation relations
\beq
\{\tcB^{A},\tcB^{B}\}~=~0~,~~~~~~~~~~
\{\tcB^{A},\tbcP_{B}\}~=~\delta^{A}_{B}~=~
-(-1)^{\eps_{B}}\{\tbcP_{B},\tcB^{A}\}~,~~~~~~~~~~
\{\tbcP_{A},\tbcP_{B}\}~=~0~,
\eeq
which have Grassman parity and new ghost number given by
\beq
\begin{array}{rccclcrcccl}
\eps(\tcB^{A})&=& \eps_{A}&=&\eps(\tbcP_{A})~,&&
\eps(\tcB^{A_{s}})&=& \eps_{A_{s}}\!+\!s&=&\eps(\tbcP_{A_{s}})~, \\
&&&&&&\ngh(\tcB^{A_{s}})&=&s\!+\!1&=&-\ngh(\tbcP_{A_{s}})~.
\end{array}
\eeq
The Grassmann parity and the old ghost number are shifted among the tilde ghost
superpartners as follows
\beq
\begin{array}{rccclcrcccl}
\eps(\tcB_{0}^{\alpha})&=&\eps_{\alpha}
&=&\eps(\tbcP^{0}_{\alpha})~, &&
\eps(\tcB_{0}^{\alpha_{s}})&=&\eps_{\alpha_{s}}\!+\!s
&=&\eps(\tbcP^{0}_{\alpha_{s}})~,  \\
\eps(\tcB_{1}^{\alpha})&=&\eps_{\alpha}\!+\!1
&=&\eps(\tbcP^{1}_{\alpha})~,&&
\eps(\tcB_{1}^{\alpha_{s}})&=&\eps_{\alpha_{s}}\!+\!s\!+\!1
&=&\eps(\tbcP^{1}_{\alpha_{s}})~,\\
&&&&&&\gh(\tcB_{0}^{\alpha_{s}})&=&s\!+\!2&=&-\gh(\tbcP^{0}_{\alpha_{s}})~, \\
&&&&&&\gh(\tcB_{1}^{\alpha_{s}})&=&s\!+\!3&=&-\gh(\tbcP^{1}_{\alpha_{s}})~.
\end{array} 
\label{newtstatistics}
\eeq
It is now straightforward to check that the weak involution relations
\e{newweaklycommute}, \es{mixedweaklycommute}{tildeweaklycommute} follow from
the \mb{\tomegaone} nilpotency \e{newtbrstcharge}. An aesthetically drawback 
of the approach, is, that the tilde constraints \mb{\tcT_{A}} appear inside
two different generating expansions, namely \mb{\cS_{1}} and \mb{\tomegaone}.

\end{document}